\documentclass[11pt]{article}

\usepackage{amsmath,amsfonts,amssymb,epsfig,wick,a4}

\renewcommand\slash[1]{\not \! #1}

\newcommand{\rightslash}{\! \stackrel{\rightarrow}{\!\slash{\partial}}}
\newcommand{\leftslash}{\! \stackrel{\leftarrow}{\!\slash{\partial}}}
\newcommand{\leftDslash}{\! \stackrel{\leftarrow}{\slash{\! D}}}
\newcommand{\rightDslash}{\! \stackrel{\rightarrow}{\slash{\! D}}}

\begin{document}

%
\def\be{\begin{equation}}
\def\ee{\end{equation}}
\def\bea{\begin{eqnarray}}
\def\eea{\end{eqnarray}}
\def\rarr{\rightarrow}
\def\C{{\rm\kern.24em
    \vrule width.02em height1.4ex depth-.05ex
    \kern-.26em C}}
\def\N{{\rm I\kern-.18em N}}
\def\R{{\rm I\kern-.21em R}}
\def\Z{{\rm\kern.26em
    \vrule width.02em height0.5ex depth 0ex
    \kern.04em
    \vrule width.02em height1.47ex depth-1ex
    \kern-.34em Z}}
\def\d{{\rm\kern.22em
    \vrule width.02em height1.0ex depth0ex
    \kern-.24em d}}
\def\nn{\nonumber}
\def\fr{\frac}
\renewcommand\slash[1]{\not \! #1}
\newcommand\qs{\!\not \! q}
\def\del{\partial}
\def\gam{\gamma}
\newcommand\vphi{\varphi}
\def\tr{\mbox{tr}\,}
\newcommand\qt{\tilde{q}}
\newcommand\ns{\!\not \! n}
\newcommand\lto{\longrightarrow}
\newcommand\real{\mbox{Re}\,}
\newcommand\imag{\mbox{Im}\,}
\newcommand\nin{\noindent}
\newcommand\lbr{\left(}
\newcommand\rbr{\right)}
\newcommand\lbk{\left[}
\newcommand\rbk{\right]}
\newcommand\lbc{\left\{}
\newcommand\rbc{\right\}}
\newcommand\bmat{\boldmath}
\newcommand\ubmat{\unboldmath}
\newcommand\mb{\mbox}
\newcommand{\vDelta}{\vec\Delta}
\newcommand{\vdelta}{\vec\delta}

\def\del{\partial}
\def\pbar{\bar{p}}
\def\zbar{\bar{z}}
\def\rhobar{\bar{\rho}}
\def\kf{{\bf k}}
\def\qf{{\bf q}}
\def\lf{{\bf l}}
\def\vf{{\bf v}}
\def\wf{{\bf w}}
\def\Af{\mbox{\bf A}} 
\def\Vf{\mbox{\bf V}} 
\def\Ff{\mbox{\bf F}} 
\def\ca{{\cal A}}
\def\cA{{\cal A}}
\def\cb{{\cal B}}
\def\ccal{{\cal C}}
\def\dcal{{\cal D}}
\def\cD{{\cal D}}
\def\cS{{\cal S}}
\def\cO{{\cal O}}
\def\cP{{\cal P}}
\def\cx{{\cal X}}
\def\cy{{\cal Y}}
\def\cz{{\cal Z}}
\def\rhobar{\bar{\rho}}
\def\C{{\rm\kern.24em
    \vrule width.02em height1.4ex depth-.05ex
    \kern-.26em C}}
\def\N{{\rm I\kern-.18em N}}
\def\O{{\rm\kern.24em
    \vrule width.02em height1.45ex depth-.05ex
    \kern-.26em O}}
\def\P{{\rm I\kern-.25em P}}
\def\R{{\rm I\kern-.21em R}}
\def\Z{{\rm\kern.26em
    \vrule width.02em height0.5ex depth 0ex
    \kern.04em
    \vrule width.02em height1.47ex depth-1ex
    \kern-.34em Z}}
\newcommand\spommi{{\mbox{\scriptsize\P}}}
\newcommand\soddi{{\mbox{\scriptsize\O}}}
\newcommand\tpommi{{\mbox{\tiny\P}}}
\newcommand\toddi{{\mbox{\tiny\O}}}
\newcommand\sbfkl{{\mbox{\scriptsize BFKL}}}
\newcommand\tbfkl{{\mbox{\tiny BFKL}}}
\def\nn{\nonumber}
\def\fr{\frac}
\renewcommand\slash[1]{\not \! #1}

\begin{titlepage}
\begin{flushright}
HD-THEP-04-18\\
IFUM-790-FT\\
ECT$^*$-05-19\\
hep-ph/0604087
\\
\end{flushright}
\vfill
\begin{center}
\boldmath
{\LARGE{\bf Towards a Nonperturbative Foundation}}\\[.2cm]
{\LARGE{\bf of the Dipole Picture:}}\\[.2cm]
{\LARGE{\bf II. High Energy Limit}}
\unboldmath
\end{center}
\vspace{1.2cm}
\begin{center}
{\bf \Large
Carlo Ewerz\,$^{a,b,c,1}$, Otto Nachtmann\,$^{a,2}$
}
\end{center}
\vspace{.2cm}
\begin{center}
$^a$
{\sl
Institut f\"ur Theoretische Physik, Universit\"at Heidelberg\\
Philosophenweg 16, D-69120 Heidelberg, Germany}
\\[.5cm]
$^b$
{\sl
Dipartimento di Fisica, Universit{\`a} di Milano and INFN, Sezione di Milano\\
Via Celoria 16, I-20133 Milano, Italy}
\\[.5cm]
$^c$
{\sl
ECT\,$^*$, Strada delle Tabarelle 286, 
I-38050 Villazzano (Trento), Italy}
\end{center}                                                                
\vfill
\begin{abstract}
\noindent
This is the second of two papers in which we study real and
virtual photon-proton scattering in a nonperturbative framework.
In the first paper we have identified the leading contributions
to this process at high energies and have derived expressions
for them which take into account the renormalisation
of the photon-quark-antiquark vertex.
In the present paper we investigate the approximations and 
assumptions that are necessary to obtain the 
dipole model of high energy scattering from
the results derived in the first paper.
We discuss the gauge invariance of different contributions 
to the scattering amplitude and point out some subtleties 
related to gauge invariance in the correct definition of a 
perturbative photon wave function. 
As a phenomenological consequence of the dipole picture 
we derive a bound on the ratio of the cross sections 
for longitudinally and transversely polarised photons. 
This bound is independent of any particular model for the 
dipole-proton cross section and allows one to test the validity 
of the assumptions leading to the dipole 
picture in particular at low photon virtualities. 
We conclude that the naive dipole model formula should be
supplemented by two additional terms which can potentially
become large at small photon virtualities.
\vfill
\end{abstract}
\vspace{5em}
\hrule width 5.cm
\vspace*{.5em}
{\small \noindent
$^1$ email: Ewerz@ect.it \\
$^2$ email: O.Nachtmann@thphys.uni-heidelberg.de
}
\end{titlepage}

\section{Introduction}
\label{sec:intro2}

In this paper we continue our study of real and virtual
photon-proton scattering in a nonperturbative framework
which we have initiated in \cite{Ewerz:2004vf}, hereafter
referred to as I. Sections of that paper will be referred to as I.2.1 
and equations as (I.1) etc. We use the same notation and conventions 
as in I, and definitions made there will not be repeated here. In the 
present paper we include only references that are directly relevant 
to the material treated here, for a more extensive list of references 
we refer the reader to I. 

In the first paper we have classified the contributions to the process 
of real or virtual photon-proton scattering, see section I.2.1, figure I.2. 
We have identified the leading contributions 
at high energies in sect.\ I.2.2 and have derived expressions
for them which take into account the renormalisation
of the photon-quark-antiquark vertex. 
In the present paper we study in detail the high energy limit. 
We investigate the approximations 
and assumptions that are necessary to derive 
the dipole model of high energy scattering from
the results obtained in the first paper.
We point out that the naive dipole model formula should be
supplemented by two additional terms which can potentially
become large at small photon virtualities.

Gauge invariance is known to place strong constraints on 
current-induced scattering amplitudes. We find it therefore 
interesting to study the gauge invariance of different contributions to
the scattering amplitude. It turns out that at finite energies 
the part of the Compton amplitude from which the dipole picture 
emerges is not gauge invariant separately. This also leads to 
subtleties in the definition of the wave function of longitudinally 
polarised photons which we discuss in some detail. 

As a key result of the present paper we identify the assumptions 
and approximations that lead us from the most general description 
of the Compton amplitude to the usual dipole picture. We consider it 
a very important task to test the validity of those 
assumptions and approximations for the experimentally accessible 
ranges of the kinematic variables. In the dipole picture the total 
photon-proton cross section is a convolution of the square of the photon 
wave function with the reduced cross 
section describing the scattering of a colour dipole off the proton. 
At present, the reduced cross section cannot be derived from first 
principles and one has to retreat to models based on ideas 
like saturation. One finds that the currently available data permit 
considerable freedom for models of the reduced cross section. If one 
is to find phenomenological tests of the dipole picture (and hence of 
the assumptions on which it is based) one should thus concentrate 
on the other convolution factor in the dipole formula, that is on the 
photon wave function. This motivates us to study in some detail the 
properties of the wave function. 

As an immediate consequence of the dipole formula for the total 
virtual photon-proton cross section we derive a bound on the ratio $R$ of 
the cross sections for longitudinally and transversely polarised 
photons that results only from the respective photon wave functions. 
Although there are only few data points available at energies 
sufficiently high for one reasonably to expect the dipole model to be 
applicable the bound turns out to give some indication as to the 
photon virtualities below which the dipole model becomes 
questionable. In this region the two additional terms mentioned above 
should become relevant. 

In the dipole picture one talks about a quark-antiquark pair scattering 
on a target. If we want to be rigorous we have to face the question how 
far we can go treating quarks as asymptotic states. Of course, quarks are 
confined and do not have a mass shell. In this paper we shall nevertheless 
assume -- as is usually done -- that we can treat quarks and antiquarks 
as asymptotic particles. As we shall show we can then make a precise 
connection between the photon induced reaction and the reaction where 
the photon is replaced by a superposition of quark-antiquark states. 
We leave the investigation of the realistic case of quarks which have 
no mass shell for future work. At present we may give the 
following simple argument. Let us assume that the two-point functions 
for quarks have a K\"allen-Lehmann representation with 
standard properties, but with no mass shell. This is indicated by calculations 
using lattice as well as Schwinger-Dyson equation methods, see for example 
\cite{Alkofer:2003jj}. 
We can then think that the mass-distribution function of a quark propagator 
may have a strong peak at some value $m_q$ which we shall call the quark mass. 
If the peak is approximated by a $\delta$-function we have effectively a quark 
mass shell.

This second paper is organised as follows. We start by discussing 
the high energy limit of the parts of the Compton amplitude 
that are leading at high energies and identify their asymptotic behaviour 
in section \ref{highenergysect}. 
In section \ref{gaugesect} we investigate the separate gauge invariance 
of different parts of the amplitude. 
The derivation of the photon wave function is performed in section 
\ref{wavefunctsec} where we also discuss the gauge non-invariance 
of this object and its consequences for the choice of the photon 
polarisation vectors. In section \ref{sec:genphotons} we generalise 
our findings to the case of more general amplitudes with incoming and 
outgoing photons and define dipole states. In section \ref{sec:dippictdis} 
we finally obtain the usual dipole picture and identify all underlying 
assumptions and approximations. Here we also study the properties 
of the photon wave function. In section \ref{sec:phencons} we 
derive a bound on the ratio of the cross sections 
for longitudinally and transversely polarised virtual photons that needs to 
be satisfied if the usual dipole picture is valid, and compare this bound 
to the available experimental data. 
Our conclusions are presented in section \ref{conclsect}. 
In appendix \ref{appuseful} we collect some technical details relevant 
to the discussion of the high energy limit in sections \ref{highenergysect} 
and \ref{wavefunctsec}. Appendix \ref{appdiffcontgauge} supplements 
section \ref{gaugesect} with further discussion of the separate gauge 
invariance of different parts of the Compton amplitude 
which are subleading at high energies. 

\section{The high energy limit}
\label{highenergysect}

In this section we shall study the high energy limit of the real or virtual 
Compton amplitude (I.\ref{2.3}), 
$|\mathbf{q}|\rightarrow\infty$ with $Q^2$ fixed.
Our aim is to obtain eventually the usual dipole picture of high energy 
photon-proton scattering. As was discussed in section
I.\ref{sec:relsize} in the high energy limit
the two amplitudes $\mathcal{M}^{(a)}$ and
$\mathcal{M}^{(b)}$ give the leading contribution
to the full amplitude. However, it is only the amplitude
$\mathcal{M}^{(a)}$ from which we expect the usual dipole
picture to emerge. In the present section we will
therefore study mainly that amplitude
$\mathcal{M}^{(a)}$ (I.\ref{2.8}), (I.\ref{2.20}). But it should be
kept in mind that at low photon virtuality an important
correction to the dipole picture can arise from the
amplitude $\mathcal{M}^{(b)}$, see the discussion in section
I.\ref{sec:relsize}.

We will mainly be concerned here with the term
$\mathcal{M}^{(a,1)}$ (I.\ref{2.33}) of $\mathcal{M}^{(a)}$
which we will find to be leading
in the high energy limit, whereas the terms
$\mathcal{M}^{(a,2)}$ to $\mathcal{M}^{(a,4)}$ are subleading.
In the amplitude $\mathcal{M}^{(a,1)}$ by definition only the
momentum $k'^{(1)}$ occurs but not the momenta
$k'^{(2)}$ to $k'^{(4)}$ which correspond to the
other three parts of amplitude $\mathcal{M}^{(a)}$, 
see (I.\ref{B.20})-(I.\ref{B.23}). 
Therefore we now simplify our notation and write
$k'$ instead of $k^{\prime (1)}$ and $\mathbf{k}'$
instead of $\mathbf{k}'^{(1)}$.

\begin{figure}[ht]
\begin{center}
\includegraphics[width=8.6cm]{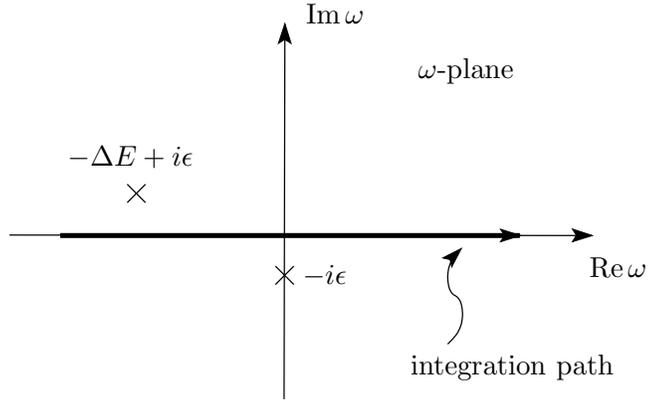}
\caption{The singularities of the integrand of $\mathcal{M}^{(a,1)}$ 
(\ref{2.33again}) in the
$\omega$-plane. There are poles at $\omega=-i\epsilon$ and
$\omega=-\Delta E +i\epsilon$.
\label{fig7}}
\end{center}
\end{figure}

Let us now consider the term $\mathcal{M}^{(a,1)}$ which 
we have obtained in (I.\ref{2.33}) in the form
\begin{eqnarray}\label{2.33again}
\mathcal{M}^{(a,1)\mu\nu}_{s^{\prime}s}(p^{\prime},p,q)
\!&=&\!
\frac{1}{2\pi} \sum_qQ^2_q
\int\frac{d\omega}{\omega+i\epsilon}\int\frac{d^3k}{(2\pi)^32k^0}
(q^0-k^0-k^{\prime 0}-\omega+i\epsilon)^{-1}
\nn \\
&&{} 
(2k^{\prime 0})^{-1}
\sum_{r^{\prime},r}\langle \gamma(q^{\prime},\mu),
p(p^{\prime},s^{\prime})|\mathcal{T}^{(a)}|\bar{q}(k^{\prime}_\omega,r^{\prime}),
q(k_\omega,r),p(p,s)\rangle
\nn \\
&&{}
\bar{u}_r(k)\left\{\Gamma^{(q)\nu}(k,-k^{\prime})
+ \sum_{q'}\int K^{(q,q')}S^{(q')}_F\Gamma^{(q')\nu}S^{(q')}_F\right\}
v_{r^{\prime}}(k^{\prime})\,,
\nn\\
&&{}
\end{eqnarray}
and let us in particular concentrate on 
the integration over $\omega$ in the complex $\omega$-plane.
Recall that $\omega$ is the off-shell part of the energy of the quark 
on the left-hand side of the diagram in figure I.\ref{fig4}a, see 
(I.\ref{B.3}). 
We expect the matrix element of $\mathcal{T}^{(a)}$ to be regular
in the vicinity of $\omega=0$.
Thus we have the two explicit pole factors in the integrand (\ref{2.33again})
with the poles in $\omega$ situated at $\omega=-i\epsilon$ and at
$\omega=-\Delta E +i\epsilon$
with
\be
\label{defDeltaE}
\Delta E =k^0+k^{\prime 0}-q^0\,,
\ee
see figure \ref{fig7}. 
One finds that always $\Delta E > 0$. 
The poles will contribute a factor of order $1/q^0$ in the amplitude unless
there is a pinching of the two singularities, that is for
\begin{equation}\label{3.1}
-\Delta E = q^0-k^0-k^{\prime 0}\rightarrow 0 \,.
\end{equation}
The conditions for this to happen are easily derived and in principle well known. To make
the article self-contained we discuss them here in detail.
We have from (I.\ref{B.20})
\begin{eqnarray}\label{3.2}
\mathbf{q}-\mathbf{k}-\mathbf{k}^{\prime}&=&0\,,
\nonumber\\
q^0&=& \sqrt{\mathbf{q}^2-Q^2}
\nonumber\\
k^0&=&\sqrt{\mathbf{k}^2+m^2_q}\,,\nonumber\\
k^{\prime 0}&=&\sqrt{\mathbf{k}'^2+m^2_q}\,.
\end{eqnarray}
We set
\begin{eqnarray}\label{3.3}
\mathbf{k}&=& \alpha \mathbf{q}+\mathbf{k}_T,\nonumber\\
\mathbf{k}^{\prime}&=&(1-\alpha)\mathbf{q}-\mathbf{k}_T\,,
\end{eqnarray}
where we suppose
\begin{equation}\label{3.4}
\mathbf{q} \, \mathbf{k}_T=0\,,
\end{equation}
that is, longitudinal and transverse directions are defined
relative to $\mathbf{q}$. We then get
\begin{eqnarray}\label{3.5}
\Delta E &=&{}
(q^0+k^0+k^{\prime 0})^{-1}
\left[(k^0+k^{\prime 0})^2-(q^0)^2\right]\nonumber\\
&=&{}
(q^0+k^0+k^{\prime 0})^{-1}
\left[Q^2+2m^2_q+2k^0k^{\prime 0}-2\mathbf{k}\,\mathbf{k}^{\prime}\right]
\nonumber\\
&=&{}
2(q^0+k^0+k^{\prime 0})^{-1}
\bigg[\,\frac{1}{2}\,Q^2+(\alpha^2\mathbf{q}^2+\mathbf{k}^2_T+m^2_q)^{1/2} 
\nonumber\\
&&{}
\left((1-\alpha)^2\mathbf{q}^2+\mathbf{k}^2_T+m^2_q\right)^{1/2}
-\alpha(1-\alpha)\mathbf{q}^2+\mathbf{k}^2_T +m_q^2 \bigg]\,.
\end{eqnarray}
For large $|\mathbf{q}|$ the energy difference $\Delta E$
becomes small only if the terms proportional to $\mathbf{q}^2$ in the square
brackets cancel. This happens for $\alpha\neq 0,1$ if and only if
\begin{equation}\label{3.6}
|\alpha(1-\alpha)|-\alpha(1-\alpha)=0\,,
\end{equation}
that is for
\begin{equation}\label{3.7}
0<\alpha<1\,.
\end{equation}
Thus we get a pinch and a large contribution only if the splitting of
the photon $q$ is into a quark-antiquark pair where the longitudinal
momenta are both in the direction $\mathbf{q}$.
Note again that we have used no arguments related to perturbation
theory to arrive at this conclusion.

\begin{figure}[ht]
\begin{center}
\includegraphics[width=8cm]{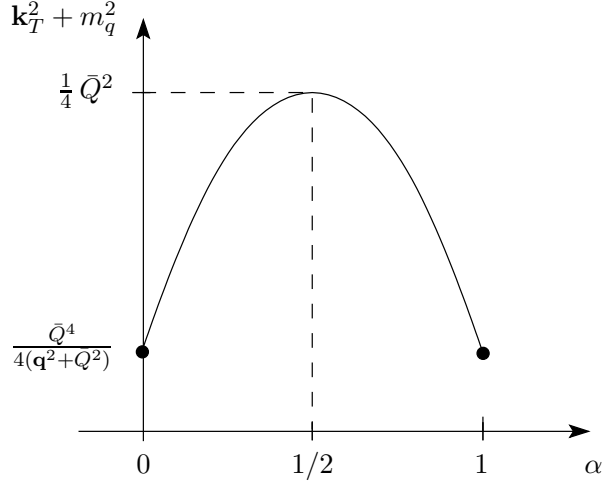}
\caption{The upper limit of {\bf k}$^2_T+m^2_q$ corresponding to
$\tilde{m}^2(\alpha,${\bf k}$ _T)\leq\bar{Q}^2$ (see (\ref{3.11}), (\ref{3.12})).
\label{fig8}}
\end{center}
\end{figure}

We can rewrite (\ref{3.5}) in yet another form:
\begin{equation}\label{3.8}
\Delta E =
\frac{Q^2+\tilde{m}^2(\alpha,\mathbf{k}_T)}{q^0+k^0+k^{\prime 0}}\,,
\end{equation}
where
\begin{eqnarray}\label{3.9}
\lefteqn{\tilde{m}^2(\alpha,\mathbf{k}_T)=2(\mathbf{k}^2_T+m^2_q)}
\nonumber
\\
&&
\left[ \, 1
+\frac{\left(\alpha^2+(1-\alpha)^2\right)\mathbf{q}^2+\mathbf{k}^2_T+m^2_q}
{(\alpha^2\mathbf{q}^2+\mathbf{k}^2_T+m^2_q)^{\frac{1}{2}}
\left((1-\alpha)^2\mathbf{q}^2+\mathbf{k}^2_T+m^2_q\right)^{\frac{1}{2}}
+\alpha(1-\alpha)\mathbf{q}^2}\right] .
\end{eqnarray}
In figure \ref{fig8} we show the range in $\mathbf{k}^2_T+m^2_q$
corresponding to
\begin{equation}\label{3.10}
\tilde{m}^2(\alpha,\mathbf{k}_T)\leq\bar{Q}^2 \,,
\end{equation}
where $\bar{Q}$ is some chosen fixed mass scale for which we always 
suppose $\bar{Q}^2 \ge 4 m^2_q$ and $0\leq \alpha\leq 1$. 
We see that (\ref{3.10}) corresponds to
\begin{equation}\label{3.11}
\mathbf{k}^2_T+m^2_q \,\leq \,
\frac{\bar{Q}^4}{4(\mathbf{q}^2+\bar{Q}^2)} 
\end{equation}
for $\alpha=0$ and $\alpha=1$ , and to  
\begin{equation}\label{3.12}
\mathbf{k}_T^2+m^2_q \,\leq\,\frac{1}{4}\bar{Q}^2
\end{equation}
for $\alpha=1/2$.
Thus the pinch condition in the form
\begin{equation}\label{3.13}
\Delta E \leq \frac{Q^2+\bar{Q}^2}{q^0+k^0+k^{\prime 0}}
\end{equation}
is only fulfilled for $\alpha\rightarrow 0$ and $1$ for very small
$\mathbf{k}^2_T+m^2_q$ whereas for 
$\alpha\approx 1/2 \quad \mathbf{k}^2_T+m^2_q$
ranges up to $\mathcal{O}(\bar{Q}^2)$.
This is discussed in more detail in appendix \ref{appIIApinch}. 

Taking now only the contribution from the pinch in (\ref{2.33again})
we obtain asymptotically for large $|\mathbf{q}|$
\begin{eqnarray}\label{3.14}
\mathcal{M}^{(a,1)\mu\nu}_{s^\prime s}(p^\prime, p,q)
&=& i \sum_qQ^2_q
\int \frac{d^3k}{(2\pi)^32k^02k^{\prime 0}} \,
(\Delta E)^{-1} \,
\theta\left(\bar{Q}^2-\tilde{m}^2(\alpha,\mathbf{k}_T)\right)
\nonumber
\\
&&{}
\sum_{r^\prime ,r} \, \langle \gamma(q^\prime,\mu),
p(p^\prime,s^\prime)|\mathcal{T}^{(a)}|\bar{q}(k^{\prime},r^\prime),
q(k,r),p(p,s) \rangle
\\
&& {}
\bar{u}_r(k)
\left\{\Gamma^{(q)\nu}(k,-k^{\prime})
+\sum _{q'} \int K^{(q,q')}S^{(q')}_F\Gamma^{(q')\nu}S^{(q')}_F \right\}
v_{r^\prime}(k^\prime )\,.
\nonumber
\end{eqnarray}
Here the four-momenta $k,k^{\prime}$ are given as in (\ref{3.2}) and
(\ref{3.3}) with $0\leq \alpha\leq 1$. We have inserted a $\theta$-function to
ensure the pinch condition in the form (\ref{3.13}).
This limits $\mathbf{k}^2_T$ as shown in figure \ref{fig8}; see also 
the discussion in appendix \ref{appIIApinch}. 

We note that in $\mathcal{M}^{(a,2)}$ to $\mathcal{M}^{(a,4)}$, 
(I.\ref{2.23})-(I.\ref{2.25}), 
no pinch of the explicit pole terms in $\omega$ occurs
and we expect these terms to be suppressed relative to the leading term
(\ref{3.14}) by a factor of the order 
\begin{equation}\label{3.15}
\frac{\Delta E}{q^0}\approx\frac{Q^2+\bar{Q}^2}{(q^0)^2}
\end{equation}
for large $|\mathbf{q}|$. 

In (\ref{3.14}) we already see something like the dipole picture emerging.
The photon splits into a $q\bar{q}$ pair with both the quark and the 
antiquark on the mass shell $m_q$. This on-shell pair interacts with 
the proton to produce the final state.
But the photon `wave function' is not simply the vertex function.
There is also the piece containing the kernel $K^{(q,q')}$ which in a way 
represents rescattering corrections of the $q\bar{q}$-pair. 

We should keep in mind also the contribution from the 
amplitude $\mathcal{M}^{(b)}$. In complete analogy to the discussion 
above we start from (I.\ref{2.45}) and 
find a pinch singularity also for that amplitude. The 
leading term in $\mathcal{M}^{(b)}$ for large $|\mathbf{q}|$ then 
asymptotically becomes 
\begin{eqnarray}\label{3.17}
\mathcal{M}^{(b,1)\mu\nu}_{s^\prime s}(p^\prime, p,q)
&=& i \sum_{q'',q} Q_{q''} Q_q 
\int \frac{d^3k}{(2\pi)^32k^02k^{\prime 0}} \,
(\Delta E)^{-1} \,
\theta\left(\bar{Q}^2-\tilde{m}^2(\alpha,\mathbf{k}_T)\right)
\nonumber
\\
&&{}
\sum_{r^\prime ,r} \, \langle \gamma(q^\prime,\mu),
p(p^\prime,s^\prime)|\mathcal{T}^{(q'',b)}|\bar{q}(k^{\prime},r^\prime),
q(k,r),p(p,s) \rangle
\\
&& {}
\bar{u}_r(k)
\left\{\Gamma^{(q)\nu}(k,-k^{\prime})
+\sum _{q'} \int K^{(q,q')}S^{(q')}_F\Gamma^{(q')\nu}S^{(q')}_F \right\}
v_{r^\prime}(k^\prime )\,.
\nonumber
\end{eqnarray}

The discussion of the high energy limit presented here is consistent 
with the well-known space-time picture of photon-hadron scattering 
at high energies that is the origin of the dipole picture. 
The quantity $\Delta E$ is the energy imbalance between the photon 
and the quark-antiquark pair, see (\ref{defDeltaE}). 
According to the energy-time uncertainty relation 
the quark-antiquark pair into which the photon fluctuates
can therefore live at most for a time $(\Delta E)^{-1}$. 
In the high energy limit we have $\Delta E \to 0$ (for fixed $Q^2$) 
such that the maximal lifetime of the pair 
increases with increasing energy. In the rest frame of the proton, 
for example, the photon typically 
splits into the quark-antiquark pair already at a large 
distance from the proton. The actual interaction of the pair with 
the proton then takes place on much shorter timescales. Some 
typical values of $\Delta E$ in realistic situations are given 
in appendix \ref{appuseful}. 

In order to extract the dipole picture explicitly from the formulae obtained 
above it will turn out to be useful to address first the problem of gauge 
invariance of various parts of the amplitude. 

\section{Gauge invariance}
\label{gaugesect}

In this section we want to study the question of electromagnetic gauge 
invariance of the real or virtual Compton amplitude. 
In particular we are interested in the
question whether the different parts of the amplitude arising from
the decompositions that we have performed up to this point 
are separately gauge invariant. In the following we will therefore 
consider the general amplitude without imposing the high 
energy limit. Only in section \ref{subsec:gaugehighenergy} 
we will come back to this limit. 

The full Compton amplitude (I.\ref{2.3}) obviously exhibits electromagnetic
gauge invariance. More precisely, electromagnetic current conservation
implies
\be
\label{generalgaugeinv}
q_\nu \mathcal{M}^{\mu\nu}_{s^{\prime}s}(p^{\prime},p,q) =0 \,.
\ee
Here and in the following we always consider current conservation 
for the incoming photon, hence the contraction with $q_\nu$. 
The outgoing photon can be treated analogously. 

\subsection{Skeleton decomposition of the full amplitude}
\label{subsec:gaugefullampl}

We first consider the general decomposition of the amplitude
derived in section I.\ref{nonpertsect} and do not yet impose the
high energy limit. 

Again we are in particular interested in the contributions 
$\mathcal{M}^{(a)}$ and $\mathcal{M}^{(b)}$ 
represented by figures I.\ref{fig2}a and I.\ref{fig2}b, which are 
not suppressed at high energies. Let us first consider the contribution 
$\mathcal{M}^{(a)}$, that is the term from which the usual dipole picture 
emerges at high energies, as we will see in the following sections. 

We start from expression (I.\ref{2.8}) in which we recognise 
that the $q$-dependence of $\mathcal{M}^{(a)}$ is fully contained in 
the term $A^{(q)\mu\nu}(q)$ given in (I.\ref{2.8a}). We therefore 
consider the expression 
\bea
\label{qtimesAa}
q_\nu A^{(q)\mu\nu}(q) &=&
\int d^4 x \, \left( q_\nu e^{-iqx} \right)
\mathrm{Tr} \left[ \gamma^\mu S_F^{(q)}(0,x;G) \gamma^\nu
S_F^{(q)}(x,0;G) \right]
\nn \\
&=&{}
\int d^4 x \, \left( i \frac{\partial}{\partial x^\nu} \, e^{-iqx} \right)
\mathrm{Tr} \left[ \gamma^\mu S_F^{(q)}(0,x;G) \gamma^\nu
S_F^{(q)}(x,0;G) \right]
\nn \\
&=&{}
- \int d^4 x \,  e^{-iqx} \, i \frac{\partial}{\partial x^\nu}
\mathrm{Tr} \left[ \gamma^\mu S_F^{(q)}(0,x;G) \gamma^\nu
S_F^{(q)}(x,0;G) \right] \,,
\eea
where we have integrated by parts. This contains the following expression, 
in which we can complete the derivatives to full covariant derivatives 
in order to obtain inverse quark propagators,  
\bea
\label{qnuAcontr}
\lefteqn{
i \frac{\del}{\del x^\nu} 
\mathrm{Tr} \left[ \gamma^\mu S_F^{(q)}(0,x;G) \gamma^\nu
S_F^{(q)}(x,0;G) \right]
}
\nn \\
&=&{}
\mathrm{Tr} \left[ \gamma^\mu S_F^{(q)}(0,x;G)
\left( i \leftslash_x + i \rightslash_x \right) S_F^{(q)}(x,0;G)
\right]
\nn \\
&=&{}
\mathrm{Tr} \left\{ \gamma^\mu S_F^{(q)}(0,x;G)
\left[ \left(i \leftDslash_x +m_q^{(0)} \right)
+ \left( i \rightDslash_x -m_q^{(0)}\right) \right] 
S_F^{(q)}(x,0;G) \right\} \,.
\eea
Inserting this back in (\ref{qtimesAa}) 
we obtain with (I.\ref{A.4}) and  (I.\ref{A.4b}) 
\bea
q_\nu A^{(q)\mu\nu}(q) &=&
- \int d^4 x \,   e^{-iqx} \,
\mathrm{Tr} \left[ \gamma^\mu \delta^{(4)}(x)
S_F^{(q)}(x,0;G) - \gamma^\mu \delta^{(4)}(x) S_F^{(q)}(0,x;G)
\right]
\nn \\
&=& 0 \,.
\eea
Consequently, the amplitude $\mathcal{M}^{(a)}$ is separately
gauge invariant, 
\be
\label{Magaugeinv}
q_\nu \mathcal{M}^{(a) \mu\nu}_{s^{\prime}s}(p^{\prime},p,q) =0 \,.
\ee

Next we turn to the amplitude $\mathcal{M}^{(b)}$. 
The factor in that amplitude which is relevant for our considerations here 
is according to (I.\ref{Mbexpl}) given by $A_b^{(q)\nu}(q)$, 
see (I.\ref{C.3}). Its contraction with $q_\nu$ gives 
\bea
\label{Abpartial}
q_\nu A_b^{(q)\nu}(q) &=& 
q_\nu \int d^4x \, e^{-iqx} \, 
\mathrm{Tr} \left[ \gamma^\nu  S_F^{(q)}(x,x;G) \right]
\nn \\
&=&{}
- \int d^4x \,  e^{-iqx} \,
 i \frac{\del}{\del x^\nu} 
\mathrm{Tr} \left[ \gamma^\nu  S_F^{(q)}(x,x;G) \right]
\,,
\eea
where we have again integrated by parts. The derivative of the trace 
then becomes 
\bea
\label{derivtraceMb}
i \frac{\del}{\del x^\nu} 
\mathrm{Tr} \left[ \gamma^\nu  S_F^{(q)}(x,x;G) \right]
&=&
\mathrm{Tr} \left[ i \rightslash_x S_F^{(q)}(x,x;G) \right]
\nn \\
&=&{}
\left.
\mathrm{Tr} \left[ i \rightslash_x S_F^{(q)}(x,x';G) + 
S_F^{(q)}(x,x';G) i \leftslash_{x'} 
\right] \right|_{x'=x} 
\eea
Here we can again complete full covariant derivatives and 
obtain for (\ref{Abpartial}) with  (I.\ref{A.4}) and  (I.\ref{A.4b}) 
\bea
\label{fullcovderinMb}
q_\nu A_b^{(q)\nu}(q) &=& 
- \int d^4x \, e^{-iqx} \, 
\mathrm{Tr} \left[ \left( i \rightDslash_x -m_q^{(0)} \right) 
S_F^{(q)}(x,x';G) 
\right.
\nn \\
&&{} \hspace*{3cm}
\left. 
+ \, S_F^{(q)}(x,x';G) 
\left. \left( i \leftDslash_{x'} + m_q^{(0)} \right) 
\right] \right|_{x'=x} 
\nn \\
&=&{}
\left.
\int d^4x \, e^{-iqx} \,
\mathrm{Tr} \left[ \delta^{(4)} (x -x') - \delta^{(4)} (x -x') \right] 
\right|_{x'=x} 
\nn \\
&=&{} 
0 \,.
\eea
Hence also the amplitude $\mathcal{M}^{(b)}$ is separately gauge 
invariant, 
\be
\label{Mbgaugeinv}
q_\nu \mathcal{M}^{(b) \mu\nu}_{s^{\prime}s}(p^{\prime},p,q) =0 \,.
\ee

Finally we consider the remaining parts of the amplitude 
in the skeleton decomposition, 
$\mathcal{M}^{(c)}$ -- $\mathcal{M}^{(g)}$, 
which are subleading in the high energy limit. 
Here we find that $\mathcal{M}^{(c)}$ taken together 
with $\mathcal{M}^{(d)}$ is gauge invariant, 
\be
\label{Mcdgaugeinv}
q_\nu \left( 
\mathcal{M}^{(c) \mu\nu}_{s^{\prime}s}(p^{\prime},p,q) 
+ \mathcal{M}^{(d) \mu\nu}_{s^{\prime}s}(p^{\prime},p,q) 
\right) 
=0 \,,
\ee
whereas each of the remaining three amplitudes $\mathcal{M}^{(e)}$, 
$\mathcal{M}^{(f)}$ and $\mathcal{M}^{(g)}$ is gauge invariant 
by itself, 
\be
\label{Mefggaugeinv}
q_\nu \mathcal{M}^{(e) \mu\nu}_{s^{\prime}s}(p^{\prime},p,q) =
q_\nu \mathcal{M}^{(f) \mu\nu}_{s^{\prime}s}(p^{\prime},p,q) =
q_\nu \mathcal{M}^{(g) \mu\nu}_{s^{\prime}s}(p^{\prime},p,q) =
0 \,.
\ee
The derivation of these results is presented in appendix 
\ref{appdiffcontgauge}. 

\subsection{Decomposition of the amplitudes ${\cal M}^{(a)}$ and ${\cal M}^{(b)}$}
\label{subsec:gaugedecMaMb}

Among the contributions to the skeleton decomposition of the full amplitude 
we have identified $\mathcal{M}^{(a)}$ and $\mathcal{M}^{(b)}$ as 
the ones that will be leading at high energies. We now turn to the spin 
sum decomposition which we have performed for these two amplitudes. 
At this point we still consider the general decomposition of 
these amplitudes performed before taking the high energy limit. 

We have decomposed $\mathcal{M}^{(a)}$ into the 
four terms $\mathcal{M}^{(a,j)}$, $j=1,\dots,4$, 
see (I.\ref{2.20})-(I.\ref{2.25}) 
and the graphical illustration in figure I.\ref{fig4}. 
According to (\ref{Magaugeinv}) we clearly have 
\be
q_\nu \sum_{j=1}^4
\mathcal{M}^{(a,j) \mu\nu}_{s^{\prime}s}(p^{\prime},p,q) = 0\,.
\ee

Since we have obtained the dipole picture only from $\mathcal{M}^{(a,1)}$, 
it is now interesting to see whether that part is gauge invariant by itself. 
We use the expression  (\ref{2.33again}) for $\mathcal{M}^{(a,1)}$ 
and remember that the terms in curly brackets in that equation equal 
$Z_q \gamma^\nu$ via (I.\ref{2.31}). Hence the contraction
$q_\nu \mathcal{M}^{(a,1)\mu\nu}_{s^{\prime}s}(p^{\prime},p,q)$
contains as the relevant factor in the integrand the expression 
\bea
\label{qZubarv}
q_\nu \bar{u}_r(k) Z_q \gamma^\nu v_{r^{\prime}}(k')
&=& Z_q \bar{u}_r(k) \! \slash{q} \, v_{r^{\prime}}(k')
\nn \\
&=&{}
Z_q \bar{u}_r(k) 
\left[ (\slash{k} -m_q ) + ( \slash{k}' +m_q) - (\Delta E )\gamma^0 \right] 
v_{r^{\prime}}(k')
\nn \\
&=&{}
- (\Delta E) Z_q \bar{u}_r(k) \gamma^0 v_{r^{\prime}}(k')
\,,
\eea
where we have used (\ref{defDeltaE}), (\ref{3.2}), 
and that by definition the spinors $u$ and $v$ 
are solutions of the free Dirac equation, 
\bea
\label{freeDirac}
\bar{u}_r(k) (\slash{k} -m_q) &=&{} 0 \,,
\nn \\
(\slash{k}' + m_q ) v_{r^{\prime}}(k') &=&{} 0\,.
\eea
Clearly, (\ref{qZubarv}) is in general different from zero. 
Inserting that term back into the full expression (\ref{2.33again}) 
for $\mathcal{M}^{(a,1)}$ we thus find that the resulting expression 
does in general not vanish. Hence 
the amplitude $\mathcal{M}^{(a,1)}$ does in 
general not fulfil electromagnetic gauge invariance separately. 

Similar considerations hold for each of the other three terms
$\mathcal{M}^{(a,j)}$, $j=2,3,4$, in (I.\ref{2.20}). None of them is 
gauge invariant separately. 

The situation for the spin sum decomposition (I.\ref{C.11}) 
of the amplitude ${\cal M}^{(b)}$ into the terms 
${\cal M}^{(b,j)}$ ($j=1,\dots,4$) is completely analogous. 
None of the terms ${\cal M}^{(b,j)}$ given in (I.\ref{C.13}) - (I.\ref{C.16}) 
is gauge invariant by itself. In particular, the term ${\cal M}^{(b,1)}$ 
(I.\ref{C.14}) gives rise to the same relevant factor 
(\ref{qZubarv}) proportional to $\Delta E$ 
when contracted with the photon momentum. 

Finally, we turn to the decomposition of the photon 
vertex $Z_q \gamma^\nu$ in the amplitude ${\cal M}^{(a,1)}$ 
into the renormalised vertex term 
and the rescattering term, see (I.\ref{2.31}) and 
the expression in curly brackets in 
(\ref{2.33again}). It is easy to see that none of the two contributions 
to ${\cal M}^{(a,1)}$ arising from these two terms is gauge 
invariant separately. Let us consider for example the renormalised 
vertex term. Contracting ${\cal M}^{(a,1)}$ of (\ref{2.33again}) 
with $q_\nu$ the term containing the renormalised vertex gives 
rise to a factor 
\bea
\label{qnurenormvertex}
\lefteqn{q_\nu \bar{u}_r(k) \Gamma^{(q)\nu} (k,-k') v_{r^{\prime}}(k') }
\nn \\
&=&{}
\bar{u}_r(k) (k+k')_\nu \Gamma^{(q)\nu} (k,-k') v_{r^{\prime}}(k') 
- (\Delta E) \bar{u}_r(k) \Gamma^{(q)0} (k,-k') v_{r^{\prime}}(k') 
\nn \\
&=&{}
\bar{u}_r(k) \left[ S_F^{(q)-1}(k) -  S_F^{(q)-1}(-k') \right]v_{r^{\prime}}(k') 
- (\Delta E) \bar{u}_r(k) \Gamma^{(q)0} (k,-k') v_{r^{\prime}}(k') 
\nn \\
&=&{}
- (\Delta E) \bar{u}_r(k) \Gamma^{(q)0} (k,-k') v_{r^{\prime}}(k') \,,
\eea
where we have used the QED Ward identity and the Dirac equation 
(\ref{freeDirac}), and have assumed the existence of a mass shell for 
quarks, that is $S_F^{(q)-1}(m_q)=0$. 
The remaining expression does not vanish and hence the 
corresponding term in ${\cal M}^{(a,1)}$ is not gauge invariant 
separately. One can show that also the rescattering term is not 
gauge invariant separately. 

The same considerations hold for the analogous decomposition 
(I.\ref{2.31}) of the amplitude ${\cal M}^{(b,1)}$ as applied in 
(I.\ref{2.45}). 

\subsection{High energy limit of ${\cal M}^{(a,1)}$ and ${\cal M}^{(b,1)}$}
\label{subsec:gaugehighenergy}

So far we have seen that the amplitudes $\mathcal{M}^{(a)}$ and 
$\mathcal{M}^{(b)}$ are gauge invariant by themselves, 
but their most interesting parts in the high energy limit, 
$\mathcal{M}^{(a,1)}$ and $\mathcal{M}^{(b,1)}$ respectively, 
are not. Now we want to see what happens to the latter in the 
high energy limit. 

We start with the formula (\ref{3.14}) 
for $\mathcal{M}^{(a,1)\mu \nu}$ in the high energy limit. 
After contracting that amplitude with $q_\nu$ the terms 
in curly brackets together with the two spinors 
give exactly the expression (\ref{qZubarv}) proportional to $\Delta E$, 
and this factor cancels against $(\Delta E)^{-1}$ in (\ref{3.14}). 
Thus, $q_\nu \mathcal{M}^{(a,1)\mu \nu}$ has no $(\Delta E)^{-1}$ 
enhancement factor and is not of leading order at high energies. 
Consequently, we have for the ratio 
\be
\label{qnuMatoqvecM0}
\frac{q_\nu \mathcal{M}^{(a,1)\mu \nu}}{|\mathbf{q}|\mathcal{M}^{(a,1)\mu 0}}
= {\cal{O}}\left(\frac{\Delta E}{|\mathbf{q}|}\right) 
\,,
\ee
where we have chosen the component with $\nu=0$ in the denominator as a 
reference for the typical magnitude of the components of 
$\mathcal{M}^{(a,1)\mu \nu}$.
Since $\Delta E = {\cal{O}}(|\mathbf{q}|^{-1})$ (see (\ref{DeltaElargeq})) 
this ratio tends to zero as $|\mathbf{q}| \to \infty$. 
This can be considered 
an approximate gauge invariance of $\mathcal{M}^{(a,1)\mu \nu}$ 
in the high energy limit in the sense that $q_\nu \mathcal{M}^{(a,1)\mu \nu}$ 
is suppressed by two powers of $|\mathbf{q}|$ compared to the nominal 
power counting of the two factors of that contraction. 

It is a separate question whether $q_\nu \mathcal{M}^{(a,1)\mu \nu}$ 
itself vanishes in the high energy limit $|\mathbf{q}| \to \infty$. 
That is in general not the case. We will see this explicitly when we 
discuss the proper choice for the polarisation vector in the photon 
wave function for longitudinally polarised photons in 
section \ref{sec:longphotons} below. 

By the same argument as just given for $\mathcal{M}^{(a,1)}$ 
also the gauge invariance of $\mathcal{M}^{(b,1)}$ is restored 
asymptotically in the high energy limit, again in the sense that 
\be
\label{qnuMbtoqvecM0}
\frac{q_\nu \mathcal{M}^{(b,1)\mu \nu}}{|\mathbf{q}|\mathcal{M}^{(b,1)\mu 0}}
= {\cal{O}}\left(\frac{\Delta E}{|\mathbf{q}|}\right)
\,.
\ee

Finally, we turn to the decomposition (I.\ref{2.31}) of the photon 
vertex $Z_q \gamma^\nu$ into renormalised vertex and rescattering 
term. In (\ref{qnurenormvertex}) above we have seen that the 
contraction of the renormalised vertex term with $q_\nu$ gives rise 
to an expression proportional to $\Delta E$. We can hence use the 
same argument as given above for the amplitude ${\cal M}^{(a,1)}$: 
again, the $(\Delta E)^{-1}$ in (\ref{3.14}) is cancelled by the factor 
$\Delta E$ from (\ref{qnurenormvertex}). 
It then immediately follows from (I.\ref{2.31}) and (\ref{qZubarv}) 
that the same happens for the rescattering term. 
We conclude that approximate gauge invariance in the sense of 
(\ref{qnuMatoqvecM0}) holds 
separately for the renormalised vertex term and for the rescattering 
term. The same conclusions hold also for the corresponding 
decomposition of the amplitude ${\cal M}^{(b,1)}$. 

Note that the photon-quark-antiquark vertex $Z_q \gamma^\nu$ 
does not involve any momentum arguments. However, its separation 
(I.\ref{2.31}) into renormalised vertex and rescattering terms makes 
it necessary to assign momentum arguments for the quark lines in 
the latter two. 
In (\ref{qnurenormvertex}) we have made use of the particular 
choice (I.\ref{2.32}). Other choices for the momentum assignment 
would have given different results not necessarily giving rise to 
the separate approximate gauge invariance for the renormalised 
vertex and the rescattering term in the high energy limit. 
Therefore the separate approximate 
gauge invariance of the two terms {\sl a posteriori } justifies the 
momentum assignment (I.\ref{2.32}). 

\section{Photon wave function at leading order}
\label{wavefunctsec}

\subsection{Definitions and general formulae}
\label{subsec:gammawavefunctgen}

In this section we stay in the high energy limit and
derive the perturbative photon wave function at leading order 
from the results 
obtained above in (\ref{3.14}) and (\ref{3.17}). We will do this 
explicitly starting from the amplitude $\mathcal{M}^{(a)}$. 
Exactly the same steps can be performed also for $\mathcal{M}^{(b)}$. 

In the following we assume that the photon virtuality is sufficiently 
large so that we can apply perturbation theory. 
We work in the high energy limit and suppose furthermore 
\bea
\label{kinlimitalphaq}
\alpha \,|\mathbf{q}| &\gg &|\mathbf{k}_T|, \,m_q  \,,\nn \\
(1-\alpha) \, |\mathbf{q}| &\gg &|\mathbf{k}_T|, \,m_q \,.
\eea
This explicitly excludes the end points in longitudinal momentum, 
$\alpha=0$ and $\alpha=1$. 

In leading order of perturbation theory we have
\be
\label{3.16}
\Gamma^{(q)\nu}=\gamma^\nu \,,
\ee
while the rescattering term vanishes in this order, 
\be
\label{Kisnull}
K^{(q,q')}=0\,.
\ee
We insert this in the leading contribution to the amplitude 
$\mathcal{M}^{(a)}$ at high energies as given in (\ref{3.14}). 
Applying (\ref{3.3}) we arrive at
\bea
\label{Masympt}
\left.
\mathcal{M}^{(a)\mu\nu}_{s^\prime s}(p^\prime, p,q)
\right|_{\mathrm{asympt}} \!&=&\!
i \sum_{q} Q_q \int_0^1 d\alpha \int \frac{d^2 k_T}{(2 \pi)^2}
\sum_{A',A}\,
\sum_{\lambda^\prime ,\lambda}
\frac{1}{N} \delta_{A'A} \,
\nn\\
&&{} 
\langle \gamma(q^\prime,\mu),
p(p^\prime,s^\prime)|\mathcal{T}^{(a)}|
\bar{q}(k^{\prime},\lambda^\prime,A'),q(k,\lambda,A),p(p,s) \rangle
\nn \\
&&{}
\tilde{\psi}^{(q)\nu}_{\gamma, \lambda \lambda'} (\alpha, \mathbf{k}_T,Q)
\,,
\eea
where we have defined 
\be
\label{defpsischlange}
\tilde{\psi}^{(q)\nu}_{\gamma, \lambda \lambda'} (\alpha, \mathbf{k}_T,Q)
=
Q_q\,\frac{N}{\Delta E} \frac{|\mathbf{q}|}{2\pi 2 k^0 2k'^0} \,
\bar{u}_\lambda(k) \gamma^\nu v_{\lambda'}(k') \,
\theta \left(\bar{Q}^2 - \tilde{m}^2(\alpha,\mathbf{k}_T)\right)
\,.
\ee
Note that we have separated the sums over $r,r'$ into
sums over Dirac indices $\lambda,\lambda'$ and colour indices $A,A'$.
In the following sections we will determine the leading contribution 
to this expression in the high energy limit. 

We have introduced a normalisation factor $N$ here which 
could in principle still be chosen freely at this stage. 
In order to conform with the usual normalisation of the photon 
wave function used in the literature we will pick the choice 
\be
\label{normalizationchoice}
N = - 2 \sqrt{N_c \pi} \, e \, \sqrt{\alpha (1 - \alpha)}
\ee
with the proton charge $e = \sqrt{4 \pi \alpha_{\rm em}}$ 
and the number of colours $N_c=3$. 
Note that the freedom in choosing $N$ at this stage also involves 
a dependence on the fractions $\alpha$ and $(1-\alpha)$ 
of the longitudinal momentum of the photon which are carried by 
the quark and antiquark, respectively. The particular dependence 
chosen here will be motivated later on by the way in which the 
remaining $T$-matrix element is related to dipole 
matrix elements and in which the latter depend on $\alpha$ 
and $(1-\alpha)$, see sections \ref{sec:realvirtleadingorder} and 
\ref{subsec:dipolepicture} below. 

Note further that (\ref{defpsischlange}) includes the theta function 
involving the mass scale $\bar{Q}$ introduced in section \ref{highenergysect}. 
We first notice that the function 
$\tilde{m}^2(\alpha, \mathbf{k}_T)$ becomes independent of 
$\mathbf{q}$ in the high energy limit $|\mathbf{q}| \to \infty$, 
see (\ref{m2atlargeq}). Therefore this theta function naturally provides 
a regularisation of any singularities that might occur in the 
photon wave function at large transverse momenta $|\mathbf{k}_T|$, 
or equivalently at small distances $|\mathbf{R}_T|$ 
in transverse position space. 
In the end of our calculation we want to choose the momentum 
scale $\bar{Q}$ asymptotically large, thereby effectively removing 
the theta function. In view of this we will drop it already from now 
on in order to simplify the notation. 
However, it is very important to keep in mind that in our derivation 
of the photon wave function all possible singularities at small 
distances in position space are naturally regularised. Therefore 
in our opinion the physical significance of any effects depending on the 
divergences of the photon wave function at small distances is doubtful.  

The Fourier transformation from transverse momentum 
to transverse coordinate space is given by 
\be
\label{fourierwavefunction}
\psi^{(q)\nu}_{\gamma, \lambda \lambda'} (\alpha, \mathbf{R}_T,Q)
= \int \frac{d^2 k_T}{(2\pi)^2} \,e^{i \mathbf{k}_T \mathbf{R}_T} \,
\tilde{\psi}^{(q)\nu}_{\gamma, \lambda \lambda'} (\alpha, \mathbf{k}_T,Q)
\,. 
\ee

We now define the photon wave function for transversely and
longitudinally polarised photons in momentum space as the contraction 
of $\tilde{\psi}^{(q)\nu}_{\gamma, \lambda \lambda'}$ in 
(\ref{defpsischlange}) with the respective photon polarisation vectors. 
For transversely polarised photons we define 
\be
\label{phischlpm}
\tilde{\psi}^{(q)\pm}_{\gamma, \lambda \lambda'} (\alpha, \mathbf{k}_T,Q) =
- \varepsilon_{\pm \nu} \,
\tilde{\psi}^{(q)\nu}_{\gamma, \lambda \lambda'} (\alpha, \mathbf{k}_T,Q)
\,,
\ee
and for longitudinally polarised photons we define 
\be
\label{phischlL}
\tilde{\psi}^{(q)L}_{\gamma, \lambda \lambda'} (\alpha, \mathbf{k}_T,Q) =
- \varepsilon'_{L \nu} \,
\tilde{\psi}^{(q)\nu}_{\gamma, \lambda \lambda'} (\alpha, \mathbf{k}_T,Q)
\,.
\ee
The corresponding wave functions 
$\psi^{(q)\pm}_{\gamma, \lambda \lambda'} (\alpha, \mathbf{R}_T,Q)$
and 
$\psi^{(q)L}_{\gamma, \lambda \lambda'} (\alpha, \mathbf{R}_T,Q)$ 
in coordinate space are obtained via the Fourier transformation 
(\ref{fourierwavefunction}). 
Suitable choices for the polarisation vectors will be discussed in the following 
sections, where we give $\varepsilon_{\pm \nu}$ explicitly in 
(\ref{defepspm}) and $\varepsilon'_{L \nu}$ in (\ref{correctepslongcomp}). 
Already here we should point out that the correct choice of a suitable 
polarisation vector in the case of longitudinally polarised photons involves 
some subtleties. As a consequence of that our definition (\ref{phischlL}) 
contains a polarisation vector $\varepsilon'_L$ which differs from 
the most commonly used polarisation vector $\varepsilon_L$, see 
(\ref{defepslong}) and (\ref{correctepslong}) below. 

For the discussion in the following sections it is convenient 
to define the coordinate unit vectors in Minkowski space 
\be
\label{defepsbasis0,1}
e^{(0)}=
\left( \begin{array}{c}
1\\
0\\
0\\
0
\end{array} \right)
\,,
\quad \quad 
e^{(1)}=
\left( \begin{array}{c}
0\\
1\\
0\\
0
\end{array} \right)
\,,
\quad \quad 
e^{(2)}=
\left( \begin{array}{c}
0\\
0\\
1\\
0
\end{array} \right)
\,,
\quad \quad 
e^{(3)}=
\left( \begin{array}{c}
0\\
0\\
0\\
1
\end{array} \right)
\ee
such that $\mathbf{q}$ points in the positive 3-direction, 
$\mathbf{q} = |\mathbf{q}| \,\mathbf{e}^{(3)}$. 

We choose conventions \cite{4} for our spinors such that
\be
\label{defu}
u_\lambda (k) = \frac{1}{\sqrt{k^0 + m_q}}
\left( \begin{array}{c}
(k^0 + m_q) \chi_\lambda \\
\mbox{\boldmath $\sigma$}\mathbf{k} \,\chi_\lambda
\end{array} \right)
\ee
and
\be
\label{defv}
v_\lambda (k) = \frac{-1}{\sqrt{k^0 + m_q}}
\left( \begin{array}{c}
\mbox{\boldmath $\sigma$}\mathbf{k} \,\epsilon \chi_\lambda^*\\
(k^0 + m_q) \epsilon \chi_\lambda^*
\end{array} \right)
\,,
\ee
where $\sigma^i$ are the Pauli matrices and 
\be
\label{defepsilon}
\epsilon =
\left( \begin{array}{cc}
0 & 1\\
-1 & 0
\end{array} \right)
\,,
\ee
and for the $2$-spinors $\chi_\lambda$ we have the basis 
\be
\chi_{+\frac{1}{2}} =
\left( \begin{array}{c}
1\\
0
\end{array} \right) \,,
\quad \quad \quad
\chi_{-\frac{1}{2}} =
\left( \begin{array}{c}
0\\
1
\end{array} \right)
\,.
\ee
It will be useful to introduce the quantity 
\be
\label{defepssubq}
\epsilon_q = \sqrt{\alpha (1-\alpha) Q^2 +m_q^2}
\,.
\ee

\subsection{Transversely polarised photons}
\label{sec:transphotons}

We start with transversely polarised photons, for which we 
choose the polarisation vectors 
\be
\label{defepspm}
\left( \varepsilon_\pm^\nu \right)=
\mp \frac{1}{\sqrt{2}}
\left( e^{(1)} \pm i e^{(2)} \right) 
= 
\mp \frac{1}{\sqrt{2}}
\left( \begin{array}{c}
0\\
1\\
\pm i\\
0
\end{array} \right)
\,.
\ee
Here the signs are chosen such that they are in agreement with 
the canonical Condon-Shortley sign conventions for angular momenta.

With these polarisation vectors and with (\ref{defpsischlange}) the wave 
functions (\ref{phischlpm}) for transversely polarised photons become 
\be
\label{explcontrpm}
\tilde{\psi}^{(q)\pm}_{\gamma, \lambda \lambda'} (\alpha, \mathbf{k}_T,Q) = 
\mp \frac{N}{\sqrt{2}}\, Q_q (\Delta E)^{-1} 
\frac{ |\mathbf{q}|} {2 \pi 2k^0 2k'^0} \,
\bar{u}_\lambda(k) (\gamma^1 \pm i \gamma^2) v_{\lambda'}(k')
\,.
\ee
Using elementary algebra of Dirac and Pauli matrices and the 
expansions (\ref{expkstr0he})-(\ref{qfrac}) given in appendix 
\ref{appIIAfurther} we obtain the leading contribution to this 
expression in the high energy limit as 
\bea
\label{psipmmomentum}
\tilde{\psi}_{\gamma, \lambda \lambda'}^{(q)\pm} (\alpha, \mathbf{k}_T,Q) 
&=&{} 
{\sqrt{2 N_c}} \, \sqrt{\alpha_{\rm em}} \, Q_q \,
\frac{1}{\alpha (1-\alpha) Q^2 + \mathbf{k}_T^2 + m_q^2} 
\nn \\
&&{}
\left\{ \pm
(k_{T1} \pm i k_{T2}) 
\left[ \alpha \, \delta_{\lambda,\pm \frac{1}{2}} \delta_{\lambda', -\lambda} 
- (1 - \alpha) \, \delta_{\lambda,\mp \frac{1}{2}} \delta_{\lambda', -\lambda} 
\right] 
\right.
\nn \\
&&{}
\left.
\hspace*{.2cm}
+ \,m_q \, \delta_{\lambda,\pm \frac{1}{2}} \delta_{\lambda', \lambda} 
\right\} \,.
\eea
Interestingly, terms proportional to $|\mathbf{q}|$ occur when 
the components of spinors and Dirac matrices are multiplied out in 
(\ref{explcontrpm}), but they cancel in the sum. We therefore have the 
important result that the photon wave function remains finite 
in the high energy limit. 

In order to obtain the wave function in transverse position space we 
apply the Fourier transformation (\ref{fourierwavefunction}) 
which gives with (\ref{BesselK0}) and (\ref{BesselK1}) 
\bea
\label{psipmcoord}
\psi_{\gamma, \lambda \lambda'}^{(q)\pm} (\alpha, \mathbf{R}_T,Q) 
&=&{} 
\frac{\sqrt{N_c}}{\sqrt{2} \pi} \sqrt{\alpha_{\rm em}} \, Q_q 
\nn \\
&&{}
\left\{ 
\pm \, i e^{\pm i \phi_R }\left[ 
\alpha \, \delta_{\lambda,\pm \frac{1}{2}} \delta_{\lambda', -\lambda} 
- (1 - \alpha) \, \delta_{\lambda,\mp \frac{1}{2}} \delta_{\lambda', -\lambda} 
\right] \epsilon_q K_1(\epsilon_q R_T) 
\right.
\nn \\
&&{}
\left.
\hspace*{.2cm}
+\, m_q \, \delta_{\lambda,\pm \frac{1}{2}} \delta_{\lambda', \lambda} 
K_0(\epsilon_q R_T) 
\right\}
\,,
\eea
where $\phi_R=\arg (R_{T1} + iR_{T2})$ and 
$K_j(x)$ are the modified Bessel functions. 
The results (\ref{psipmmomentum}) and (\ref{psipmcoord}) are 
in agreement with those found in the literature up to the use 
of different phase conventions, see for example \cite{Bartels:2003yj}. 

\subsection{Longitudinally polarised photons}
\label{sec:longphotons}

Now we turn to longitudinally polarised photons. For these one often 
chooses the polarisation vector 
\be
\label{defepslong}
\left( \varepsilon_L^\nu \right) =
\frac{1}{Q}
\left( |\mathbf{q}|e^{(0)} + q^0 e^{(3)} \right) 
=
\frac{1}{Q}
\left( \begin{array}{c}
|\mathbf{q}|\\
0\\
0\\
q^0
\end{array} \right)
\,.
\ee
As we will explain momentarily, this choice is not appropriate for 
calculating the photon wave function in the high energy 
limit starting from the amplitude $\mathcal{M}^{(a,1)}$
found above. The correct result for the wave function of longitudinally 
polarised photons is obtained if one instead chooses the polarisation vector 
\be
\label{correctepslong}
\varepsilon_L^{\prime\nu}=
\varepsilon_L^\nu - \frac{q^\nu}{Q}
\,,
\ee
or explicitly in components
\be
\label{correctepslongcomp}
\left( \varepsilon_L^{\prime\nu} \right) =
\frac{|\mathbf{q}| -q^0}{Q}
\left( e^{(0)} -e^{(3)} \right) 
= 
\frac{Q}{|\mathbf{q}| +q^0}
\left( e^{(0)} -e^{(3)} \right) 
= 
\frac{1}{Q}
\left( \begin{array}{c}
|\mathbf{q}| -q^0\\
0\\
0\\
q^0 - |\mathbf{q}| 
\end{array} \right)
\,.
\ee
With this choice for the polarisation vector and 
with (\ref{defpsischlange}) 
the wave function for the longitudinal photon (\ref{phischlL}) 
in transverse momentum space becomes 
\be
\label{explcontrL}
\tilde{\psi}^{(q)L}_{\gamma, \lambda \lambda'} (\alpha, \mathbf{k}_T,Q) =
- N Q_q (\Delta E)^{-1} 
\frac{ |\mathbf{q}|} {2 \pi 2k^0 2k'^0} \,
\frac{|\mathbf{q}|-q^0}{Q}  \,
\bar{u}_\lambda(k) (\gamma^0  +  \gamma^3 ) v_{\lambda'}(k')
\,.
\ee
Using again the expansions (\ref{expk0he})-(\ref{DeltaElargeq}) and 
\be
\label{expandq0inqvec}
q^0 = |\mathbf{q}| + {\cal{O}} \left(\frac{Q^2}{|\mathbf{q}|}\right)
\ee
we obtain for the wave function for a longitudinally polarised photon 
in transverse momentum space  
\be
\label{psiLmomentum}
\tilde{\psi}_{\gamma, \lambda \lambda'}^{(q)L} (\alpha, \mathbf{k}_T,Q) = 
- 2 \sqrt{N_c} \, \sqrt{\alpha_{\rm em}} \, Q_q \,
\frac{\alpha (1- \alpha) Q}{\alpha (1-\alpha) Q^2 + \mathbf{k}_T^2 + m_q^2} 
\, \delta_{\lambda', -\lambda} 
\,.
\ee
Hence the wave function remains finite for $|\mathbf{q}| \to \infty$ 
also in the case of longitudinally 
polarised photons. In position space we obtain using (\ref{BesselK0}) 
\be
\label{psiLposition}
\psi_{\gamma, \lambda \lambda'}^{(q)L} (\alpha, \mathbf{R}_T,Q) = 
- \frac{\sqrt{N_c}}{\pi} \sqrt{\alpha_{\rm em}} \, Q_q 
Q \, \alpha (1- \alpha)  \, \delta_{\lambda', -\lambda}  \,
K_0(\epsilon_q R_T) \,,
\ee
again in agreement with the well-known result, see for example 
\cite{Bartels:2003yj}. 

We now come to the problem of correctly choosing a polarisation vector 
for longitudinally polarised photons which we have already mentioned 
before. While the simplest possible choice for such a polarisation vector 
is $\varepsilon_L^\nu$ as given by (\ref{defepslong}) one can easily 
see that also 
\be
\label{possibleepslong}
\left( \varepsilon_L^{\prime \prime \nu} \right) =
\left( \varepsilon_L^\nu + \kappa \, \frac{q^\nu}{Q} \right) = 
\frac{|\mathbf{q}| + \kappa q^0}{Q} \, e^{(0)}
+ \frac{q^0 + \kappa |\mathbf{q}|}{Q} \, e^{(3)}
\ee
describes a longitudinally polarised photon for any real or complex 
number $\kappa$. Indeed, since $\varepsilon_L^{\prime \prime \nu}$ 
and $\varepsilon_L^\nu$ differ only by a multiple of $q^\nu$ one 
can invoke electromagnetic current conservation (\ref{generalgaugeinv}) 
to show that the contraction of the full Compton amplitude 
$\mathcal{M}^{\mu\nu}$ with $\varepsilon_L^{\prime \prime \nu}$ 
gives the same result as the contraction with $\varepsilon_L^\nu$, 
\be
\label{contrepsgleichcontreps2str}
\varepsilon_{L \nu}^{\prime \prime} 
\mathcal{M}^{\mu\nu}_{s^{\prime}s}(p^{\prime},p,q) 
= 
\varepsilon_{L\nu}
\mathcal{M}^{\mu\nu}_{s^{\prime}s}(p^{\prime},p,q) 
\,.
\ee
The same is true for the contraction of the two choices of a 
polarisation vector with the amplitude $\mathcal{M}^{(a) \mu\nu}$, 
\be
\label{contrepsgleichcontreps2strforMa}
\varepsilon_{L \nu}^{\prime \prime} 
\mathcal{M}^{(a) \mu\nu}_{s^{\prime}s}(p^{\prime},p,q) 
=
\varepsilon_{L\nu}
\mathcal{M}^{(a) \mu\nu}_{s^{\prime}s}(p^{\prime},p,q) 
\,,
\ee
because of (\ref{Magaugeinv}). Due to (\ref{Mbgaugeinv}) 
the same holds also for the amplitude $\mathcal{M}^{(b) \mu\nu}$. 
Therefore the polarisation vector $\varepsilon_{L \nu}^{\prime \prime}$ 
is completely equivalent to $\varepsilon_{L \nu}$ as long as it is 
contracted with a gauge invariant amplitude, as expected 
on general grounds. The crucial point is now that in the definition of the 
wave function of longitudinally polarised photons (see (\ref{defpsischlange}) 
and (\ref{phischlL}) for the correct choice) the polarisation vector is 
contracted with an expression that is not gauge invariant, 
namely with $\bar{u}_\lambda(k) \gamma^\nu v_{\lambda'}(k')$ 
which  occurs in the amplitude $\mathcal{M}^{(a,1) \mu\nu}$.  
We have in fact seen explicitly in (\ref{qZubarv}) that 
gauge invariance does not hold for this expression separately. 

Let us see which concrete consequences this general observation has 
for the wave function of longitudinally polarised photons. For that 
we calculate the analogue of (\ref{phischlL}) for 
$\varepsilon_{L \nu}^{\prime \prime}$, 
\be
\label{pseudodefpsiL}
- \varepsilon_{L \nu}^{\prime \prime} \,
\tilde{\psi}^{(q)\nu}_{\gamma, \lambda \lambda'} (\alpha, \mathbf{k}_T,Q)
\,,
\ee
which would naively appear to be a reasonable definition of 
the photon wave function for general $\kappa$. 
With (\ref{defpsischlange}) and (\ref{possibleepslong}) we find 
for this contraction 
\be
\label{contrpseudopsiL}
- N Q_q (\Delta E)^{-1} 
\frac{ |\mathbf{q}|} {2 \pi 2k^0 2k'^0} \,
\,\frac{1}{Q} \,
\bar{u}_\lambda(k) \left[ 
\left(|\mathbf{q}| + \kappa \,q^0 \right) \gamma^0 
- \left(q^0 + \kappa |\mathbf{q}| \right) \gamma^3
\right]
v_{\lambda'}(k') 
\,.
\ee
Now the product of the two spinors $\bar{u}_\lambda(k)$ 
and $v_{\lambda'}(k')$ with $\gamma^0$ occurring here gives 
the leading term 
\be
\label{ubargamma0v}
\bar{u}_\lambda(k) \gamma^0 v_{\lambda'}(k') = 
-\,\frac{1}{\sqrt{k^0+m_q}} \frac{1}{\sqrt{k'^0+m_q}}
\left[\,2 \alpha (1-\alpha) |\mathbf{q}|^2 \, \delta_{\lambda',-\lambda} 
+ {\cal{O}} (m_q |\mathbf{q}|, |\mathbf{k}_T||\mathbf{q}|) \right]
\,,
\ee
and interestingly their product with $\gamma^3$ gives exactly 
the same leading term, 
\be
\label{ubargamma3v}
\bar{u}_\lambda(k) \gamma^3 v_{\lambda'}(k') = 
-\,\frac{1}{\sqrt{k^0+m_q}} \frac{1}{\sqrt{k'^0+m_q}}
\left[\,2 \alpha (1-\alpha) |\mathbf{q}|^2 \,\delta_{\lambda',-\lambda} 
+ {\cal{O}} (m_q |\mathbf{q}|, |\mathbf{k}_T||\mathbf{q}|) \right]
\,,
\ee
where we have again used (\ref{expk0he})-(\ref{k0plusmkstr0plusm}). 
One finds in fact that even the terms proportional to $m_q |\mathbf{q}|$ 
and to $|\mathbf{k}_T||\mathbf{q}|$, that is the next-to-leading terms 
in the expansion in $1/|\mathbf{q}|$, are identical in (\ref{ubargamma0v}) 
and (\ref{ubargamma3v}), but we will not need those terms here. 
For the square brackets and the spinors in (\ref{contrpseudopsiL}) 
one then obtains 
\be
\label{simpleqkappaq}
\left(|\mathbf{q}| + \kappa \,q^0 \right) - \left(q^0 + \kappa |\mathbf{q}| \right) 
= (1-\kappa) \left(|\mathbf{q}| - q^0 \right)
\ee
times the expression (\ref{ubargamma0v}). 
Applying (\ref{expk0he})-(\ref{DeltaElargeq}) and (\ref{expandq0inqvec}) 
and taking into account the prefactors in (\ref{contrpseudopsiL}) 
we finally have 
\be
\label{pseudoepsLfinal}
- \varepsilon_{L \nu}^{\prime \prime} \, 
\tilde{\psi}^{(q)\nu}_{\gamma, \lambda \lambda'} (\alpha, \mathbf{k}_T,Q)
= 
- (1-\kappa)  \sqrt{N_c} \, \sqrt{\alpha_{\rm em}} \, Q_q \,
\frac{\alpha (1- \alpha) Q}{\alpha (1-\alpha) Q^2 + \mathbf{k}_T^2 + m_q^2} 
\, \delta_{\lambda', -\lambda} \,.
\ee
Obviously, this expression strongly depends on $\kappa$ which reflects 
the fact that $\tilde{\psi}^{(q)\nu}_{\gamma, \lambda \lambda'}$ is not 
gauge invariant. Choosing arbitrary values for $\kappa$ here would 
lead to completely arbitrary wave functions for longitudinally polarised 
photons. 

What happens here is that cancellations take place between terms 
coming from different components in the contraction with the polarisation 
vector. If the chosen polarisation vector contains large terms proportional 
to $|\mathbf{q}|$ or $q^0$, subleading terms in the amplitude (or in the 
product of spinors) can at the same time be enhanced due to these factors 
when the high energy limit $|\mathbf{q}|\to \infty$ is taken. 
Those subleading terms, however, cannot be correctly determined only 
from $\tilde{\psi}^{(q)\nu}_{\gamma, \lambda \lambda'}$ which was 
extracted only from the amplitude $\mathcal{M}^{(a,1)}$. Instead, 
one would have to take into account also the subleading terms 
$\mathcal{M}^{(a,2)}$ to $\mathcal{M}^{(a,4)}$. Those terms 
do not have a pinch singularity (see the discussion in 
section \ref{highenergysect}) and are therefore difficult to identify 
in a nonperturbative framework using the methods that we have used 
here. We will comment on the situation in lowest order of perturbation 
theory at the end of the section.

If one wants to determine the correct wave function from the 
leading term at high energies alone, that is only from 
$\mathcal{M}^{(a,1)}$ and hence from 
$\tilde{\psi}^{(q)\nu}_{\gamma, \lambda \lambda'}$, one must 
therefore choose a polarisation vector which remains finite in the 
high energy limit. Using (\ref{expandq0inqvec}) one sees that the 
only possible choice for $\kappa$ in (\ref{possibleepslong}) 
consistent with this requirement is 
\be
\label{corrchoicekappa}
\kappa = -1 + \mathcal{O} \left( \frac{Q}{|\mathbf{q}|} \right) 
\,,
\ee
up to subleading terms as indicated. The leading term in this 
expression turns (\ref{possibleepslong}) into the correct choice 
$\varepsilon_L^{\prime\nu}$ (\ref{correctepslong}) all components 
of which remain finite in the high energy limit $|\mathbf{q}|\to \infty$. 
With that choice one obtains in fact the correct wave function 
(\ref{psiLmomentum}) for longitudinally polarised photons as 
described above. 

Note that a similar problem with the choice of a polarisation 
vector would also occur for transversely polarised photons 
if one decided to  add terms proportional to $q^\nu$ to 
$\varepsilon_\pm^\nu$ (\ref{defepspm}). 
There, however, the simplest choices for the polarisation vectors 
do not involve any factors of $|\mathbf{q}|$ or $q^0$. 
Therefore the gauge invariance problem is usually not discussed 
in the case of transversely polarised photons. 

Let us finally comment on the situation in lowest order of 
perturbation theory where the interaction of the quark-antiquark 
pair with the proton is given by two-gluon exchange, possibly 
including the resummation of soft or collinear logarithms. 
Here the problem of choosing appropriate polarisation vectors 
can be studied explicitly, and the above results are found to be confirmed.  
In the perturbative situation it is possible to compute the photon 
wave function to lowest order starting from a gauge invariant 
expression, as it has been done for example in \cite{Bartels:2003yj}. 
There one starts with a set of diagrams in which the incoming 
and outgoing photons couple to a quark loop with two gluons attached 
in all possible ways. That amounts to four diagrams one of which 
is shown in figure  I.\ref{fig200}. This set of diagrams is exactly the 
expansion of our amplitude $\mathcal{M}^{(a) \mu\nu}$ (see figure 
I.\ref{fig2}a) 
to lowest nontrivial order and constitutes a gauge invariant 
expression. For such a gauge invariant set 
of diagrams due to (\ref{contrepsgleichcontreps2strforMa}) 
and its perturbative expansion the different choices of photon 
polarisation vectors are fully equivalent. One can then compute the 
leading contribution to the quark loop at high energy and 
subsequently split the result into factors corresponding to 
the incoming and outgoing photon. More precisely, it is then possible 
to identify the perturbative photon wave functions from the 
different combinations of transversely and longitudinally polarised 
photons by invoking conservation of quark helicities at high energy. 
The resulting photon wave functions are in complete agreement 
with the results obtained above. 

\section{Real and virtual photons in the initial and final states }
\label{sec:genphotons}

\subsection{General formulae}
\label{sec:realvirtgenform}

\begin{figure}[ht]
\begin{center}
\includegraphics[width=8.5cm]{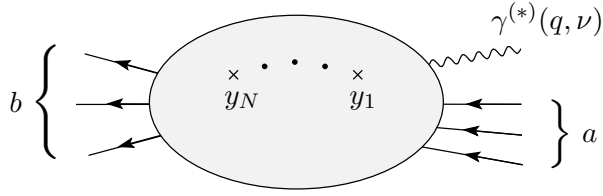}
\caption{Diagram for the matrix element (\ref{II1}).  
\label{figagammab}}
\end{center}
\end{figure}

In this section we summarise our findings for the transition from photon to 
dipole scattering in the high energy limit. We give general formulae for real 
and virtual photons in the initial and final states. Let $|a, \mbox{in}\rangle$ 
be some initial, $|b, \mbox{out} \rangle$ some final state and let 
$R_1(y_1),\dots,R_N(y_N)$ be some local operators. 
We consider the matrix element of figure \ref{figagammab}, 
\be\label{II1}
{\cal M}^\nu(q,y_1, \dots, y_N)=\int d^4x~ e^{-iqx}
\langle b,\textup{out}|{\rm T}^*R_1(y_1)\dots R_N(y_N)
J^\nu(x)|a,\textup{in}\rangle \,.
\ee 
Here ${\rm T}^*$ is the covariantised T-product. 
In the following we will for the sake of brevity suppress 
the arguments $y_i$ and denote the above expression 
simply by ${\cal M}^\nu(q)$. 
For the case that no operators $R_i$ are present ${\cal M}^\nu(q)$ 
is the amplitude for
\be\label{II2}
\gamma^{(*)}(q)+a\rightarrow b \,.
\ee
As in previous sections we suppose
\be\label{II2a}
q^2=-Q^2\leq 0 \,.
\ee

We should point out that in general we do not have $q_\nu {\cal M}^\nu =0$ 
for the amplitude (\ref{II1}) due to the T-ordering. 
Instead, $q_\nu {\cal M}^\nu$ vanishes only if for all $i=1,\dots,N$ 
\be
\label{condTRicomm}
 \left[ J^0(x), R_i(y)\right] \delta(x^0 - y^0) = 0 \,. 
\ee
If the $R_i$ are interpolating field operators for asymptotic 
states though, the full amplitude will be described by the LSZ 
formula which then contains additional factors. If for example 
$R_i$ corresponds to an outgoing particle with spin 1/2 and mass $m_i$, 
the LSZ formula will contain the additional integration 
\be
\label{lszforgeneralRi}
i \int d^4y_i \, e^{ip_i y_i} \bar{u}_{s_i}(p_i) 
(-i\rightslash_{y_i}+m_i) \,
{\cal M}^\nu(q, \dots, y_i, \dots) \,. 
\ee
The contraction of that full amplitude with $q_\nu$ will then 
vanish. An example for this general mechanism is the separate 
gauge invariance of the amplitude ${\cal M}^{(f)}$ which is 
discussed in detail in appendix \ref{appdiffcontgauge}.  
For the case that $R_i$ is an electromagnetic current, on the 
other hand, the condition 
(\ref{condTRicomm}) holds in any case. Hence for interesting 
operators $R_i$ current conservation for ${\cal M}^\nu$ 
(or for the full scattering amplitude with LSZ factors for the 
case of interpolating fields $R_i$) is fulfilled. 

Our discussion in I and in section \ref{highenergysect} 
can easily be adapted to the study of the 
high energy limit of the matrix element (\ref{II1}). We get for the leading 
term of ${\cal M}^\nu(q)$ for $|\mathbf{q}|\rightarrow\infty$ and fixed $Q^2$ 
\bea\label{II3}
{\cal M}^\nu(q) 
&\xrightarrow[|\mathbf{q}|\rightarrow\infty]~{} &
i \sum_q \sum_{r',r}
\int\frac{d^3k}{(2\pi)^32k^02k^{\prime 0}}\frac{1}{\Delta E}
\nonumber\\
&&{} \hspace*{1cm}
\langle b,\textup{out}|{\rm T}^* R_1(y_1)\dots R_N(y_N)
|\bar{q}(k',r'),q(k,r),a,\textup{in}\rangle
\nonumber\\
&&{} \hspace*{1cm}
Q_q\bar{u}_r(k)\left[\Gamma^{(q)\nu}(k,-k')
+L^{(q)\nu}(k,-k')\right]v_{r'}(k') \,.
\eea
Here $k,k',\Delta E$ are as in (\ref{defDeltaE}), (\ref{3.2}), (\ref{3.3}), 
and $L^{(q)\nu}(k,-k')$ denotes the rescattering term as contained 
in (I.\ref{2.31}), 
\be
\label{defLrescatt}
L^{(q)\nu}(k,-k') = 
\sum_{q'}\int K^{(q,q')}S^{(q')}_F \Gamma^{(q')\nu} S^{(q')}_F
\,,
\ee
with the momentum assignments as in figure I.\ref{fig6}. 
Note that the quark-antiquark state in the matrix element contains 
the appropriate renormalisation factor $Z_q^{-1}$ as required 
by the LSZ formula as was made explicit in (I.\ref{2.26})-(I.\ref{2.29}) 
for the case of the Compton amplitude. 

We have -- of course having in mind the caveats pointed out in 
section \ref{sec:intro2} -- also made the assumption that the quarks 
$q$ have a mass shell. Furthermore a cutoff in transverse 
momenta is understood to ensure the pinch condition (\ref{3.13}), 
see also the discussion concerning the corresponding theta function 
in section \ref{subsec:gammawavefunctgen}. 

Now we will use $CPT\equiv \Theta$ invariance to derive the analogue 
of (\ref{II3}) for real or virtual photons in the final state. 
Here and in the following we use the notation and general formulae for 
$\Theta$-invariance as in \cite{4}. 
With the antiunitary operator $V(\Theta)$ representing a 
$\Theta$-transformation we have for the electromagnetic current 
\be\label{II4}
\left(V(\Theta)J^\nu(x)V^{-1}(\Theta)\right)^\dagger=-J^\nu(-x)\,.
\ee
Let us define local operators $R^\Theta_j(y)$ by 
\be
\label{II5}
R^\Theta_j(y)=\left(V(\Theta)R_j(-y)V^{-1}(\Theta)\right)^\dagger 
\hspace{1.5cm}
(j=1,\dots,N)\,.
\ee
For state vectors we set
\bea\label{II6}
V(\Theta)|a,\textup{in}\rangle &=&|a^\Theta, \textup{out}\rangle \,,
\nn\\
V(\Theta)|b,\textup{out}\rangle &=& |b^\Theta, \textup{in}\rangle \,.
\eea
For -- hypothetical -- asymptotic quark and antiquark states we have
\bea\label{II7}
V(\Theta)|q(k,r),\textup{out}\rangle 
&=&{}
- \sum_t |\bar{q}(k,t),\textup{in}\rangle \, E_{tr}  \,,
\nonumber\\
V(\Theta)|\bar{q}(k,r),\textup{out}\rangle 
&=&{}
\sum_t |q(k,t),\textup{in}\rangle \, E_{tr} \,.
\eea
Here $r= (\lambda ,A)$, $t=(\tau,B)$ denote again the combination of 
spin and colour indices as in I, and we have 
\be\label{II8}
E_{tr} = \epsilon_{\tau \lambda} \, \delta_{BA} \,.
\ee

Our starting point is (\ref{II1}), but with $|a,\textup{in} \rangle$ 
replaced by $|b^\Theta,\textup{in} \rangle$ and 
$|b,\textup{out} \rangle$ by $|a^\Theta,\textup{out} \rangle$. 
Furthermore we replace the $R_j(y_j)$ by 
$R^\Theta_j(y'_j)$ and take them in reversed order. 
This choice is possible since the states $a$ and $b$ and the 
operators $R_j(y_j)$ in the general formulae are arbitrary. 
We get for arbitrary points $y'_1,\dots,y'_N$ using (\ref{II3}) 
and (I.\ref{2.31}) 
\bea\label{II9}
\lefteqn{
\int d^4x'~e^{-iqx'} 
\langle a^\Theta,\textup{out}|
{\rm T}^* R^\Theta_N(y^\prime_N)\dots R^\Theta_1(y'_1)J^\mu(x')
|b^\Theta,\textup{in} \rangle
}
\nonumber\\
&\xrightarrow[|\mathbf{q}|\rightarrow\infty]~{}&{}
i \sum_q \sum_{r',r}\int 
\frac{d^3k}{(2\pi)^32k^02k^{\prime 0}}\frac{1}{\Delta E}
\,Q_qZ_q\bar{u}_r(k)\gamma^\mu v_{r'}(k') 
\\
&&{}\hspace*{.5cm}
\langle a^\Theta,\textup{out}|
{\rm T}^* R^\Theta_N(y'_N)\dots R^\Theta_1(y'_1) 
|\bar{q}(k',r'),q(k,r),b^\Theta,\textup{in} \rangle \,.
\nonumber
\eea
From (\ref{II4}) and (\ref{II5}) we have
\be
\label{II10}
{\rm T}^* R^\Theta_N(y'_N)\dots R^\Theta_1(y'_1)
=\left(V(\Theta) [{\rm T}^*  R_1(y_1)\dots R_N(y_N) ]
V^{-1}(\Theta)\right)^\dagger
\ee
and
\be
\label{II11}
{\rm T}^* R^\Theta_N(y'_N)\dots R^\Theta_1(y'_1)J^\mu(x')
=\left(V(\Theta) [{\rm T}^* (-J^\mu(x)) R_1(y_1)\dots R_N(y_N)]
V^{-1}(\Theta)\right)^\dagger\,,
\ee
where $y_j=-y'_j$ and $x=-x'$. Inserting this in (\ref{II9}) and using the 
rules of antiunitary operators we get
\bea\label{II12}
\lefteqn{
\int d^4x~e^{iqx}\langle b, \textup{out}|
{\rm T}^* (-J^\mu(x) )R_1(y_1)\dots R_N(y_N)|
a, \textup{in}\rangle }
\nonumber\\
&\xrightarrow[|\mathbf{q}|\rightarrow\infty]~{}&{}
i\sum_q \sum_{r',r} \sum_{t',t}
\int \frac{d^3k}{(2\pi)^32k^02k^{\prime 0}}\frac{1}{\Delta E}
\, (-E_{t'r'}E_{tr}) \,Q_qZ_q \bar{u}_r(k)\gamma^\mu v_{r'}(k') 
\nonumber\\
&&{}\hspace*{1cm}
\langle q(k',t'),\bar{q}(k,t),b, \textup{out}| 
{\rm T}^*R_1(y_1)\dots R_N(y_N)|a, \textup{in}\rangle \,.
\eea
Now we use
\bea\label{II13}
\sum_{r}E_{tr} \,u_r(k)&=&\gamma_5 v_t(k) \,,
\nonumber\\
\sum_{r'}E_{t'r'} \, v_{r'}(k')&=&-\gamma_5 u_{t'}(k')
\eea
for the Dirac times colour spinors, see (I.\ref{colorspinoru}) and 
(I.\ref{colorspinorv}), and get from (\ref{II12}) 
\bea\label{II14}
\lefteqn{
\int d^4x~e^{iqx}\langle b, \textup{out}| 
{\rm T}^* (-J^\mu(x))R_1(y_1)\dots R_N(y_N)|a, \textup{in}\rangle 
}
\hspace*{0.5cm}
\nonumber\\
&\xrightarrow[|\mathbf{q}|\rightarrow\infty]~{}&{}
i \sum_q \sum_{t',t}\int\frac{d^3k}{(2\pi)^32k^02k^{\prime 0}}\frac{1}{\Delta E} \,
Q_qZ_q\bar{v}_t(k)\gamma^\mu u_{t'}(k')
\nonumber\\
&&{} \hspace*{1cm}
\langle q(k',t'),\bar{q}(k,t),b,\textup{out}| 
{\rm T}^* R_1(y_1)\dots R_N(y_N)|a,\textup{in}\rangle \,.
\eea
We interchange $q$ and $\bar{q}$ in the final state, relabel the summation 
and integration variables, and use (I.\ref{2.31}) in order to obtain 
\bea\label{II15}
\lefteqn{
\int d^4 x~e^{iqx}\langle b, \textup{out}| 
{\rm T}^*J^\mu(x)R_1(y_1)\dots R_N(y_N)|a,\textup{in}\rangle
}
\nonumber\\
&\xrightarrow[|\mathbf{q}|\rightarrow\infty]~{} &{}
i \sum_q \sum_{r',r}\int \frac{d^3k}{(2\pi)^32k^02k^{\prime 0}}\frac{1}{\Delta E} 
\, Q_q\bar{v}_{r'}(k')\left[\Gamma^{(q)\mu}(-k',k) 
+L^{(q)\mu}(-k',k)\right] u_r(k)
\nonumber\\
&&{}\hspace*{1cm}
\langle \bar{q}(k',r'),q(k,r),b, \textup{out}| 
{\rm T}^* R_1(y_1)\dots R_N(y_N)|a,\textup{in}\rangle \,.
\eea
This is our general result for the high energy limit of an amplitude 
with a real or virtual photon in the final state.

\subsection{Leading order formulae and dipole states}
\label{sec:realvirtleadingorder}

In applications of the dipole formalism one frequently uses the photon 
wave function only to lowest order in $\alpha_s$. That is one makes the 
following replacements in (\ref{II3})
\bea\label{V100}
\Gamma^{(q)\nu}(k,-k') &\rightarrow& \gamma^\nu,
\nonumber\\
L^{(q)\nu}(k,-k')&\rightarrow& 0 \,,
\eea
and similarly in (\ref{II15}):
\bea\label{V101}
\Gamma^{(q)\mu}(-k',k)&\rightarrow& \gamma^\mu\,,
\nonumber\\
L^{(q)\mu}(-k',k)&\rightarrow& 0 \,.
\eea
This is justified for high enough $Q^2$ where $\alpha_s(Q^2)$ is small. 

Using now the definition of the lowest order photon wave function, 
see (\ref{defpsischlange}) and (\ref{normalizationchoice}), 
we get from (\ref{II3})
\bea\label{V102}
e{\cal M}^\nu(q)
&\xrightarrow[|\mathbf{q}|\rightarrow\infty]~{}&
- i \sum_q \sum_{\lambda',\lambda} \sum_{A',A}
\int_0^1 d\alpha\int\frac{d^2 k_T}{(2\pi)^2}
\nonumber\\
&&{} \hspace*{1cm}
\langle b,\textup{out}|{\rm T}^* R_1(y_1)\dots R_N(y_N) |
\bar{q}(k',\lambda',A'),q(k,\lambda,A),a,\textup{in}\rangle
\nonumber\\
&&{}\hspace*{1cm}
\delta_{A'A}\,\frac{1}{2 \sqrt{N_c \pi}}
\frac{1}{\sqrt{\alpha(1-\alpha)}}\,
\tilde{\psi}^{(q)\nu}_{\gamma,\lambda\lambda'}(\alpha,\mathbf{k}_T,Q)\,.
\eea
With the Fourier transform (\ref{fourierwavefunction}) this reads
\bea\label{V103}
e{\cal M}^\nu(q)  &\xrightarrow[|\mathbf{q}|\rightarrow\infty]~{}&
-i \sum_q \sum_{\lambda',\lambda}\int^1_0 d\alpha \int d^2R_T
\nn \\
&&{} \hspace*{1cm}
\langle b,\textup{out}|{\rm T}^* R_1(y_1)\dots R_N(y_N)
|D^{(q)}(\mathbf{q},\alpha,\mathbf{R}_T,\lambda',\lambda),a,
\textup{in}\rangle
\nn \\
&&{} \hspace*{1cm}
\psi^{(q)\nu}_{\gamma,\lambda\lambda'}(\alpha,\mathbf{R}_T,Q)
\,,
\eea
where we have defined dipole states 
\bea\label{V104}
|D^{(q)}(\mathbf{q},\alpha,\mathbf{R}_T,\lambda',\lambda)\rangle &=& 
\sum_{A',A}\int\frac{d^2k_T}{(2\pi)^2} \,e^{-i\mathbf{k}_T\mathbf{R}_T}
\frac{1}{2 \sqrt{N_c \pi}}
\frac{1}{\sqrt{\alpha(1-\alpha)}}
\nn \\
&&{} \hspace*{0.5cm}
\delta_{A'A} \,|\bar{q}(k',\lambda',A'),q(k,\lambda,A)\rangle \,.
\eea
Treating quark and antiquark as asymptotic states we find that these dipole 
states are eigenstates of three-momentum with eigenvalue $\mathbf{q}$, 
\bea\label{V105}
\mathbf{P} \,
|D^{(q)}(\mathbf{q},\alpha,\mathbf{R}_T,\lambda',\lambda)\rangle=
\mathbf{q}\,
|D^{(q)}(\mathbf{q},\alpha,\mathbf{R}_T,\lambda',\lambda)\rangle
\,,
\eea
and satisfy for the kinematic region (\ref{kinlimitalphaq}) 
the normalisation condition 
\bea\label{V106}
\langle D^{(\tilde{q})}
(\tilde{\mathbf{q}},\tilde{\alpha},\tilde{\mathbf{R}}_T,
\tilde{\lambda}',\tilde{\lambda})|
D^{(q)}(\mathbf{q},\alpha,\mathbf{R}_T,\lambda',\lambda)\rangle 
\! &=& \!
\delta_{\tilde{q}q}\, \delta_{\tilde{\lambda}'\lambda'} \, 
\delta_{\tilde{\lambda}\lambda}
(2\pi)^3 \, 2|\mathbf{q}|\,
\delta(\tilde{\alpha}-\alpha)
\nn\\
&&{}
\delta^{(3)}(\tilde{\mathbf{q}}-\mathbf{q}) \,
\delta^{(2)}(\tilde{\mathbf{R}}_T-\mathbf{R}_T)
\,.
\eea
A dipole state of the form (\ref{V104}) is for fixed $\mathbf{R}_T$ 
and $\alpha$ a superposition of quark-antiquark states with squared 
invariant masses $(k+k')^2$ for which we have at high energy due 
to (\ref{k0kstr0he}) 
\be
\label{massesqqbarstates}
(k+k')^2 \xrightarrow[|\mathbf{q}|\rightarrow\infty]~{}
\,\frac{\mathbf{k}_T^2 + m_q^2}{\alpha (1-\alpha)} \,
\ge \,
\frac{m_q^2}{\alpha (1-\alpha)}\,.
\ee
Note that the mass of the quark-antiquark states is independent of 
$Q^2$ in the high energy limit. The action 
of the squared four-momentum operator $P^2$ on the dipole states 
then gives in the high energy limit $|\mathbf{q}| \to \infty$ 
\bea
\label{P2aufdipolhe}
P^2 
|D^{(q)}(\mathbf{q},\alpha,\mathbf{R}_T,\lambda',\lambda)\rangle 
&=&{}
\sum_{A',A}\int\frac{d^2k_T}{(2\pi)^2} \,e^{-i\mathbf{k}_T\mathbf{R}_T}
\frac{1}{2 \sqrt{N_c \pi}}
\frac{1}{\sqrt{\alpha(1-\alpha)}} \,\delta_{A'A} 
\nn 
\\
&&{}\hspace*{0.5cm}
\frac{\mathbf{k}_T^2 + m_q^2}{\alpha (1-\alpha)} \,
|\bar{q}(k',\lambda',A'),q(k,\lambda,A)\rangle \,. 
\eea
Note that also the mass of the dipole state is independent of $Q^2$ 
at high energy. 

For photons in the final state we get a relation to dipole states analogous 
to (\ref{V103}) from (\ref{II15}) using the assumption (\ref{V101}): 
\bea\label{V111}
\lefteqn{
e\int d^4x~ e^{iqx}~\langle b, \textup{out}|{\rm T}^* J^\mu(x)
R_1(y_1)\dots R_N(y_N)|a,\textup{in}\rangle}
\nonumber\\
&\xrightarrow[|\mathbf{q}|\rightarrow\infty]~{}&{}
-i \sum_q \sum_{\lambda',\lambda}
\int^1_0 d\alpha\int d^2R_T
\left(\psi^{(q)\mu}_{\gamma,\lambda\lambda'}(\alpha,\mathbf{R}_T,Q)\right)^*
\\
&&{} \hspace*{1cm}
\langle D^{(q)}(\mathbf{q},\alpha,\mathbf{R}_T,\lambda',\lambda),b,\textup{out}
|{\rm T}^* R_1(y_1)\dots R_N(y_N)|a, \textup{in}\rangle \,.
\nonumber
\eea
In the following section we will apply the general formulae (\ref{V103}) 
and (\ref{V111}) to the case of deep inelastic scattering. 

At this point a remark is in order concerning the interpretation of the 
dipole states introduced above. 
Frequently one thinks of dipole states as hadron states. However, 
there is still a small problem with the states (\ref{V104}) in this respect 
since they do not have the usual normalisation condition for hadrons. 
In particular, their normalisation depends on transverse position and 
the longitudinal momentum fraction of the quark and antiquark. 
But when considering the colour dipole as a hadron, these are 
continuous internal degrees of freedom which should not occur in 
the normalisation of the full hadronic state. This situation can be remedied 
by using dipole states smeared out in $\alpha$ and $\mathbf{R}_T$. 
Consider smearing functions
\bea\label{V107}
f_T(\mathbf{R}_T) \!&\geq&\! 0 \,,
\nonumber\\
f_L(\alpha)\!&\geq&\! 0 \,,
\eea
which are strongly peaked around 
$\mathbf{R}_T=0$ and $\alpha=0$, respectively, and satisfy 
\bea\label{V108}
\int d^2R_T \,|f_T(\mathbf{R}_T)|^2 \!&=&\! 1 \,,
\nonumber\\
\int d\alpha\, |f_L(\alpha)|^2 \!&=&\! 1 \,.
\eea
We can then define smeared dipole states as
\bea\label{V109}
|\bar{D}^{(q)}(\mathbf{q},\bar{\alpha}, 
\bar{\mathbf{R}}_T,\lambda',\lambda)\rangle
&=&
\int d^2R_T \, f_T(\mathbf{R}_T-\bar{\mathbf{R}}_T)
\int^1_0 d\alpha  \, f_L(\alpha-\bar{\alpha}) 
\\
&&{} \hspace*{.2cm}
|D^{(q)}(\mathbf{q},\alpha,\mathbf{R}_T,\lambda',\lambda)\rangle \,.
\nonumber
\eea
These states still satisfy (\ref{V105}), are normalised to 
\be\label{V110}
\langle \bar{D}^{(\tilde{q})}(\tilde{\mathbf{q}},\bar{\alpha},\mathbf{\bar{R}}_T,
\tilde{\lambda}',\tilde{\lambda})|
\bar{D}^{(q)}(\mathbf{q},\bar{\alpha},\mathbf{\bar{R}}_T,\lambda',\lambda)
\rangle
= \delta_{\tilde{q} q} \,
\delta_{\tilde{\lambda}'\lambda'}\,
\delta_{\tilde{\lambda}\lambda}(2\pi)^3
\,2|\mathbf{q}|\,
\delta^{(3)}(\tilde{\mathbf{q}}-\mathbf{q})\,,
\ee
and can be thought of as hadron analogues. 
For the matrix element of the squared four-momentum operator 
between two such states we have 
\bea
\label{meanP2smeareddip}
\lefteqn{
\langle \bar{D}^{(\tilde{q})}(\tilde{\mathbf{q}},\bar{\alpha},\mathbf{\bar{R}}_T,
\tilde{\lambda}',\tilde{\lambda})| P^2 |
\bar{D}^{(q)}(\mathbf{q},\bar{\alpha},\mathbf{\bar{R}}_T,\lambda',\lambda)
\rangle
}
\nn \\
&&{}
\hspace*{1cm}
=
\delta_{\tilde{\lambda}'\lambda'}\,
\delta_{\tilde{\lambda}\lambda}(2\pi)^3
\,2|\mathbf{q}|\,
\delta^{(3)}(\tilde{\mathbf{q}}-\mathbf{q}) \,
\langle M^2 \rangle_{\bar{\alpha}, \mathbf{\bar{R}}_T} \,.
\eea
Comparing with the normalisation (\ref{V110}) we see that 
$\langle M^2 \rangle_{\bar{\alpha}, \mathbf{\bar{R}}_T}$ 
is the mean squared invariant mass of the smeared dipole state 
(\ref{V109}), and in the high energy limit we obtain 
using (\ref{P2aufdipolhe}) 
\be
\label{meanM2smeareddip}
\langle M^2 \rangle_{\bar{\alpha}, \mathbf{\bar{R}}_T} 
= \int^1_0 d\alpha  \, |f_L(\alpha-\bar{\alpha})|^2 \,
\frac{1}{\alpha(1-\alpha)}
\int \frac{d^2k_T}{(2 \pi)^2} \, |\tilde{f}_T(\mathbf{k}_T)|^2 
(\mathbf{k}_T^2 + m_q^2) 
\,,
\ee
where we have defined the Fourier transform $\tilde{f}_T$ 
of the the smearing function $f_T$ as 
\be
\label{deffTtilde}
\tilde{f}_T(\mathbf{k}_T) = 
\int d^2R_T \,e^{-i\mathbf{k}_T\mathbf{R}_T} f_T(\mathbf{R}_T)\,. 
\ee
The latter satisfies with (\ref{V108}) 
\be
\label{normfTtilde}
\int \frac{d^2k_T}{(2\pi)^2} \,|\tilde{f}_T(\mathbf{k}_T)|^2 = 
\int d^2R_T \,|f_T(\mathbf{R}_T)|^2 = 1 \,.
\ee
From (\ref{meanM2smeareddip}) we conclude that at high energy 
the mean invariant mass of the smeared dipole states (\ref{V109}) depends 
only on $\bar{\alpha}$, $m_q$ and on the shape of the smearing functions 
$f_T$ and $f_L$, but is independent of $Q^2$. 

\section{Dipole picture for deep inelastic scattering}
\label{sec:dippictdis}

\subsection{Deep inelastic scattering}
\label{subsec:dis}

As an important application of the general formulae (\ref{II3}) 
and (\ref{II15}) we consider the Compton amplitude as in (I.\ref{2.3}) 
which we write now as follows
\bea\label{II16}
\lefteqn{
(2\pi)^4\delta^{(4)}(p'+q'-p-q) {\cal M}^{\mu\nu}_{s's}(p',p,q)}
\nonumber\\
&&{}
= \frac{i}{2\pi m_p}\int d^4x'~ d^4x ~e^{iq'x'}e^{-iqx} 
\langle p(p',s')|{\rm T}^* J^\mu(x') J^\nu(x)|p(p,s)\rangle \,.
\eea
This amplitude should be understood as defined in the nonforward 
direction $p' \neq p$ and the limit $p' \to p$ is understood in 
the following when we write ${\cal M}^{\mu\nu}_{s's}(p,p,q)$. 
That amounts to taking into account only the connected part of 
the matrix element in (\ref{II16}). 

The usual hadronic tensor of deep-inelastic electron-proton scattering is
\bea\label{VI55e}
\lefteqn{
W^{\mu\nu}(p,q)=\frac{1}{2}\sum_{s',s}\frac{1}{2i}
\left[ {\cal M}^{\mu\nu}_{s's} (p,p,q)
-\left( {\cal M}^{\nu\mu}_{ss'}(p,p,q)\right)^*\right] 
\,\delta_{s's}
}
\\
&&{}
=-W_1(\nu,Q^2)\left(g^{\mu\nu}-\frac{q^\mu q^\nu}{q^2}\right)
+\frac{1}{m^2_p}W_2(\nu,Q^2)\left(p^\mu-\frac{(pq)q^\mu}{q^2}\right)
\left(p^\nu-\frac{(pq)q^\nu}{q^2}\right) 
\nn
\eea
with $\nu=pq/m_p$. 
With Hand's convention \cite{Hand:1963bb} the transverse and longitudinal 
$\gamma^*$-proton cross sections are \cite{Miller:1971qb} 
\bea\label{VI55fa}
\sigma_T(s,Q^2)
&=&{}
\frac{2\pi m_p}{s-m^2_p} \,
\varepsilon^{\mu *}_+e^2W_{\mu\nu}~\varepsilon^\nu_+
\nonumber\\
&=&{}
\frac{2\pi m_p}{s-m^2_p}\,\varepsilon^{\mu *}_- 
e^2W_{\mu\nu}~\varepsilon^\nu_-
\nonumber\\
&=&{}
\frac{2\pi m_p}{s-m^2_p} \, e^2W_1(\nu,Q^2) \,,
\\
\label{VI55fb}
\sigma_L(s,Q^2)
&=&{}
\frac{2\pi m_p}{s-m^2_p} \,
\varepsilon'^{\mu *}_L e^2W_{\mu\nu} \, \varepsilon'^\nu_L
\nonumber\\
&=&{}
\frac{2\pi m_p}{s-m^2_p}\,
\varepsilon^{\mu*}_L e^2W_{\mu\nu} \, \varepsilon^\nu_L
\nonumber\\
&=&{}
\frac{2\pi m_p}{s-m^2_p}\,
\left[e^2W_2(\nu,Q^2) \,
\frac{\nu^2+Q^2}{Q^2}-e^2W_1(\nu,Q^2)\right] \,. 
\eea
Here the $\gamma^*$-polarisation vectors are as in (\ref{defepspm}), 
(\ref{defepslong}), (\ref{correctepslongcomp}). 

From (\ref{II3}) and (\ref{II15}) we get for the high energy limit 
of the Compton amplitude 
\bea\label{II17}
\lefteqn{ \!\!\!\!\!\!
(2\pi)^4\delta^{(4)}(p'+q'-p-q){\cal M}^{\mu\nu}_{s's}(p',p,q) }
\nonumber\\
&\xrightarrow[|\mathbf{q}|,|\mathbf{q}'|\rightarrow\infty]~{}
&{}
\frac{i}{2\pi m_p} \,i \sum_{q'}\sum_{t',t}\int \frac{d^3l}{(2\pi)^32l^02l^{\prime 0}}
\frac{1}{\Delta E'}~ 
i \sum_q \sum_{r',r}\int \frac{d^3k}{(2\pi)^32k^02k^{\prime 0}}\frac{1}{\Delta E}
\nonumber\\
&&{} \hspace*{2cm}
Q_{q'}\bar{v}_{t'}(l')\left[\Gamma^{(q')\mu}(-l',l)+
L^{(q')\mu}(-l',l)\right]u_t(l)
\nonumber\\
&&{}\hspace*{2cm}
\langle\bar{q}'(l',t'),q'(l,t),p(p',s'),\textup{out}|~\bar{q}(k',r'),q(k,r),p(p,s),
\textup{in}\rangle
\nonumber\\
&&{}\hspace*{2cm}
Q_q\bar{u}_r(k)\left[\Gamma^{(q)\nu}(k,-k')+
L^{(q)\nu}(k,-k')\right]v_{r'}(k') \,.
\eea
Here $\Delta E$, $k$ and $k'$ are as in (\ref{defDeltaE}), (\ref{3.2}), 
and similarly we define 
\bea\label{II18}
l&=& \left(\begin{array}{c}
\sqrt{\mathbf{l}^2+m^2_{q'}}\\\mathbf{l}\end{array}\right)\,,
\nonumber\\
l' &=& \left(\begin{array}{c}
\sqrt{(\mathbf{q}'-\mathbf{l})^2+m^2_{q'}}\\
\mathbf{q}'-\mathbf{l}\end{array}\right) \,,
\nonumber\\
\Delta E' &=& l^0+l^{\prime 0}-q^{\prime 0} \,.
\eea
Treating quarks and antiquarks in (\ref{II17}) as asymptotic states we can write
\bea\label{II19}
\lefteqn{\hspace*{-1cm}
\langle\bar{q}'(l',t'),q'(l,t),p(p',s'),\textup{out}|~
\bar{q}(k',r'),q(k,r),p(p,s),\textup{in}\rangle }
\nonumber\\
&=&{}
\delta_{fi}+i(2\pi)^4\delta^{(4)}
(l'+l+p'-k'-k-p)\nonumber\\
&&{}
\langle\bar{q}'(l',t'),q'(l,t),p(p',s')|{\cal T}|
\bar{q}(k',r'),q(k,r),p(p,s)\rangle \,, 
\eea
where ${\cal T}$ represents the full $T$-matrix element for this 
scattering process with three incoming and three outgoing particles. 
The contribution $\delta_{fi}$ gives rise to the disconnected part of 
the matrix element in (\ref{II16}). It does not contribute to the 
cross section and will not be considered further here. 
In the high energy limit we have
\be\label{II20}
\delta^{(4)}(l'+l+p'-k-k'-p)
\, \longrightarrow \, 
\delta^{(4)}(p'+q'-p-q) 
\ee
neglecting the terms $\Delta E$ and $\Delta E'$ in the energy sum on the l.h.s. 
With this we get from (\ref{II17})
\bea\label{II21}
{\cal M}^{\mu\nu}_{s's}(p',p,q) \!\!\!\!&&\!\!\!\!
\xrightarrow[|\mathbf{q}|,|\mathbf{q}'|\rightarrow\infty]~{}\frac{1}{2\pi m_p}
\sum_{q'} \sum_{t',t}\int\frac{d^3l}{(2\pi)^32l^02l^{\prime0}}\frac{1}{\Delta E'}
\sum_q \sum_{r',r}\int\frac{d^3k}{(2\pi)^32k^02k^{\prime0}}\frac{1}{\Delta E}
\nonumber\\
&&{} \hspace*{2.2cm}
Q_{q'}\bar{v}_{t'}(l')
\left[\Gamma^{(q')\mu}(-l',l)+L^{(q')\mu}(-l',l)\right] u_t(l)
\nonumber\\
&&{} \hspace*{2.1cm}
\left.
\langle \bar{q}'(l',t'),q'(l,t),p(p',s')|{\cal T}
|\bar{q}(k',r'),q(k,r),p(p,s)\rangle \right|_{\rm asympt}
\nonumber\\
&&{} \hspace*{2.2cm}
Q_q\bar{u}_r(k)
\left[\Gamma^{(q)\nu}(k,-k')+L^{(q)\nu}(k,-k')\right] v_{r'}(k') \,.
\eea
Note that when taking the high energy limit of the Compton amplitude 
we want to keep only the contributions to the $T$-matrix element of 
(\ref{II19}) which are leading at high energies, as indicated by the 
subscript. 
In section I.\ref{sec:relsize} we have considered the high energy 
limit $|\mathbf{q}| \to \infty$ for the incoming photon and have 
found that the leading contributions are contained in the amplitudes 
${\cal M}^{(a)}$ and ${\cal M}^{(b)}$ of our skeleton decomposition 
of the Compton amplitude. In both of these amplitudes the incoming 
and the outgoing photon can be treated in a completely symmetric 
way. That symmetry can be seen from the diagrams representing 
these classes in figure I.\ref{fig2} and has also been indicated 
at the end of appendix I.\ref{appA} for the example of 
${\cal M}^{(a)}$. It is therefore straightforward to establish that 
the leading contribution to the matrix element in (\ref{II21}) 
is contained in the diagram classes corresponding to the amplitudes 
${\cal M}^{(a)}$ and ${\cal M}^{(b)}$ also for the simultaneous 
limit $|\mathbf{q}|,|\mathbf{q}'|\to \infty$. 

The result (\ref{II21}) is valid for forward as well as non-forward and 
real as well as virtual Compton scattering. 

\subsection{The usual dipole picture}
\label{subsec:dipolepicture}

To finally arrive at the standard formulae for the dipole model 
we consider forward Compton scattering. We need to make several 
assumptions (i)-(v) which we now discuss in detail. 

We start with two assumptions which we have already used before: 
\begin{itemize}
\item [(i)] 
Quarks of flavour $q$ have a mass shell $m_q$ and can 
be considered as asymptotic states. 
\item [(ii)]
The rescattering terms $L^{(q')\mu}$ and $L^{(q)\nu}$ in (\ref{II21}) 
are dropped and the vertex functions $\Gamma^{(q')\mu}$ and 
$\Gamma^{(q)\nu}$ are replaced by the lowest order terms in perturbation 
theory, that is $\Gamma^{(q')\mu}\rightarrow \gamma^\mu,\Gamma^{(q)\nu}
\rightarrow\gamma^\nu$.
\end{itemize} 
We have made use of assumption (i) when we defined dipole 
states in section \ref{sec:genphotons} above. Assumption (ii) has been used 
for deriving the photon wave function at leading order in section 
\ref{wavefunctsec}, and was applied also when we obtained the corresponding 
formulae for photons in the final state in section \ref{sec:realvirtleadingorder}. 
Since the rescattering terms $L^{(q)}$ start at order $\alpha_s$ in 
perturbation theory dropping them is consistent with keeping only the 
leading order contribution to the vertex function $\Gamma^{(q)}$. 
In higher orders there will be correction terms to both the 
vertex functions and the rescattering terms. 

\begin{figure}[ht]
\begin{center}
\includegraphics[width=11.5cm]{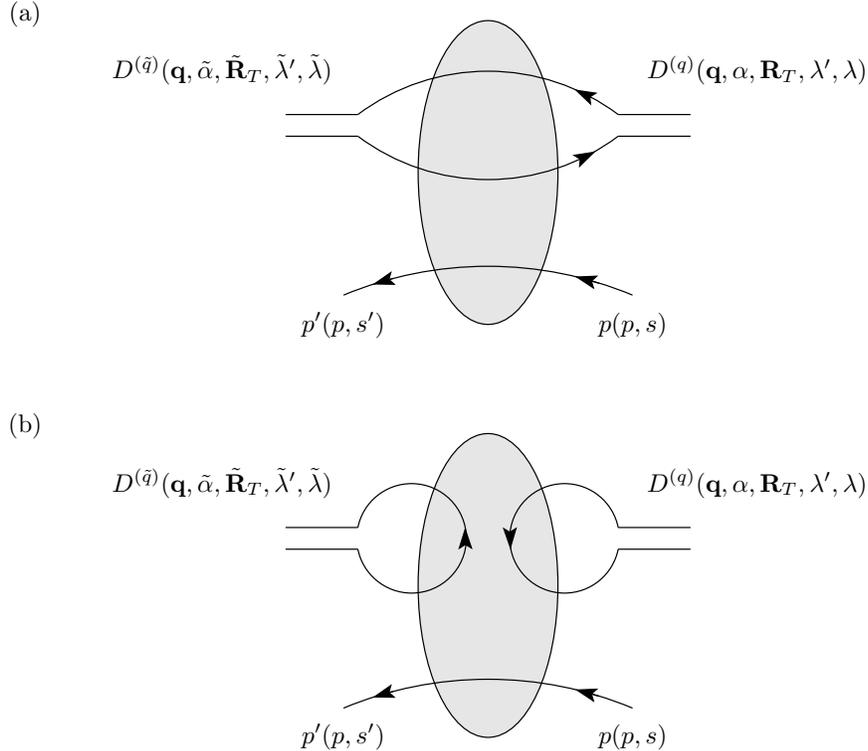}
\caption{The diagram classes containing the leading contribution 
to the $T$-matrix element of dipole states in (\protect\ref{VI55a}) 
in the high energy limit. The classes (a) and (b) result from the 
amplitudes ${\cal M}^{(a)}$ and ${\cal M}^{(b)}$ of figure I.\ref{fig2}, 
respectively. 
\label{fig:Tab}}
\end{center}
\end{figure}

With these assumptions we find from (\ref{II21}) with the photon wave 
function to lowest order as given in (\ref{defpsischlange}), 
(\ref{fourierwavefunction}) and with (\ref{V103}) and (\ref{V111}) 
\bea\label{VI55a}
\lefteqn{
\left. e^2{\cal M}^{\mu\nu}_{s' s}(p,p,q)\right|_{\textup{asympt}} }
\nn \\
&=&{}
\frac{1}{2\pi m_p}\sum_{\tilde{q}} \sum_{\tilde{\lambda}',\tilde{\lambda}}
\int^1_0 d\tilde{\alpha}\int d^2\tilde{R}_T
\sum_q \sum_{\lambda',\lambda}\int^1_0 d\alpha \int d^2R_T
\left(\psi^{(\tilde{q})\mu}_{\gamma,\tilde{\lambda}\tilde{\lambda}'}
(\tilde{\alpha},\tilde{\mathbf{R}}_T,Q)\right)^*
\nonumber\\
&&{}
\langle D^{(\tilde{q})}(\mathbf{q},\tilde{\alpha},\tilde{\mathbf{R}}_T,
\tilde{\lambda}',\tilde{\lambda}),p(p,s')|
{\cal T}|D^{(q)}(\mathbf{q},\alpha,\mathbf{R}_T,\lambda',\lambda),p(p,s)\rangle
\nonumber\\
&&{}
\psi^{(q)\nu}_{\gamma,\lambda\lambda'}(\alpha,\mathbf{R}_T,Q) \,,
\eea
where the subscript on the l.h.s.\ stands for the high energy 
limit, $|\mathbf{q}| \to \infty$, and $D$ are the dipole states defined in 
(\ref{V104}). At this stage only the leading contributions in the 
high energy limit should be included in the $T$-matrix element. 
As pointed out in section \ref{subsec:dis} above they are contained 
in the amplitudes ${\cal M}^{(a)}$ and ${\cal M}^{(b)}$ corresponding 
to the classes (a) and (b) of our skeleton decomposition of the Compton 
amplitude introduced in I. These two classes are shown again in 
figure \ref{fig:Tab}, now for incoming and outgoing dipole states. 
The shaded blobs indicate again the functional integration over all 
gluon potentials with a functional measure including the fermion 
determinant. The discussion of typical diagrams and of the relative 
size of the diagram classes presented in section I.\ref{sec:relsize} 
for incoming and outgoing photons holds in exactly the same way 
here for dipole states. In the following we will generically denote 
the two contributions to the $T$-matrix for dipole states 
corresponding to the two diagram 
classes of figure \ref{fig:Tab} by ${\cal T}^{(a)}$ and ${\cal T}^{(b)}$, 
respectively. 

The following assumptions will concern the $T$-matrix element of dipole 
states. Since the dipole picture is usually applied only to the scattering of 
photons off unpolarised protons we will in fact need to make these 
assumptions only for the proton spin averaged $T$-matrix element 
\be
\label{spinavertmatrixel}
\frac{1}{2} \sum_{s',s} \delta_{s's}
\langle D^{(\tilde{q})}(\mathbf{q},\tilde{\alpha},\tilde{\mathbf{R}}_T,
\tilde{\lambda}',\tilde{\lambda}),p(p,s')|{\cal T}|
D^{(q)}(\mathbf{q},\alpha,\mathbf{R}_T,\lambda',\lambda),p(p,s)\rangle
\,.
\ee
We will nevertheless formulate the next two assumptions for the $T$-matrix 
element before averaging and point out some interesting issues concerning 
the proton spin in our discussion. 

Our third assumption is the following:
\begin{itemize}
\item [(iii)] 
The $T$-matrix element occurring in (\ref{VI55a}) is diagonal 
in the quark flavour, in $\alpha$ and $\mathbf{R}_T$, and is proportional 
to the unit matrix in the space of spin orientations of the quark and 
antiquark in the dipole. 
\end{itemize}
As we will see this assumption is crucial for the dipole picture, 
but it is far from obvious whether it can be derived on general grounds. 

Our assumption (iii) is known to hold for the simplest perturbative 
diagrams (one of them is shown in figure I.\ref{fig200}), 
both for the case without or with the resummation of leading 
logarithms of the energy, see for example \cite{Forshaw:1997dc}. 
These diagrams also illustrate well the intuitive picture 
of high energy scattering on which the more general assumption (iii) 
is based. According to this picture a highly energetic quark 
or antiquark follows a straight trajectory and does not change its transverse 
position or its longitudinal momentum by a sizable amount while being 
scattered in the very forward direction, and similarly one expects its 
helicity to be conserved. For a discussion of this in a nonperturbative 
framework see for instance \cite{Nachtmann:1991ua,Nachtmann:1996kt}. 
The interaction would correspondingly be diagonal 
in flavour, in $\alpha$ and $\mathbf{R}_T$, and in the helicity. This picture 
has been quite successful phenomenologically; it for example forms the 
basis of all eikonal formulations of high energy scattering. 
A fully nonperturbative derivation of this picture including an estimate 
of possible corrections could presumably be done along the lines 
of \cite{Nachtmann:1991ua,Nachtmann:1996kt}, but this remains 
to be worked out. 

Let us for instance consider the diagonality of the interaction in the 
transverse position of the quark. The idea that hadronic states 
in which the partons have fixed transverse positions become 
eigenstates of the $T$-matrix in the high energy limit has been 
put forward already in \cite{Miettinen:1978jb}. In the particular 
case of the dipole picture one often argues that the typical lifetime 
of the quark-antiquark pair into which the photon fluctuates 
is much longer than the typical timescale of the interaction with 
the proton and that therefore the transverse positions (or other properties) 
of the quark and antiquark should not change during that interaction. 
But this argument does not make any reference to the actual 
dynamics of the interaction and in our opinion would need more 
rigorous support. While for soft gluon exchanges that argument 
can be made more concrete \cite{Nachtmann:1991ua} the 
situation remains to be clarified for general interactions. 
Interestingly, the results found in \cite{Bialas:2000xs} indicate 
that the diagonality in  $\mathbf{R}_T$ is violated already 
in perturbation theory if one goes beyond the leading logarithmic 
approximation by taking into account exact gluon kinematics in the 
framework of $k_T$-factorisation \cite{Catani:1990eg}. It would 
be very interesting to study this issue in more detail in the framework 
of a full calculation in next-to-leading logarithmic approximation. 
This could lead to a more solid test of the validity of assumption (iii) 
and hence of the foundation of the dipole picture at least in perturbation 
theory. 

An interesting issue is also the assumption that the helicity 
of the quark and the antiquark are conserved in the interaction. 
This helicity conservation has already been found in the 
perturbative study of various QED processes at high energy, 
see for example \cite{Cheng:1969bf,Abarbanel:1969ek,Chang:1970by}. 
The results obtained there for the leading terms at high energy 
can be applied also for gluon exchanges between quarks and 
hence also hold in QCD. In perturbation theory one finds that there 
are also contributions involving a helicity flip if one goes to the 
next-to-leading logarithmic approximation, see for example 
\cite{Fadin:1999df} and references therein. It hence remains 
an open question to which accuracy the conservation of 
quark helicities holds in general. An interesting possibility to 
address this question might be the study of spin-dependent 
structure functions as we will explain in section \ref{sec:spindependence}. 

As we have pointed out the intuitive picture of high energy scattering 
described above is reflected in the simplest perturbative diagrams, 
see for example figure I.\ref{fig200}. 
But one should remember that those diagrams contribute only 
to the amplitude ${\cal M}^{(a)}$ and hence to the part 
${\cal T}^{(a)}$ of the full $T$-matrix element, see figure \ref{fig:Tab}a.  
One can easily see that the diagonality in assumption (iii) is 
plausible only for that part ${\cal T}^{(a)}$. Let us consider 
the second contribution ${\cal T}^{(b)}$ to the $T$-matrix element 
(corresponding to the part ${\cal M}^{(b)}$ of the amplitude), 
see figure \ref{fig:Tab}b. One of the simplest diagrams contributing 
to  ${\cal T}^{(b)}$ is shown in figure I.\ref{fig200b}. 
The flavours in the two quark loops in this diagram are independent 
of each other, and the contribution of identical quark flavours in the 
loops constitutes only the smaller part of the amplitude. 
Further it appears extremely unlikely that the outgoing quark and antiquark 
should have a high probability to be produced at the same positions in 
transverse space and to carry the same longitudinal momentum fractions 
as the incoming quark and antiquark. Similar considerations can be 
applied to the full matrix element represented by figure \ref{fig:Tab}b. 
Hence we can at most expect that 
the part of the $T$-matrix element corresponding to 
the amplitude ${\cal M}^{(a)}$ satisfies the assumption (iii). 

In the light of this discussion of the simplest perturbative diagrams 
corresponding to ${\cal T}^{(b)}$ we can say that the above 
assumption (iii) is likely to exclude most contributions to the scattering 
amplitude contained in the amplitude ${\cal M}^{(b)}$. 
However, we cannot make this statement rigorous based 
on fully nonperturbative considerations. Thus, although it is most 
probably a consequence of assumption (iii) we formulate it 
as a separate assumption of the dipole model: 
\begin{itemize}
\item [(iv)]
In the $T$-matrix element in (\ref{VI55a}) only the contribution 
${\cal T}^{(a)}$ is kept while ${\cal T}^{(b)}$ is neglected. 
\end{itemize}
Recall that the amplitude ${\cal M}^{(b)}$ 
contributes only at higher orders of $\alpha_s$ 
compared to ${\cal M}^{(a)}$, but at low photon virtualities its 
contribution can well be important. 

Finally we need to make the following assumption. 
\begin{itemize}
\item [(v)]
The proton spin averaged reduced matrix element 
depends only on $\mathbf{R}^2_T$ and on $s=(p+q)^2$. 
\end{itemize}
This assumption implies in particular that the reduced matrix 
element is independent of the longitudinal momentum fractions 
of the quark and antiquark given by $\alpha$. That reflects 
the usual factorisation of transverse and longitudinal degrees of 
freedom in high energy reactions. But also for this assumption 
a general proof is not known. 

Note that we have assumed here a dependence of the reduced 
matrix element on the squared energy $s$ rather than on 
Bjorken-$x$. We will explain our choice and 
discuss it in more detail in section \ref{sec:energyvariable} below. 

With the assumptions (iii)-(v) we find for the spin averaged 
$T$-matrix element at high energy 
\bea\label{VI55b}
\lefteqn{
\hspace*{-1.5cm}
\frac{1}{2} \sum_{s',s} \delta_{s's}
\langle D^{(\tilde{q})}(\mathbf{q},\tilde{\alpha},\tilde{\mathbf{R}}_T,
\tilde{\lambda}',\tilde{\lambda}),p(p,s')|{\cal T}|
D^{(q)}(\mathbf{q},\alpha,\mathbf{R}_T,\lambda',\lambda),p(p,s)\rangle
}
\nonumber\\
&=&{}
\delta_{\tilde{q}q}\, \delta_{\tilde{\lambda}'\lambda'} \,
\delta_{\tilde{\lambda}\lambda} \,
\delta(\tilde{\alpha}-\alpha) \,
\delta^{(2)}(\tilde{\mathbf{R}}_T-\mathbf{R}_T) \,
{\cal T}^{(q)}_{\rm red}(\mathbf{R}^2_T,s) \,, 
\eea
where ${\cal T}^{(q)}_{\rm red}$ is the reduced $T$-matrix element 
and involves only contributions coming from ${\cal T}^{(a)}$. The fact that 
${\cal T}^{(q)}_{\rm red}$ is a function of $\mathbf{R}^2_T$ and $s$ 
only is an immediate consequence of assumption (v). 
The more conventional definition of the reduced $T$-matrix element 
would include all factors depending on the continuous parameters of 
the $T$-matrix element. Here we have explicitly taken out the 
two singular factors given by the two delta-functions involving 
$\alpha$ and $\mathbf{R}_T$ in order to have a reduced 
$T$-matrix element that can as usual be assumed to be a 
smooth function of its arguments. 
Alternatively, we could have chosen to approximate the delta-functions 
involving $\alpha$ and $\mathbf{R}_T$ in (\ref{VI55b}) by strongly 
peaked but smooth functions and to absorb them into the definition 
of the reduced matrix element. That would correspond to requiring 
an approximate diagonality in assumption (iii). 

We can now obtain the reduced cross section for the scattering of 
a dipole state on an unpolarised proton from the reduced 
$T$-matrix element via the optical theorem, 
\be\label{VI55cunsmeared}
\sigma^{(q)}_{\rm red}(\mathbf{R}^2_T,s)=
\frac{1}{s} \,{\rm Im}\, {\cal T}^{(q)}_{\rm red}
(\mathbf{R}^2_T,s) \ge 0
\,.
\ee
Recall that we have assumed that quarks have a mass-shell 
in assumption (i). Therefore we can now conclude that this 
reduced cross section is non-negative since it can due to that 
assumption be related to a physical three-to-three scattering 
process $q\bar{q}p \to q\bar{q}p$. In this sense we can further 
below interpret the reduced cross section $\sigma^{(q)}_{\rm red}$ 
as the cross section for a scattering of a dipole on an unpolarised proton. 
Note that the reduced cross section can naturally be expected to 
depend on the quark flavour $q$, for instance via the quark mass $m_q$. 

Instead of considering the original dipole states $D$ we could also 
consider the corresponding $T$-matrix element of smeared 
dipole states $\bar{D}$ as introduced in (\ref{V109}). 
Then we obtain from (\ref{VI55b}) for large $s$ 
\bea\label{VI55bsmeared}
\lefteqn{
\hspace*{-3cm}
\frac{1}{2} \sum_{s',s} \delta_{s's}
\langle \bar{D}^{(\tilde{q})}(\mathbf{q},\bar{\alpha},\bar{\mathbf{R}}_T,
\tilde{\lambda}',\tilde{\lambda}),p(p,s')|{\cal T}|
\bar{D}^{(q)}(\mathbf{q},\bar{\alpha},\bar{\mathbf{R}}_T,
\lambda',\lambda),p(p,s)\rangle
}
\nonumber\\
&=&{}
\delta_{\tilde{q}q}\, \delta_{\tilde{\lambda}'\lambda'} \,
\delta_{\tilde{\lambda}\lambda} \,
\bar{{\cal T}}^{(q)}_{\rm red}(\bar{\mathbf{R}}^2_T,s) 
\eea
with 
\be
\label{Tredsmeared}
\bar{{\cal T}}^{(q)}_{\rm red}(\bar{\mathbf{R}}^2_T,s) 
=
\int d^2 R_T \,
|f_T (\mathbf{R}_T-\bar{\mathbf{R}}_T) |^2\,
{\cal T}^{(q)}_{\rm red}(\mathbf{R}^2_T,s) \,.
\ee
We can interpret $\bar{{\cal T}}^{(q)}_{\rm red}$ as the genuine 
proton spin averaged $T$-matrix element for smeared dipole states 
and can assume that it is a smooth function of its arguments. 
Note that in the matrix element (\ref{VI55bsmeared}) 
the incoming and outgoing dipole states need to have the same 
$\bar{\alpha}$ and $\bar{\mathbf{R}}_T$ in order to arrive 
at (\ref{Tredsmeared}). We can then again invoke the optical 
theorem to obtain the total cross section for a smeared dipole state, 
a hadron analogue, scattering on an unpolarised proton for large $s$ as 
\be\label{VI55c}
\bar{\sigma}^{(q)}_{\rm red}(\mathbf{\bar{R}}^2_T,s)
=\frac{1}{s} \,{\rm Im}\,
\bar{{\cal T}}^{(q)}_{\rm red}
(\mathbf{\bar{R}}^2_T,s) 
= \int d^2 R_T \, 
|f_T (\mathbf{R}_T-\bar{\mathbf{R}}_T) |^2\,
\sigma^{(q)}_{\rm red}(\mathbf{R}^2_T,s)
\,.
\ee
For a narrow smearing function $f_T$ we get 
$\sigma^{(q)}_{\rm red}(\mathbf{R}^2_T,s) \approx 
\bar{\sigma}^{(q)}_{\rm red}(\mathbf{R}^2_T,s)$, 
that is (\ref{VI55cunsmeared}) gives a correctly normalised 
hadron-analogue cross section. 

Returning now to the unsmeared dipole states we obtain with 
(\ref{VI55b}) from (\ref{VI55a}) 
\bea\label{VI55d}
e^2 \frac{1}{2} \sum_{s',s} \delta_{s's} 
\left.
{\cal M}^{\mu\nu}_{s's}(p,p,q)
\right|_{\textup{asympt}} \!&=&{}\!
\frac{1}{2\pi m_p}
\sum_q \sum_{\lambda,\lambda'}
\int^1_0 d\alpha\int d^2R_T
\left(\psi^{(q)\mu}_{\gamma,\lambda\lambda'}(\alpha,\mathbf{R}_T,Q)\right)^*
\nn \\
&&{} \hspace*{1cm}
{\cal T}^{(q)}_{\rm red}(\mathbf{R}^2_T,s) \,
\psi^{(q)\nu}_{\gamma,\lambda\lambda'}(\alpha,\mathbf{R}_T,Q) \,.
\eea
This gives us for the hadronic tensor of (\ref{VI55e}) at high energy 
\bea\label{VI55h}
\left. e^2W^{\mu\nu}(p,q)\right|_{\textup{asympt}}
&=& 
\frac{1}{2\pi m_p}\sum_q \sum_{\lambda,\lambda'}
\int^1_0d\alpha \int d^2R_T 
\left(\psi^{(q)\mu}_{\gamma,\lambda\lambda'}(\alpha,\mathbf{R}_T,Q)\right)^*
\nonumber\\
&&{}\hspace*{1cm}
s \,\sigma^{(q)}_{\rm red}(\mathbf{R}^2_T,s) \,
\psi^{(q)\nu}_{\gamma,\lambda\lambda'}(\alpha,\mathbf{R}_T,Q)
\,.
\eea
Accordingly, we find with (\ref{VI55fa}) and (\ref{VI55fb}) 
for $\sigma_T$ and $\sigma_L$ at high energy 
\bea\label{VI55g}
\sigma_T(s,Q^2)&=&{}
\sum_q\int d^2 R_T \,
w^{(q)}_T(R_T,Q^2)\,
\sigma^{(q)}_{\rm red}(\mathbf{R}^2_T,s) \,,
\\
\label{VI55i}
\sigma_L(s,Q^2) &=& {}
\sum_q\int d^2R_T \,
w^{(q)}_L(R_T,Q^2)\,
\sigma^{(q)}_{\rm red}(\mathbf{R}^2_T,s) \,,
\eea
where we have defined 
\bea
\label{VI55j}
w^{(q)}_T(R_T,Q^2) &=&{}
\sum_{\lambda,\lambda'}\int^1_0 d\alpha \,
\left|
\psi^{(q)\mu}_{\gamma,\lambda\lambda'}(\alpha,\mathbf{R}_T,Q)
\,\varepsilon_{+\mu} \right|^2 
\nn \\
&=&{}
\sum_{\lambda,\lambda'}\int^1_0 d\alpha \,
\left|
\psi^{(q)\mu}_{\gamma,\lambda\lambda'}(\alpha,\mathbf{R}_T,Q)
\,\varepsilon_{- \mu} \right|^2 \,,
\\
\label{VI55k}
w^{(q)}_L(R_T,Q^2) &=&{}
\sum_{\lambda,\lambda'}\int^1_0 d\alpha\,
\left| 
\psi^{(q)\mu}_{\gamma,\lambda\lambda'}(\alpha,\mathbf{R}_T,Q)
\,\varepsilon'_{L\mu} \right|^2 \,.
\eea
We recall that in (\ref{VI55k}) we have to use the polarisation vector 
$\varepsilon'_L$ (\ref{correctepslong}) since only then do we get 
the correct asymptotic expression, as was explained in 
section \ref{wavefunctsec}. Here and in the rest of the paper we only 
deal with the asymptotic expressions and no longer indicate this 
by a subscript. Note that $w_L^{(q)}$ and $w_T^{(q)}$ depend only 
on $R_T$ but not on the orientation of $\mathbf{R}_T$ 
due to (\ref{psipmcoord}) and (\ref{psiLposition}). 

With (\ref{VI55g})-(\ref{VI55k}) we have finally arrived at the standard 
formulae for the dipole picture used extensively in the literature, 
see for example \cite{Donnachie:en}. We have tried to spell 
out all the assumptions that enter and that should be carefully 
examined when one wants to draw conclusions, for instance concerning 
saturation, from a supposed `unitarity limit' for 
$\sigma_{\rm red}(\mathbf{R}^2_T,s)$. 
We consider it an important task for the future to test the assumptions and 
to quantify their accuracy. 

Note that in the dipole formulae (\ref{VI55g}) and (\ref{VI55i}) 
$R_T$ appears formally as an integration variable. But the interpretation 
of these formulae as a factorisation into the square of a photon wave function 
and a cross section for the scattering of a dipole of transverse size $R_T$ 
on a proton is very natural in view of the relation of the latter to 
the corresponding $T$-matrix element, see (\ref{VI55b}) and 
(\ref{VI55cunsmeared}). We will therefore simply call 
$\sigma^{(q)}_{\rm red}$ a dipole cross section in the following. 
Due to the same relation to a $T$-matrix element involving 
a particular quark flavour it is also natural that this dipole cross section 
still depends on that quark flavour $q$. 

Next we discuss two issues that we have postponed earlier in this section. 

\subsection{Proton spin dependence of the dipole cross section}
\label{sec:spindependence}

The first point is the dependence on the proton spin. So far we 
have considered only the scattering of transversely or longitudinally 
polarised photons on an unpolarised proton, and we have obtained 
the usual formulae of the dipole picture for the total cross sections of 
these processes. As a consequence, we needed to make our assumption (iii) 
only for the proton spin averaged $T$-matrix element 
(\ref{spinavertmatrixel}) rather than for the actual $T$-matrix 
element which occurs in (\ref{VI55a}). Nevertheless we have 
formulated our assumption (iii) for the latter, 
which amounts to making a stronger assumption. 
It now seems very interesting to think about possible tests of the 
dependence of that matrix element on the proton spin. 
To our knowledge, all models proposed so far for the dipole 
cross section $\sigma_{\rm red}$ have been used only for 
calculating scattering processes on unpolarised protons. 
Some of these models are based on an additional assumption 
about the proton, namely that it can be considered 
at high energies as a superposition of colour dipoles described 
by a suitable weight function for their transverse size and 
orientation. The scattering process can then 
be expressed in terms of the cross section of two colour dipoles. 
In phenomenological applications one hence only needs to 
make a model for that simpler cross section. For a 
typical approach along these lines see for example \cite{Shoshi:2002in}. 
We emphasise that this assumption goes beyond the dipole picture as 
we have derived it here, and it is in our opinion far from obvious 
that the proton can be approximated by a collection of dipoles. 
It would of course be very interesting to see if that assumption 
can be justified using the techniques developed in the present work. 
Let us now for a moment adopt that additional assumption. 
One expects the dipole-dipole cross section and the corresponding 
$T$-matrix element 
to be symmetric under the exchange of the two dipoles at least 
in the large-$N_c$ limit in which a colour dipole can be viewed 
as a quark-antiquark pair. In assumption (iii) we have assumed 
that the $T$-matrix element in (\ref{VI55a}) is proportional to 
the unit matrix in the space of spin orientations of the quark and 
antiquark. Accordingly, one would now have to assume that the 
dipole-dipole $T$-matrix element is also proportional to the 
unit matrix in the space of spin orientations corresponding 
to the dipole in the proton. As an immediate consequence one 
finds that in such a picture the polarised proton structure 
function $g_1$ vanishes. Since this is in contradiction with 
the experimental findings (for a review see \cite{Anselmino:1994gn}) 
one might conclude that at least the additional assumption does not 
hold, but that would be premature. On the contrary, that result would be 
in agreement with the general finding that the polarised structure 
function $g_1$ is suppressed at high energy with respect to the 
unpolarised structure function $F_2$. It has been established for 
example in Regge theory that $g_1$ is suppressed with respect to 
$F_2$ by a power of the energy \cite{Heimann:1973hq}, and a similar 
picture emerges also in perturbation theory both in extrapolations 
of Altarelli-Parisi evolution \cite{Altarelli:1977zs} to small Bjorken-$x$ 
\cite{Ahmed:1975tj,Einhorn:1985dy} and in calculations taking into 
account double-logarithmic contributions beyond Altarelli-Parisi evolution 
\cite{Bartels:1995iu,Bartels:1996wc}. In many steps of our arguments leading 
to the dipole picture subleading terms at high energy have been neglected. 
Thus it is consistent that the leading level result for 
an observable that starts at a subleading level is zero. 
Turning this argument around 
the polarised structure function $g_1$ might offer an opportunity to 
study the structure and the size of subleading terms at high energies. 
The most important question is of course whether it is possible to 
find a consistent description of $g_1$ (or of any other observable that 
starts at a subleading level in the expansion in inverse powers of the energy) 
in the dipole picture. This problem could be studied independently of the 
additional assumption that the proton can be considered as a superposition 
of colour dipoles. (Note, however, that that additional assumption is at least 
consistent with the known results about $g_1$.) 
Further studies along these lines might either show that also subleading 
terms can be factorised according to the dipole picture, or tell us about 
a potential breakdown of the dipole picture for subleading terms. 
In the latter case one could even hope to quantify the size of potential 
corrections to the dipole picture in general. 

\subsection{The energy variable of the dipole cross section}
\label{sec:energyvariable}

The other important issue that we have not yet discussed in detail 
is the question whether the dipole cross section $\sigma_{\rm red}$ 
should depend on the squared energy $s$ or on Bjorken-$x$ in 
addition to the dependence on the dipole size $R_T$. Since at high 
energies $x_{\rm Bj}=Q^2/s$ the latter possibility would imply a 
dependence of the dipole cross section on $Q^2$ while such a dependence 
would be absent if the former possibility were chosen. 
Phenomenological models for the dipole cross section have been 
proposed based on both possibilities, and both can lead to a satisfactory 
description of the available data. Overall, models based on an 
$x_{\rm Bj}$-dependent dipole cross section appear to be more popular 
at present. A model with an $s$-dependent dipole cross section has been 
constructed for instance in \cite{Forshaw:1999uf} 
while a prominent example for an $x_{\rm Bj}$-dependent dipole cross 
section is the model of \cite{Golec-Biernat:1998js}. 
In our approach we have obtained the dipole cross section from a 
$T$-matrix element for the scattering of a dipole state on a proton, 
see (\ref{VI55b}) and (\ref{VI55cunsmeared}). But as we have 
discussed in section \ref{sec:realvirtleadingorder} the dipole states 
entering this matrix element are independent of the photon virtuality 
$Q^2$ at high energy as can be seen from the expression for their 
invariant mass (\ref{P2aufdipolhe}), see also the corresponding 
formulae (\ref{meanP2smeareddip}) and (\ref{meanM2smeareddip}) 
for smeared dipole states. 
The relevant $T$-matrix element thus corresponds to a scattering 
process in which the initial and final state is independent of $Q^2$. 
According to this observation we have chosen our dipole cross 
section $\sigma_{\rm red}$ to depend only on $s$ rather than on 
$x_{\rm Bj}$ in our assumption (v). The main motivation for choosing an 
$x_{\rm Bj}$-dependence in many phenomenological models comes 
from studying the simultaneous limit of large $s$ and large $Q^2$. 
In this limit the cross section for photon-proton scattering should 
agree with the perturbative result obtained in the double-leading 
logarithmic approximation (DLLA) \cite{DeRujula:1974rf}. As was 
discussed already in \cite{Nikolaev:1993th} the dipole cross section 
can in that limit be identified up to factors with the gluon density of 
the proton. The latter naturally depends on $x_{\rm Bj}$ and $Q^2$ 
and therefore suggests to choose the dependence on $x_{\rm Bj}$ 
also at lower $Q^2$. Can we decide which variable is the correct 
one for the dipole cross section? 

Let us first recall that $\mathbf{R}_T$ in (\ref{VI55g}) and 
(\ref{VI55i}) is an integration variable. We can therefore perform 
a change of variables 
\be
\label{changeofintvar}
\mathbf{R}_T = \frac{Q_0}{Q}\,\mathbf{R}'_T 
\,,
\ee
where $Q_0$ is some fixed momentum scale. Expressing the 
integrands of (\ref{VI55g}) and (\ref{VI55i}) in terms of the new 
integration variable $\mathbf{R}'_T$ we obtain formulae similar 
to the dipole model that contain a factor 
\be
\label{defsigafterchange}
\hat{\sigma}^{(q)}_{\rm red}(\mathbf{R}'^2_T,Q,x_{\rm Bj}) 
= \sigma^{(q)}_{\rm red}
\left(\frac{Q_0^2}{Q^2}\,\mathbf{R}'^2_T,s\right)
\ee
which we could define as a reduced cross section that depends 
on $\mathbf{R}'^2_T$, $Q$ and $x_{\rm Bj}$ instead of 
$\mathbf{R}^2_T$ and $s$. However, in this case $\mathbf{R}'_T$ 
is no longer the physical size parameter of the colour dipole. Moreover, 
the substitution (\ref{changeofintvar}) would affect also the other 
factor in the integrands of (\ref{VI55g}) and (\ref{VI55i}) and hence 
spoil the interpretation of the integrand as a product of an actual 
dipole-proton cross section with the square of the perturbative photon 
wave function. We therefore discard the mathematical possibility 
of changing the integration variable from $\mathbf{R}_T$ to 
$\mathbf{R}'_T$. 

A discussion of $s$ versus $x_{\rm Bj}$ dependence of the dipole cross 
section was given in chapter 9 of \cite{Donnachie:en} and our remarks 
below follow the same lines. For large photon virtualities $Q^2$ 
one finds that the integrals in (\ref{VI55g}) and (\ref{VI55i}) 
receive the dominant contribution from small $R_T$ due to 
the shape of the weight factors $w^{(q)}_{T,L}$. For small $R_T$ 
the dipole-proton cross section  should become small due to colour 
transparency, and the typical behaviour to be expected at small 
$R_T$ is $\sigma_{\rm red} (\mathbf{R}_T^2,s) \propto R_T^2$. 
Hence a typical ansatz for small $R_T$ is 
\be
\label{sigsmallRexpfactR2}
\sigma^{(q)}_{\rm red}  (R_T^2,s) 
= R_T^2 \, f \left(\frac{1}{s R_T^2} \right) 
\,,
\ee
where $f$ is a dimensionless scaling function. Naive scaling arguments 
suggest this and in QCD we expect (\ref{sigsmallRexpfactR2}) 
to hold up to logarithmic corrections in $R_T$. Inserting 
(\ref{sigsmallRexpfactR2}) in (\ref{VI55g}) and (\ref{VI55i}) 
we find that the explicit factor $R_T^2$ together with the weight 
functions $w^{(q)}_{T,L}(R_T,Q^2)$ and an additional factor 
$R_T$ from the integral measure $d^2R_T$ leads to a pronounced 
maximum in the integrands situated at 
\be
\label{intmaxpos}
R^2_{T\,{\rm max}} = \frac{C^2}{Q^2+4m_q^2} 
\ee
with $C^2\simeq 4$, see p.\ 273 of \cite{Donnachie:en} and 
figure \ref{fig:spikes} in section \ref{densitysect} below. That is, 
for large $Q^2$ mostly dipoles of size $R_{T\,{\rm max}}$
contribute. We may therefore make the replacement 
\bea
\label{replaceinint}
\sigma^{(q)}_{\rm red} (R_T^2,s) 
&\longrightarrow& 
R_T^2 \, f\left(\frac{1}{s R^2_{T\,{\rm max}} }\right) 
\simeq 
R_T^2 \, f\left(\frac{Q^2+4 m^2_q}{s C^2}\right) 
\nn \\
&\longrightarrow& 
R_T^2 \, f\left(\frac{x_{\rm Bj}}{C^2}\right) 
\equiv \tilde{\sigma}_{\rm red}^{(q)} (R_T^2,x_{\rm Bj}) 
\eea
for $Q^2 \gg 4 m_q^2$. 
Hence it is only due to the behaviour of the other factors in the 
integrands and the explicit factor $R_T^2$ in 
(\ref{sigsmallRexpfactR2}) that an $s$-dependent 
$\sigma_{\rm red}$ and an $x_{\rm Bj}$-dependent 
$\sigma_{\rm red}$ lead to the same result at large $Q^2$. 
It is therefore not possible to establish an $x_{\rm Bj}$-dependence 
of the dipole cross section in general based on this argument. 
As we have described above, in our approach an $s$-dependence 
occurs naturally. We cannot exclude that a different derivation of the 
dipole picture based on a different set of assumptions is possible in which 
an $x_{\rm Bj}$-dependence occurs. If one wants to maintain 
the interpretation of $\sigma_{\rm red}$ as an actual cross section 
involving a quark-antiquark colour dipole state, however, one would 
need to define the dipole states in such a way that they depend on 
$Q^2$ in leading order in the expansion in inverse powers of $s$. 
This would in particular exclude the natural definition (\ref{V104}) 
which we have used. 

In this context we should add a further remark. 
It would of course be interesting to establish the relation of our 
derivation of the dipole picture with the perturbative picture of the 
DLLA step by step. But there is a potential problem. In the present 
paper we study the limit 
\be
\label{limitsinfQfixed}
s \to \infty\,, \:\:\:\:\: 
Q^2 \,\,{\rm fixed}\,,
\ee
whereas in the DLLA one takes 
\be
\label{limitsinfQinf}
s \to \infty\,, \:\:\:\:\: 
Q^2 \to \infty \,, \:\:\:\:\: 
\frac{Q^2}{s} \,\,{\rm fixed}\,. 
\ee
It is not at all clear and not necessary that first taking the 
limit (\ref{limitsinfQfixed}) and then letting $Q^2$ become large 
will lead to the same result as taking the limit (\ref{limitsinfQinf}). 
It remains to be seen whether this makes the comparison of our 
derivation with the perturbative approach intrinsically difficult. 

\subsection{Density for the photon wave function in leading order}
\label{densitysect}

In the dipole picture the photon-proton cross section is obtained as the 
convolution of the square of the photon wave function with 
the dipole cross section. The latter cannot be calculated 
from first principles at present and needs to be described by models 
in phenomenological applications of the dipole picture. 
The photon wave function, on the other hand, is known explicitly 
in perturbation theory. We therefore find it useful to discuss here some 
known aspects of the photon wave function and to illustrate some of 
its properties. In the next section we will then derive some phenomenological 
consequences of the dipole picture which will be based only on the 
properties of the perturbative photon wave function but will not 
depend on any particular model assumptions about the dipole cross 
section. 

In the following we consider the densities 
\bea
\label{sumpsi+dens}
v^{(q)}_T (\alpha,R_T,Q^2) &=&
\sum_{\lambda, \lambda'} \left| 
\psi_{\gamma, \lambda \lambda'}^{(q) +} (\alpha, \mathbf{R}_T,Q) \right|^2 
\\
&=&{}
\frac{N_c}{2 \pi^2} \, \alpha_{\rm em} Q_q^2 
\left\{ \left[ \alpha^2 + (1-\alpha)^2 \right] 
\epsilon_q^2 [K_1(\epsilon_q R_T) ]^2 
+ m_q^2 [K_0(\epsilon_q R_T) ]^2 
\right\}
\nn 
\eea
and 
\bea
\label{sumpsiLdens}
v^{(q)}_L(\alpha, R_T,Q^2) &=& 
\sum_{\lambda, \lambda'} \left| 
\psi_{\gamma, \lambda \lambda'}^{(q) L}(\alpha, \mathbf{R}_T,Q) \right|^2 
\nn \\
&=&{} 
\frac{2 N_c}{\pi^2} \, \alpha_{\rm em} Q_q^2 
Q^2 [\alpha (1-\alpha)]^2 [K_0(\epsilon_q R_T) ]^2 
\eea
for transversely and longitudinally polarised photons as obtained 
from the respective photon wave functions in leading order, see 
(\ref{psipmcoord}), (\ref{psiLposition}) and (\ref{defepssubq}). 
Upon integration over the longitudinal momentum fraction $\alpha$ 
these densities give the functions $w_T$ and $w_L$ in (\ref{VI55j}) 
and (\ref{VI55k}) that occur in the dipole formulae (\ref{VI55g}) 
and (\ref{VI55i}), 
\be
\label{wfromv}
w^{(q)}_{T/L}(R_T,Q^2) = \int_0^1 d\alpha \,
v^{(q)}_{T/L} (\alpha,R_T,Q^2) 
\,.
\ee
For our numerical results we assume the light quarks ($q=u,d,s$) to be 
massless while for the heavy quark masses we use $m_c = 1.3\,\mbox{GeV}$ 
and $m_b = 4.6\,\mbox{GeV}$. 

\begin{figure}[htbp]
\begin{center}
\includegraphics[width=14.5cm]{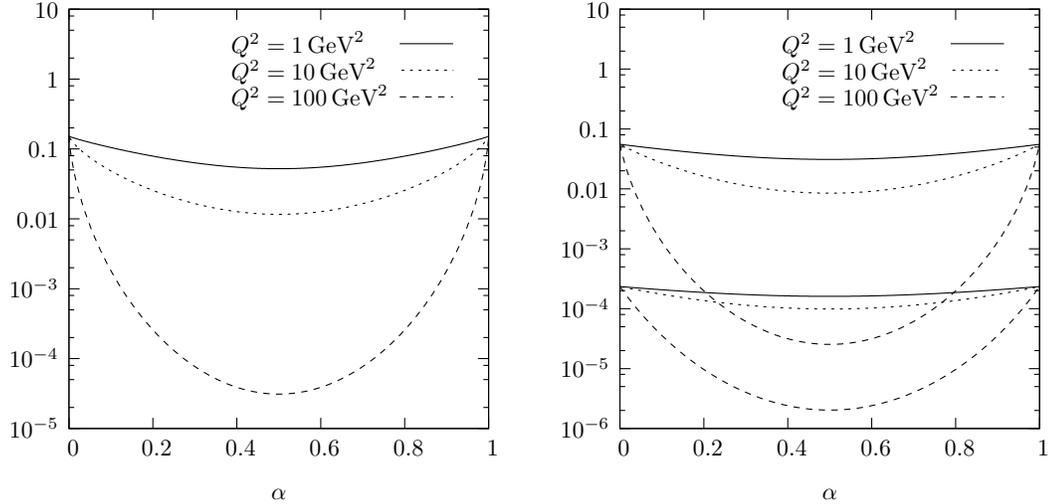}
\end{center}
\caption{Dependence of the density 
$\frac{1}{\alpha_{\rm em} Q_q^2} \, v^{(q)}_T (\alpha, R_T, Q^2)$ 
of transversely polarised photons (see (\ref{sumpsi+dens})) on 
$\alpha$ for $R_T\!=\!1 \,\mbox{GeV}^{-1}$ and for three different 
values $Q^2=1,10,100\,\mbox{GeV}^2$, plotted in units of $\mbox{GeV}^2$; 
on the left for massless ($q=u,d,s$) quarks, 
on the right for $c$-quarks (upper lines) and $b$-quarks (lower lines). 
\label{fig:vTR1Qdep}}
\end{figure}

\begin{figure}[htbp]
\begin{center}
\includegraphics[width=14.5cm]{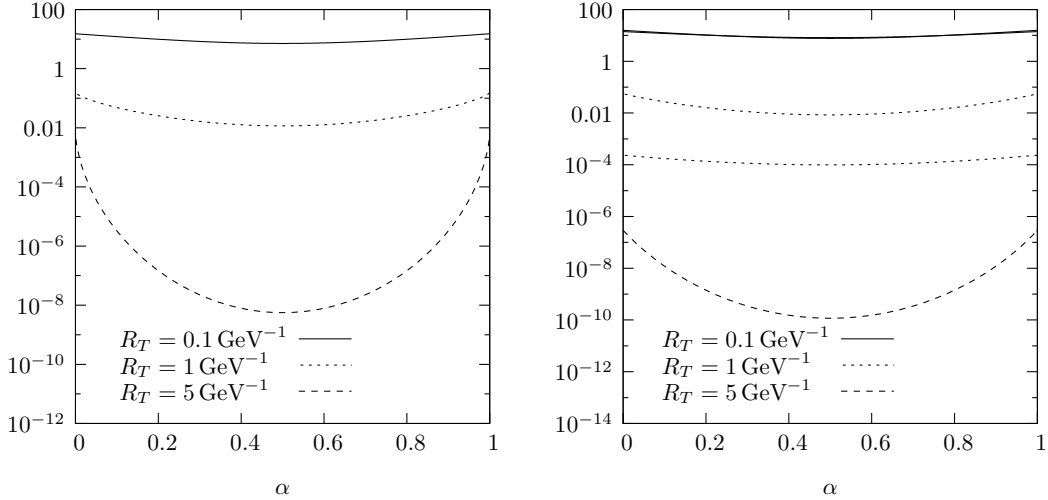}
\end{center}
\caption{Dependence of the density 
$\frac{1}{\alpha_{\rm em} Q_q^2} \, v^{(q)}_T (\alpha, R_T, Q^2)$ 
of transversely polarised photons (see (\ref{sumpsi+dens})) 
on $\alpha$ for $Q^2\!=\!10 \,\mbox{GeV}^2$ and for three different 
values $R_T=0.1,1,5\,\mbox{GeV}^{-1}$, plotted in units of $\mbox{GeV}^2$; 
on the left for massless ($q=u,d,s$) quarks, 
on the right for $c$-quarks (upper lines for each line type) and 
$b$-quarks (lower lines for each line type). 
\label{fig:vTQ10Rdep}}
\end{figure}

We start with transversely polarised photons. Figure \ref{fig:vTR1Qdep} 
shows the dependence of the density 
$v^{(q)}_T/(\alpha_{\rm em} Q_q^2)$ on the longitudinal 
momentum fraction $\alpha$ for fixed $R_T\!=\!1 \,\mbox{GeV}^{-1}$ 
and for three different values of the photon virtuality 
$Q^2=1,10,100\,\mbox{GeV}^2$. The left panel is for massless 
($u,d,s$) quarks while the right panel is for $c$- and $b$-quarks. 
Note that for transversely polarised photons there is a nonvanishing 
contribution from the end points $\alpha=0$ and $\alpha=1$ even 
for very large photon virtualities $Q^2$. As can be seen also from 
(\ref{sumpsi+dens}) the value of $v^{(q)}_T$ at these end points 
depends on the mass of the quark flavour under consideration. 
The density has a minimum at $\alpha=1/2$ for all values of $Q^2$ 
and the relative size of the density at this minimum compared to the 
end points shrinks with increasing $Q^2$. That means that 
a symmetric distribution of the longitudinal momentum 
of the photon between 
the quark and antiquark is the least likely configuration, and this 
effect becomes more pronounced for larger $Q^2$. 
The size of the density decreases rapidly with increasing $Q^2$ 
for all $\alpha$ with the exception of the end points. 
Figure \ref{fig:vTQ10Rdep} shows the dependence 
of $v^{(q)}_T/(\alpha_{\rm em} Q_q^2)$ on $\alpha$ for fixed 
photon virtuality $Q^2\!=\!10 \,\mbox{GeV}^2$ and for three 
different values of the dipole size $R_T=0.1,1,5\,\mbox{GeV}^{-1}$, 
again for massless quarks on the left and for massive quarks on the right. 
While for $R_T=0.1\,\mbox{GeV}^{-1}$ the densities for $c$- 
and $b$-quarks cannot be distinguished when plotted on a logarithmic 
scale like in figure \ref{fig:vTQ10Rdep} their sizes 
become very different for larger $R_T$. Already for 
$R_T=5 \,\mbox{GeV}^{-1}$ the density for the $b$-quark 
is so small that it is far below the range shown in the figure. 
For all quark flavours the dependence of the 
density on $\alpha$ changes with $R_T$ in 
a similar way as it changes with $Q$, in particular it decreases 
rapidly with increasing $R_T$ for all momentum fractions $\alpha$. 
For massless quarks this similarity can be understood immediately 
based on dimensional arguments. Namely, for massless quarks 
the dependence of the density on $Q^2$ is related to the dependence 
on $R_T$ because the dimensionless quantity $R_T^2 v^{(q)}_T$ 
can depend only on the product $Q R_T$ by dimensions. 
However, we have plotted the dimensionful density $v_T$ 
in our figures and hence the values of the density at 
the end points $\alpha=0$ and $\alpha=1$ are different for different 
values of $R_T$. 

\begin{figure}[htbp]
\begin{center}
\includegraphics[width=14.5cm]{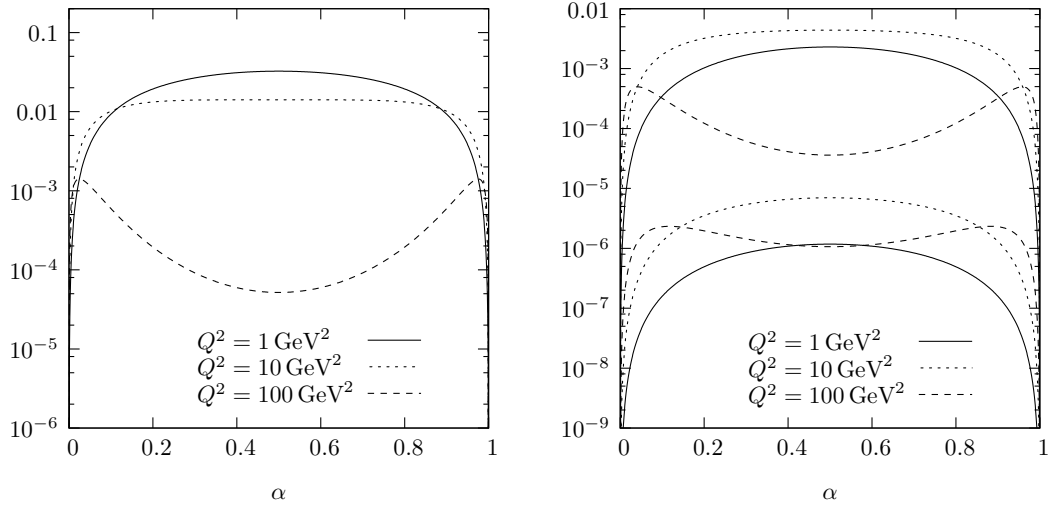}
\end{center}
\caption{Dependence of the density 
$\frac{1}{\alpha_{\rm em} Q_q^2} \, v^{(q)}_L (\alpha, R_T, Q^2)$ 
of longitudinally polarised photons (see (\ref{sumpsiLdens})) 
on $\alpha$ for $R_T\!=\!1 \,\mbox{GeV}^{-1}$ and for three different 
values $Q^2=1,10,100\,\mbox{GeV}^2$, plotted in units of $\mbox{GeV}^2$; 
on the left for massless ($q=u,d,s$) quarks, on the right for 
$c$-quarks (upper lines) and $b$-quarks (lower lines). 
\label{fig:vLR1Qdep}}
\end{figure}

\begin{figure}[htbp]
\begin{center}
\includegraphics[width=14.5cm]{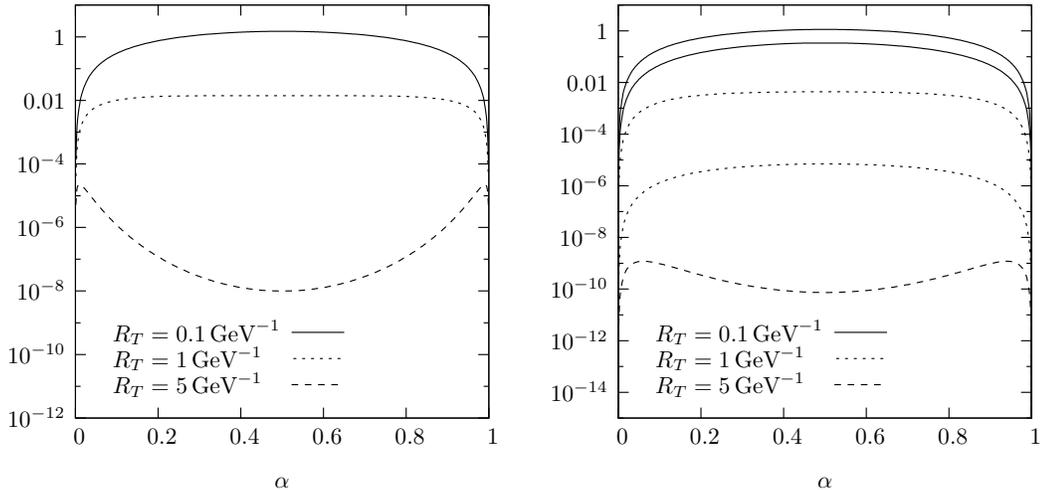}
\end{center}
\caption{Dependence of the density 
$\frac{1}{\alpha_{\rm em} Q_q^2} \, v^{(q)}_L (\alpha, R_T, Q^2)$ 
of longitudinally polarised photons (see (\ref{sumpsiLdens})) 
on $\alpha$ for $Q^2\!=\!10 \,\mbox{GeV}^2$ and for three different 
values $R_T=0.1,1,5\,\mbox{GeV}^{-1}$, plotted in units of $\mbox{GeV}^2$; 
on the left for massless ($q=u,d,s$) quarks, on the right for 
$c$-quarks (upper lines for each line type) and 
$b$-quarks (lower lines for each line type). 
\label{fig:vLQ10Rdep}}
\end{figure}

Next we turn to longitudinally polarised photons. In 
figure \ref{fig:vLR1Qdep} we show the dependence of the density 
$v^{(q)}_L/(\alpha_{\rm em} Q_q^2)$ on the longitudinal 
momentum fraction $\alpha$ for fixed $R_T\!=\!1 \,\mbox{GeV}^{-1}$ 
and for three different values of the photon virtuality 
$Q^2=1,10,100\,\mbox{GeV}^2$. The left panel is again for massless 
($u,d,s$) quarks while the right panel is for $c$- and $b$-quarks. 
In the case of longitudinally polarised photons the density vanishes 
at the end points $\alpha=0$ and $\alpha=1$ for all photon virtualities 
$Q^2$ and for all quark flavours. 
For small photon virtualities the density has a maximum at $\alpha=1/2$. 
For larger $Q^2$ the symmetric distribution of the longitudinal momentum 
of the photon between the quark and the antiquark becomes less likely and 
there appears a local minimum in the density at $\alpha=1/2$. 
At the same time two maxima of the density develop close to the end points 
$\alpha=0$ and $\alpha=1$. For massless quarks the overall size of the 
density decreases rapidly with increasing $Q^2$, except for small regions 
close to the end points where the maxima appear at larger $Q^2$. 
For massive quarks that overall decrease with $Q^2$ sets in only 
for $Q^2>m_q^2$. 
Figure \ref{fig:vLQ10Rdep} shows the dependence of the density 
on $\alpha$ for fixed photon virtuality $Q^2\!=\!10 \,\mbox{GeV}^2$ 
and for three different values $R_T=0.1,1,5\,\mbox{GeV}^{-1}$, again 
for massless quarks on the left and for massive quarks on the right. 
The size of the density decreases with increasing $R_T$, and 
for $R_T = 5\,\mbox{GeV}^{-1}$ the density for $b$-quarks is already so 
small that it is far below the range shown in the figure. For massless quarks 
the dependence of the density on $R_T$ can again be deduced from the 
dependence on $Q^2$ via the dimensionless quantity $R_T^2 v^{(q)}_L$ 
which depends only on $Q R_T$. For large $R_T$ the dependence 
of the density for massive quarks on $R_T$ follows a pattern similar 
to that of massless quarks. 

\begin{figure}[htbp]
\begin{center}
\includegraphics[width=13.8cm]{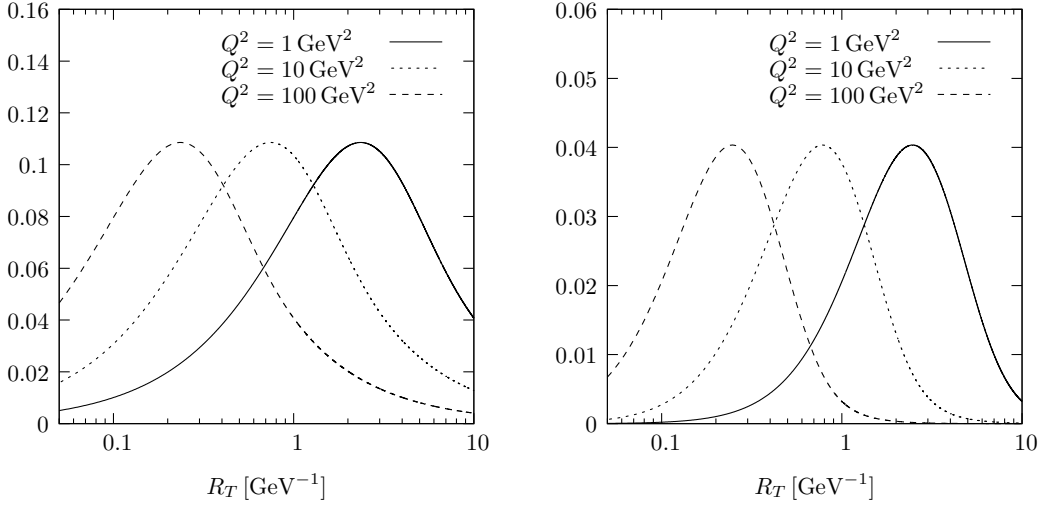}
\end{center}
\caption{Dependence of the dimensionless quantity 
$Q R_T^3 w^{(q)}_{T,L}$ on $R_T$ for massless ($q=u,d,s$) quarks 
for three different values $Q^2=1,10,100\,\mbox{GeV}^2$; 
on the left for transversely polarised photons, on the right for 
longitudinally polarised photons. 
\label{fig:spikes}}
\end{figure}

Finally, we consider the dimensionless quantity $Q R_T^3 w^{(q)}_{T,L}$ 
for transversely and longitudinally polarised photons 
which occurred in our discussion of the energy variable of the dipole 
cross section in section \ref{sec:energyvariable}. There we pointed out 
that the densities $w^{(q)}_{T,L}$ occur in the dipole formulae 
(\ref{VI55g}) and (\ref{VI55i}) together with a factor $R_T$ from 
the integral measure and another factor of $R_T^2$ coming from 
the dipole cross section at small $R_T$, see (\ref{sigsmallRexpfactR2}). 
Multiplying by $Q$ leads us to the dimensionless quantity 
$Q R_T^3 w^{(q)}_{T,L}$. We recall that 
$w^{(q)}_{T,L}$ is obtained from $v^{(q)}_{T,L}$ in (\ref{sumpsi+dens}) 
and (\ref{sumpsiLdens}) by integrating over $\alpha$, see (\ref{wfromv}). 
Figure \ref{fig:spikes} 
shows the dependence of $Q R_T^3 w^{(q)}_{T,L}$ on $R_T$ for 
massless quarks ($q=u,d,s$) and for 
three different photon virtualities $Q^2=1,10,100\,\mbox{GeV}^2$. 
The left panel is for transversely polarised photons and the right panel 
for longitudinally polarised photons. The figure shows that there 
are indeed pronounced maxima, and their position $R_{T {\rm max}}$ 
scales with $1/Q$, see also (\ref{intmaxpos}). For transversely polarised 
photons the maxima turn out to be wider than for longitudinally polarised 
photons. Since for massless quarks the dimensionless 
$Q R_T^3 w^{(q)}_{T,L}$ can depend only on $Q R_T$ the value of this 
quantity at the maximum is the same for all $Q^2$ here. 

\boldmath
\section{A bound on $R=\sigma_L/\sigma_T$ from the 
dipole picture\,\protect\footnote{Some of the results in this section have 
been presented in abridged form also in the Letter \cite{Ewerz:2006an}.}}
\unboldmath
\label{sec:phencons}

In the previous section we have obtained the dipole picture of 
high energy scattering based on a number of assumptions. 
It is obviously important to test these assumptions and 
to find the range of kinematical parameters in which they hold, 
or in other words, the range in which the dipole picture can be applied. 

In the dipole picture the cross section for photon-proton scattering 
is written as a convolution of the square of the perturbative 
photon wave function with the dipole-proton cross section, 
see (\ref{VI55g}) and (\ref{VI55i}). The former can be determined 
explicitly in perturbation theory (see section \ref{wavefunctsec}) 
while the latter cannot be calculated from first principles at present. 
Especially in the interesting region of intermediate and low photon 
virtualities one therefore has to retreat to models for the dipole cross 
section which can then be tested against the available data. 
Although the required agreement with the data places constraints 
on the possible models for the dipole cross section there is still 
considerable freedom in these models. Based on such models it 
is therefore difficult to test the assumptions on which the dipole picture 
is based and to explore their potential limits. It would hence appear 
favourable to find observables which can be used to test the 
assumptions of the dipole picture in a way which does not 
depend on any particular model for the dipole cross section. 

In the present section we want to study such an observable, 
namely the ratio $R$ of the cross sections for longitudinally and 
transversely polarised photons, 
\be
\label{defR}
R(s,Q^2) = \frac{\sigma_L(s,Q^2)}{\sigma_T(s,Q^2)}
\,.
\ee
In the following we will derive a bound on $R$ from the 
dipole picture which will involve only the properties of the 
perturbative wave functions of longitudinally and transversely 
polarised photons. 

First we recall that the quantities $w^{(q)}_T$  and $w^{(q)}_L$ in 
(\ref{VI55j}) and (\ref{VI55k}) are positive since they are obtained 
as integrals over the squared modulus of the perturbative photon 
wave function. Further we recall that the dipole cross section is 
non-negative according to our assumptions made in section 
\ref{subsec:dipolepicture}, see (\ref{VI55cunsmeared}). 
We then start from the obvious relation 
\be 
\label{trivialmaxmin}
\frac{w^{(q)}_L(R_T,Q^2)}{w^{(q)}_T(R_T,Q^2)} 
\, \leq \, 
\max_{q,R_T} \,
\frac{w^{(q)}_L(R_T,Q^2)}{w^{(q)}_T(R_T,Q^2)} \,,
\ee
where the maximum is taken over all dipole sizes $R_T$ and over 
all quark flavours $q$. We then multiply both sides of this relation by the 
positive quantity $w^{(q)}_T(R_T,Q^2)$ and by the dipole cross section 
$\sigma^{(q)}_{\rm red}(R^2_T,s)$ which is non-negative. 
Next we integrate over $\mathbf{R}_T$ and sum over all quark flavours 
which gives with the dipole formulae (\ref{VI55g}) and (\ref{VI55i}) 
\be
\label{zwischenschrittmax}
\sigma_L(s,Q^2) \, \le \, 
\sigma_T(s,Q^2) \, 
\max_{q,R_T} \,\frac{w^{(q)}_L(R_T,Q^2)}{w^{(q)}_T(R_T,Q^2)} 
\,.
\ee
An analogous relation is readily obtained for the minimum of 
$w^{(q)}_L/w^{(q)}_T$, such that we have 
\be
\label{VI55l}
\min_{q,R_T} \,
\frac{w^{(q)}_L(R_T,Q^2)}{w^{(q)}_T(R_T,Q^2)}
\, \leq \, R(s,Q^2)
\, \leq \,\max_{q,R_T} \,
\frac{w^{(q)}_L(R_T,Q^2)}{w^{(q)}_T(R_T,Q^2)} \,.
\ee
With this relation we have obtained an upper and a lower bound on 
the ratio $R(s,Q^2)$ which should be respected for all $s$ and $Q^2$ 
for which the dipole picture is valid or -- phrased differently -- for which 
the assumptions and approximations (i)-(v) hold which we have discussed 
in section \ref{subsec:dipolepicture}. Note that the bounds (\ref{VI55l}) 
depend only on the photon virtuality $Q^2$ but do not depend on $s$. 
This is of course expected since the bounds involve only the photon 
wave function but not the dipole cross section $\sigma_{\rm red}(R_T^2,s)$, 
and only the latter contains the energy dependence of 
$\sigma_T$ and $\sigma_L$ according to the dipole formulae 
 (\ref{VI55g}) and (\ref{VI55i}). 

Similarly, we can obtain bounds for the ratio of the 
cross sections for the production of a heavy quark flavour, $q=c$ 
or $q=b$, in deep inelastic scattering, 
\be
\label{boundRheavy}
\min_{R_T} \,
\frac{w^{(q)}_L(R_T,Q^2)}{w^{(q)}_T(R_T,Q^2)}
\, \leq \, 
R_q (s, Q^2) = \frac{\sigma^{(q)}_L(s,Q^2)}{\sigma^{(q)}_T(s,Q^2)}
\, \leq \,\max_{R_T} \,
\frac{w^{(q)}_L(R_T,Q^2)}{w^{(q)}_T(R_T,Q^2)} \,.
\ee

\begin{figure}[htbp]
\begin{center}
\includegraphics[width=11.4cm]{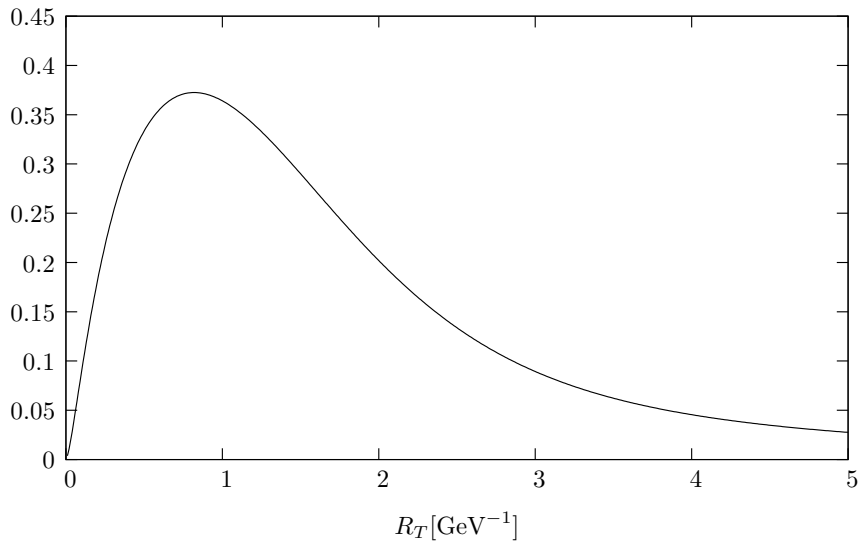}
\end{center}
\caption{The ratio $w_L^{(q)}(R_T,Q^2)/w_T^{(q)}(R_T,Q^2)$ 
as a function of $R_T$ for $Q^2 \!=\! 10\,\mbox{GeV}^2$ for 
massless ($q= u, d, s$) quarks. 
\label{fig:intrat0}}
\end{figure}

Let us now determine the numerical values of the bounds derived here. 
In figure \ref{fig:intrat0} we show the ratio 
$w_L^{(q)}(R_T,Q^2)/w_T^{(q)}(R_T,Q^2)$ for massless ($q=u,d,s$) quarks 
as a function of $R_T$ for a particular value $Q^2 \!=\! 10\,\mbox{GeV}^2$. 
This ratio has a maximum at which its value is $0.37248$, and asymptotically 
it approaches zero. Since $w^{(q)}_L/w^{(q)}_T$ is a dimensionless quantity it can 
for massless quarks depend only on $Q R_T$ such that also for different photon 
virtualities $Q^2$ we find a maximum with the same value. 
For massive quarks ($q=c, b$) the ratio $w^{(q)}_L/w^{(q)}_T$ as a 
function of $R_T$ also exhibits a maximum with a value which now 
depends on $Q^2$, but for all $Q^2$ we find that this value is 
smaller than for massless quarks. 

\begin{figure}[htbp]
\begin{center}
\includegraphics[width=11.4cm]{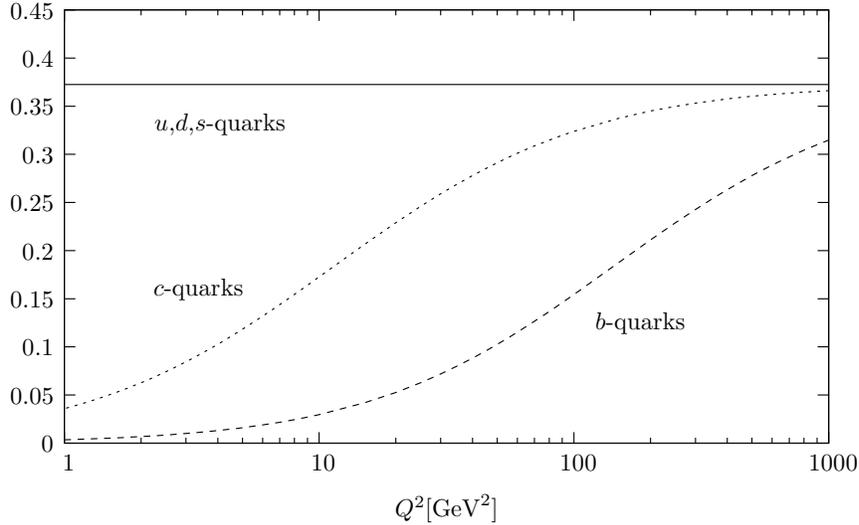}
\end{center}
\caption{The upper bound on $R=\sigma_L/\sigma_T$ 
resulting from the dipole model as a function of $Q^2$. 
The solid line is for massless ($u$, $d$, $s$) 
quarks and constitutes the bound (\ref{VI55l}). The lower 
lines are upper bounds on $R_c= \sigma^{(c)}_L/\sigma^{(c)}_T$ 
and $R_b= \sigma^{(b)}_L/\sigma^{(b)}_T$, respectively, 
see (\ref{boundRheavy}). 
\label{fig:maxrat}}
\end{figure}

We can therefore conclude that the upper bound on $R$ in (\ref{VI55l}) 
gives 
\be
\label{boundonR}
R(s,Q^2) \le 0.37248
\,,
\ee
while the lower bound on $R$ in (\ref{VI55l}) reduces to the trivial 
statement that $R$ is non-negative. Also the lower bound on the 
ratios $R_c$ and $R_b$ in (\ref{boundRheavy}) is trivial. 
The upper bounds on $R_c$ and $R_b$ resulting from (\ref{boundRheavy}), 
on the other hand, are nontrivial and depend on $Q^2$. 
This dependence is shown in figure \ref{fig:maxrat} together with the 
upper bound (\ref{boundonR}) on $R$. 

In many practical applications of the dipole picture 
one chooses to introduce phenomenological masses for the 
light ($q=u,d,s$) quarks. In \cite{Golec-Biernat:1998js} 
for example a light quark mass of $m_q=140\,\mbox{MeV}$ 
is chosen, and this mass is implemented in the perturbative 
photon wave function. Such a modified photon wave function 
results in a corresponding modification of the bound on $R$ 
which then becomes $Q^2$-dependent. For $m_q=140\,\mbox{MeV}$ 
this bound is shown on a linear scale in figure \ref{fig:RGBWmodpsi} below. 
With the logarithmic $Q^2$-scale in figure \ref{fig:maxrat} it would 
correspond to a curve of the same shape as the bound on $R_c$ or $R_b$ 
but shifted to the left such that its value at $Q^2=1\,\mbox{GeV}^2$ 
would be $0.318$. 

A popular extension of the dipole picture as we have discussed it so far 
is to diffractive deep inelastic scattering. In this extension one usually 
writes for the diffractive photon-proton cross section \cite{Nikolaev:et} 
\be
\label{dipcrossdiff}
\left.\frac{d\sigma^{\rm diff}_{T/L}}{dt} \right|_{t=0}
\!\!(s,Q^2) =
\frac{1}{16 \pi} \sum_q\int d^2R_T \,
w^{(q)}_{T/L}(R_T,Q^2)\,
\left( \sigma^{(q)}_{\rm red}(\mathbf{R}^2_T,s) \right)^2 \,,
\ee
where the dipole cross section $\sigma_{\rm red}$ is the same as 
in the usual dipole picture. The dipole formula for diffractive 
scattering (\ref{dipcrossdiff}) therefore relates diffractive to 
inclusive photon-proton scattering. Accepting (\ref{dipcrossdiff}) 
as a valid description of the diffractive cross section 
one finds that exactly the same bounds as obtained above 
apply also to diffractive photon-proton scattering, both for 
the total diffractive cross section and for diffractive production 
of heavy quarks. In particular, we have 
\be
\label{boundonRdiff}
\frac{\left.\frac{\sigma^{\rm diff}_L}{dt} \right|_{t=0} 
\!\!(s,Q^2)}{\left. \frac{\sigma^{\rm diff}_T}{dt}\right|_{t=0}
\!\!(s,Q^2)}
\le 0.37248
\,.
\ee

We emphasise, however, that the extension of the dipole picture 
to diffractive scattering (\ref{dipcrossdiff}) goes beyond what we 
have discussed so far in the present work. It is in particular not 
clear and in fact not necessary that the dipole formula 
(\ref{dipcrossdiff}) for diffractive scattering follows from the 
same assumptions (i)-(v) which we have used to obtain the dipole 
picture for inclusive deep inelastic scattering. 
It is an open question whether it is possible to establish the dipole 
picture of diffractive photon-proton scattering 
based on the techniques presented here. 

The dipole picture has been applied also to a number of exclusive 
reactions, a typical example is diffractive vector meson 
production. Here the wave function of the outgoing photon is 
replaced by the wave function of the produced vector meson, 
and the latter usually involves some model assumptions. 
We should point out that our bounds on $R$ cannot be 
applied directly to such processes. This is because our derivation of the 
bounds relies on the positivity of both factors under the integrals 
in the dipole formulae (\ref{VI55g}) and (\ref{VI55i}). In particular, 
it relies on the positivity of the square of the photon wave function 
which occurs due to the incoming and the outgoing photon. 
In other reactions that factor is changed and positivity is in 
general not guaranteed. 

We now return to inclusive deep inelastic photon-proton 
scattering and discuss the experimental data on $R$. 
The presently available data on $R$ at high energies 
have been obtained by the NMC \cite{Arneodo:1996qe}, 
CCFR \cite{Yang:2001xc}, E143 \cite{Abe:1998ym}, 
EMC \cite{Aubert:1985fx} and CDHSW \cite{Berge:1989hr} 
collaborations. The scattering processes used to extract $R$ 
include not only $e^{\pm} p$ scattering but also a variety of 
other scattering processes including muon and neutrino 
scattering on nuclear targets. 
In some of these processes one might in general 
expect additional caveats concerning the applicability of the 
dipole picture. For lack of further experimental data we 
nevertheless include the corresponding data points here. 
At high energies $R$ is expected 
to be independent of the target, as is also confirmed by the data 
within the experimental errors. For further details we refer 
the reader to the experimental publications. 

\begin{figure}[htbp]
\begin{center}
\includegraphics[width=12cm]{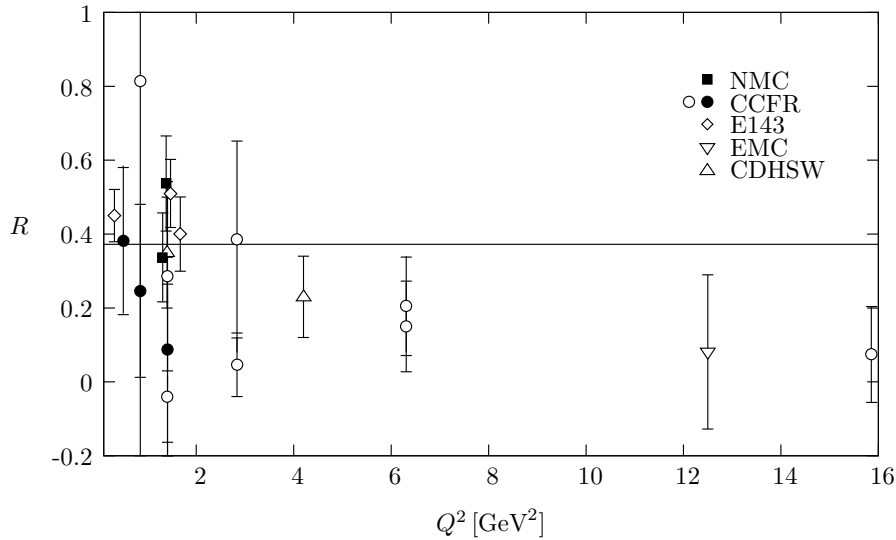}
\end{center}
\caption{Comparison of experimental data for $R=\sigma_L/\sigma_T$ in the 
region $x_{\rm Bj} <0.05$ with the bound (\ref{boundonR}) resulting from 
the dipole picture. Full points correspond to data with $x_{\rm Bj} <0.01$, 
open points are data with $0.01< x_{\rm Bj}  \le 0.05$. 
\label{fig:bound}}
\end{figure}

We want to consider only data points at sufficiently high energy 
which we choose to be data with $x_{\rm Bj} < 0.05$. 
Figure \ref{fig:bound} shows all available data in this energy 
region together with the bound (\ref{boundonR}) resulting 
from the dipole picture. 
The few data points available at $x_{\rm Bj} < 0.01$ are shown as full 
points while those with $0.01 \le x_{\rm Bj} < 0.05$ are shown as open 
points. The data have rather large errors, but by and large we can say 
that they respect the bound resulting from the dipole picture. 
For $Q^2$ below $2\,\mbox{GeV}^2$, however, there appears to be 
the tendency that the data come close to the bound. 
Note the interesting fact that data very close to the bound 
(\ref{boundonR}) could be accommodated in the dipole picture only if 
the dipole cross section $\sigma_{\rm red}$ were strongly peaked 
in $R_T$ around the maximum of $w^{(q)}_L/w^{(q)}_T$ -- which 
would appear to be very 
unlikely in reality. Hence already data close to the bound 
could be interpreted as an indication of the breakdown of the dipole 
picture. The errors of presently available data are 
far too large to draw any firm conclusion about the assumptions 
underlying the dipole picture here, but we might take them as 
an indication for the values of $Q^2$ below which we should be 
careful in interpreting scattering processes in terms of the dipole 
picture only. 

Unfortunately, no direct measurements of $F_L$ and $R$ 
have been performed at HERA so far. A discussion of the available 
indirect determinations of $F_L$ in view of our bound will be 
given elsewhere. Direct measurements of $F_L$ at HERA 
are planned and will hopefully lead to a better  determination of $R$, 
allowing for a stringent test of the validity of the dipole 
picture at low $Q^2$. 
Also a potential future upgrade of RHIC to an electron-ion collider 
eRHIC would offer the possibility to obtain a clear picture of 
$R$ at high energies and low $Q^2$, and thus to establish a conclusive 
test of the dipole picture in this region. 

Let us finally see which behaviour of $R(s,Q^2)$ is predicted in the 
framework of a typical model for the dipole cross section $\sigma_{\rm red}$. 
As a prominent example we choose the Golec-Biernat-W\"usthoff model 
\cite{Golec-Biernat:1998js}. In that model the dipole cross section is 
assumed to be a function of $R_T^2$, $x_{\rm Bj}$ and $Q^2$ and not only 
of $R_T^2$ and $s$. This might pose a problem in the context of our 
discussion in section \ref{sec:energyvariable} but for the purpose of 
illustrating the behaviour of $R$ in a typical dipole model we can 
disregard this issue. We should point out in this context that our 
bound (\ref{boundonR}) is completely independent of the variables on 
which the dipole cross section depends. 
In the Golec-Biernat-W\"usthoff model the 
dipole cross section $\sigma_{\rm red}$ is given by 
\be
\label{GBWformel}
\sigma^{(q)}_{\rm GBW} = 
\sigma_0 
\left[ 1 - \exp \left( - \frac{R_T^2}{4 R_0^2} \right) \right] 
\ee
with 
\be
\label{GBWR0}
R_0^2 (\tilde{x}) = \left( \frac{\tilde{x}}{x_0} \right)^\lambda 
\, \mbox{GeV}^{-2}
\,,
\ee
where the three parameters of the model are chosen to be 
$\sigma_0 = 23 \, \mbox{mb}$, $\lambda = 0.29$, $x_0 = 3 \cdot 10^{-4}$. 
The variable $\tilde{x}$ in (\ref{GBWR0}) is obtained from $x_{\rm Bj}$ via 
\be
\label{GBWmodx}
\tilde{x}= x_{\rm Bj} \left( 1 + \frac{4 m_q^2}{Q^2} \right)\,,
\ee
due to which the dependence of $\sigma_{\rm GBW}$ on the quark 
flavour and on $Q^2$ enters. We use again $m_c = 1.3\,\mbox{GeV}$ 
for the charm quark mass. 
In order to study the photoproduction limit $Q^2 \to 0$ a phenomenological 
mass of $m^{\rm GBW}_q=140 \, \mbox{MeV}$ is chosen for the light quarks 
($q=u,d,s$) which serves as a regulator for logarithmic divergences 
that would otherwise occur in this limit. In the Golec-Biernat-W\"usthoff 
model \cite{Golec-Biernat:1998js} that phenomenological light quark mass 
is implemented not only in the dipole cross section via (\ref{GBWmodx}) 
but also in the photon wave function. 
Note, however, that after inclusion of the effects of DGLAP evolution 
a better fit to the data is obtained with a vanishing light quark 
mass \cite{Bartels:2002cj}. 

\begin{figure}[htbp]
\begin{center}
\includegraphics[width=12cm]{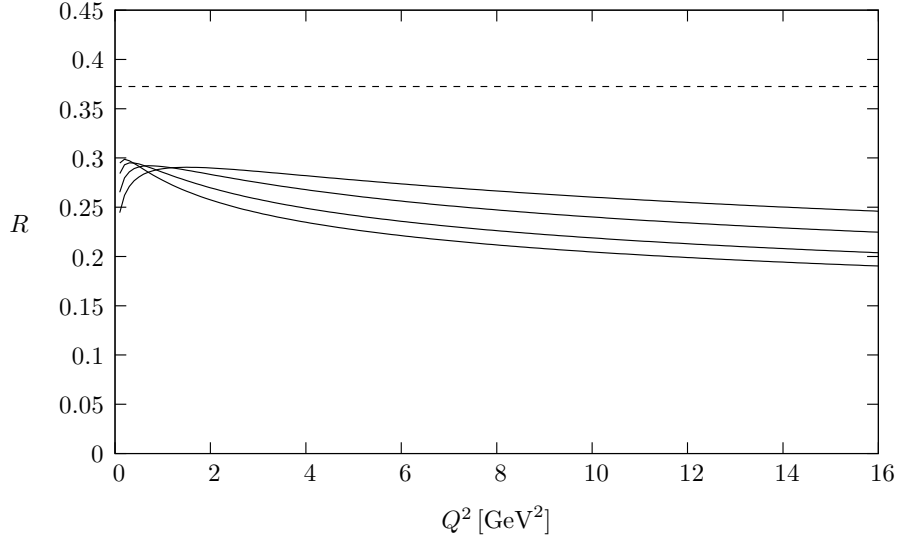}
\end{center}
\caption{
The ratio $R=\sigma_L/\sigma_T$ as obtained from the 
Golec-Biernat-W\"usthoff model as given by 
(\ref{GBWformel})-(\ref{GBWmodx}) but with vanishing 
light quark mass in the photon wave function. The solid lines 
are for four different values $x_{\rm Bj}=0.05, 0.01, 10^{-3}, 10^{-4}$ 
from bottom to top at large $Q^2$. The dashed line is the 
bound (\ref{boundonR}) derived from the dipole picture. 
\label{fig:RGBW}}
\end{figure}

In order to compare with our bound (\ref{boundonR}) we first 
insert the dipole cross section $\sigma_{\rm GBW}$ 
(\ref{GBWformel})-(\ref{GBWmodx}) with the light quark 
mass $m^{\rm GBW}_q=140 \, \mbox{MeV}$ 
into the dipole formulae (\ref{VI55g}) and (\ref{VI55i}) but keep 
the light quarks massless in the photon wave functions. 
The resulting behaviour of $R$ as a function of $Q^2$ is shown 
in figure \ref{fig:RGBW} for four different values of $x_{\rm Bj}$,  
$x_{\rm Bj}=0.05, 0.01, 10^{-3}, 10^{-4}$, together with the 
bound (\ref{boundonR}). 
The prediction for $R$ depends only weakly on $x_{\rm Bj}$ 
and stays significantly below the bound. 

Next we consider the original Golec-Biernat-W\"usthoff model, 
that is we implement the phenomenological light quark mass 
$m^{\rm GBW}_q=140 \, \mbox{MeV}$ not only in the dipole 
cross section but also in the photon wave function. The resulting 
$Q^2$-dependence of $R$ for $x_{\rm Bj}= 0.01$ is shown as 
the solid line in figure \ref{fig:RGBWmodpsi}. 
\begin{figure}[htbp]
\begin{center}
\includegraphics[width=12cm]{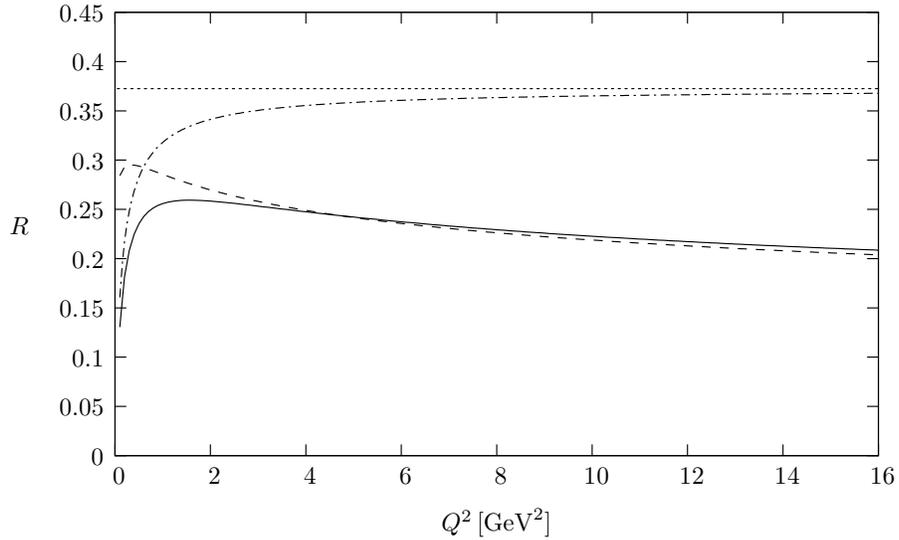}
\end{center}
\caption{The ratio $R=\sigma_L/\sigma_T$ as obtained from 
the Golec-Biernat-W\"usthoff model in comparison with the 
bound on $R$. The solid line results from the model 
(\ref{GBWformel})-(\ref{GBWmodx}) for $x_{\rm Bj}= 0.01$ 
with the light quark mass $140 \, \mbox{MeV}$ in the dipole 
cross section and in the photon wave function, the dashed line 
is obtained if that light quark mass is used only in the dipole 
cross section. The dotted line is the bound (\ref{boundonR}) 
and the dot-dashed line is the modified bound as obtained for 
a light quark mass of $140 \, \mbox{MeV}$. 
\label{fig:RGBWmodpsi}}
\end{figure}
The dashed line in that figure is the one which we obtained before in 
figure \ref{fig:RGBW} by using the light quark mass 
$m^{\rm GBW}_q$ only in the dipole cross section. At low 
$Q^2$ the phenomenological quark mass leads to a significant 
suppression of $R$ while for $Q^2 > 3 \,\mbox{GeV}^2$ the light 
quark mass in the photon wave function does not have a 
significant effect. A comparison with figure \ref{fig:bound} 
shows that, given the large errors of the data, both curves are 
in agreement with the measurements of $R$. 
The dot-dashed line in figure \ref{fig:RGBWmodpsi} is 
the bound obtained from a photon wave function with a light 
quark mass of $140 \, \mbox{MeV}$ as we discussed it already 
earlier in this section. For photon virtualities $Q^2$ below 
about $2 \,\mbox{GeV}^2$ the bound is significantly below the 
original bound (\ref{boundonR}) which was obtained for massless
light quarks and which is shown in the figure as the dotted line. 

More precise data on $R$ might in the future serve as an additional 
test of the quality of various models for the dipole cross 
section. It would therefore be interesting to study in more detail the 
behaviour of $R$ also in other models for the dipole cross section. 
An interesting aspect of such studies could be in particular the 
choice of a suitable phenomenological light quark mass which is 
required for the photoproduction limit or is motivated by the expected 
breakdown of the perturbative description of the photon wave 
function at large dipole sizes. For a consistent description 
one should clearly use the same light quark mass in the photon 
wave function and in the model for the dipole cross section. 
Since the bound on $R$ depends on the mass of the light quarks 
in the photon wave function especially at low $Q^2$ it would even 
be possible to use more precise data on $R$ in that region to obtain 
bounds on possible choices of a phenomenological mass for light 
quarks in the photon wave function and hence also in models for 
the dipole cross section. 

\section{Conclusions and outlook}
\label{conclsect}

In this paper and in the companion paper I 
we have studied how the dipole picture of high energy scattering 
can be derived in the framework of a genuinely nonperturbative 
formulation of photon-nucleon scattering. We have analysed in 
detail the Compton amplitude for real and virtual photons 
and have isolated the contributions to it which are leading 
at high energies. We have taken into account properly the 
renormalisation of these contributions and have studied 
their gauge invariance emphasising in particular its 
relevance for the definition of the perturbative photon 
wave function. We have then identified the approximations 
and assumptions which are necessary to arrive at the usual 
dipole picture. 

At high energies we obtain indeed the dipole picture, 
but also find two important additional contributions. One of them 
corresponds to rescattering effects of the quark and antiquark forming 
the colour dipole. The other additional contribution corresponds to 
diagrams in which the incoming photon and the outgoing photon 
do not couple to the same quark loop. Both of these terms are of 
higher order in the coupling constant $\alpha_s$ but are not 
suppressed by powers of the energy. In lowest order in perturbation 
theory these two additional terms do not contribute, and 
one is left with only the usual dipole picture. 
At low photon virtualities on the other hand we expect 
that both additional contributions can have sizable effects. 
We expect for example that the rescattering term will play 
an important role in the transition to low photon virtualities 
for which vector meson dominance applies. The 
rescattering term should in fact generate the relevant diagrams 
for the formation of quark-antiquark bound states. It would 
therefore be very interesting to see whether our results can 
be used to obtain a better picture of the transition from a 
perturbative photon wave function at high virtualities 
to vector meson dominance at low virtualities. 

A crucial question in all phenomenological applications of 
the dipole picture 
concerns the range of its applicability, or in other words the 
kinematical region in which the approximations and 
assumptions leading to the dipole model hold. 
We have derived a bound on the ratio $R$ of the cross sections 
for the scattering of longitudinally and transversely polarised 
photons off hadrons which is relevant in this 
context. The bound is obtained only from the 
perturbative photon wave function and is independent of 
any particular model for the dipole-proton cross section. 
Therefore it allows for a model independent test of the 
dipole picture. 
Unfortunately the presently available data are not precise 
enough to restrict the range of applicability of the dipole 
picture. However, the data appear to come close to the 
bound below photon virtualities of about $2\,\mbox{GeV}^2$, 
suggesting that at least in that region one should be careful in 
interpreting high energy scattering data in terms of the 
dipole picture only. In \cite{Ewerz:2006an} we have presented 
further bounds on ratios of nucleon structure functions $F_2$ 
taken at the same energy $\sqrt{s}$ but at different photon 
virtualities $Q^2$ resulting from the dipole picture. They are 
suitable for further constraining the range of applicability 
of the dipole picture. 

The dipole picture is often discussed in the context of 
the space-time picture of high energy photon-hadron 
scattering in the target rest frame. Here the photon splits 
into a long-lived quark-antiquark pair which then interacts 
with the hadron. Our study confirms this 
space-time picture and supplements it with the 
renormalisation of the photon-quark-antiquark vertex. 
The space-time picture is closely related to the notion 
of the light-cone wave function of the photon, suggesting 
a Hamiltonian picture of the evolution of the photon 
wave function and of the scattering process. We emphasise 
that we have obtained the dipole picture in a purely 
Lagrangian approach here. This allowed us in particular to 
implement the renormalisation procedure in a simple and 
transparent way. 
In a Hamiltonian approach a natural extension of the dipole 
picture is to include higher Fock states in the photon wave 
function. In a first step one would include for example the 
quark-antiquark-gluon component of that wave function. 
As we have already pointed out in I we do not expect to 
see such higher Fock states in the decompositions of the 
Compton amplitude which we have performed. We emphasise 
that they are not missing in our approach. Gluon emissions 
from the quark and antiquark are in our approach contained 
both in the rescattering term and in the dipole-nucleon cross 
section. It would be interesting to study in detail whether and 
how a consistent separation of such contributions into parts 
belonging to the photon wave function and to the dipole-nucleon 
scattering amplitude can be done. Naturally one would expect 
that this separation would require to introduce a factorisation 
scale. We recall that in the usual dipole picture the factorisation of the 
cross section into the square of the photon wave function and 
a dipole-proton cross section does not involve such a factorisation 
scale. 

We consider it an important task for future work to establish 
a closer relation of our approach with perturbation theory. 
It should be possible to follow our derivation of the dipole 
picture step by step in a perturbative setting. We have for example 
checked our results concerning the different choices for the 
polarisation vector for longitudinally polarised photons 
in the abelian gluon model of \cite{Nachtmann:2000gm} where 
the cancellations between different terms become very transparent. 
In the comparison with the perturbative framework it would be 
especially interesting to see how the picture of higher Fock 
states of the photon is related to our results. 

Another very important problem is to understand the relation
between inclusive and diffractive photon-nucleon scattering.
This problem has also been addressed in the framework of the 
dipole picture, see for example \cite{Nikolaev:et}. 
In our opinion it is a nontrivial question whether
the same approximations and assumptions leading to the
dipole model in inclusive scattering
also lead to the usual description of diffractive scattering 
in terms of the dipole-nucleon cross section as given 
by (\ref{dipcrossdiff}).  This question
becomes particularly challenging when one wants to include
corrections from higher Fock states in the photon wave function.
We expect that the study of diffractive scattering in the approach
presented here can provide more insight into these problems. 

We hope that our approach can 
also help to address other important questions concerning 
high energy lepton-nucleon scattering. We have pointed out 
for example that a study of polarised structure functions 
in the framework of the dipole picture might allow one to 
investigate and to quantify contributions that are subleading 
at high energies in unpolarised scattering. 
Another interesting issue is the choice of the energy variable 
in the dipole-proton cross section. Our results strongly suggest that the 
correct energy variable should be the total energy $\sqrt{s}$ rather 
than Bjorken-$x$. We cannot exclude, however, that different 
approaches to deriving the dipole picture in a nonperturbative 
framework lead to a different result concerning the energy 
variable. We consider this to be an interesting question for 
future studies. 
A third problem which we 
consider to be important is the use of the mass-shell 
condition for quarks. Motivated by local parton-hadron duality one 
often treats quarks produced into the final state as on-shell particles, 
especially in perturbative calculations. The assumption of a quark 
mass shell is also important for the usual formulation of high energy 
scattering in a Hamiltonian approach where the cross sections 
of higher Fock states of the photon are only well-defined for on-shell 
partons. Due to confinement, 
however, a mass-shell for quarks actually does not exist. 
We expect that by carefully tracing the effect of using 
the mass-shell-condition for quarks in our framework one can 
learn something about the validity of this assumption. 

The approach presented here might also offer the possibility 
to study in another way the transition region from 
high to low photon virtualities which has been a key question 
in many applications of the dipole picture to deep inelastic 
scattering. Most of those studies try to use the framework of 
the dipole picture and introduce modifications of the 
dipole-nucleon cross section at low virtualities. 
The formulae derived here appear to be suited also 
for an implementation of nonperturbative Green's functions 
as obtained for instance in studies of nonperturbative 
Dyson-Schwinger equations. In this way one could hope to 
approach the transition region also from the side of 
low virtualities. In practice that would certainly 
require a number of additional approximations, and would 
hence be less rigorous. Nevertheless, we think that it would 
be very important from a phenomenological point of view 
to study the transition region between small and large 
photon virtualities using all available methods. 

The general methods developed here can also be applied to other current 
induced reactions at high energies, including exclusive reactions. 
An interesting application would be the study of reactions 
induced by electroweak currents. The perturbative wave 
functions of $W$ and $Z$ bosons have recently been discussed in 
\cite{Fiore:2005yi}. We expect that the derivation of the dipole picture 
for electroweak currents in a nonperturbative framework 
can be performed using essentially the 
same approximations and assumptions that were needed for 
photon induced reactions. 

Let us finally give a brief summary of the main findings of the 
present paper and  of the companion paper I. 
\begin{itemize}
\item 
In I we used functional methods to classify the contributions to the Compton 
amplitude in terms of nonperturbative quark-skeleton diagrams, 
see figure I.\ref{fig2}. The important diagrams at high energies were 
found to be those of figures I.\ref{fig2}a and I.\ref{fig2}b. 
\item 
Cutting the diagrams (a) and (b) of figure I.\ref{fig2} next to the 
incoming photon vertex and inserting suitable factors of one we could 
identify parts corresponding to the photon vertex and to 
quark-antiquark scattering off the proton. 
Proper account was taken of renormalisation and this led to the 
introduction of the rescattering terms, see 
section I.\ref{sec:currenttoquarkqft}.
\item 
In the high-energy limit as studied in the present paper the Compton 
amplitude was split into a genuine quark-antiquark-proton scattering 
amplitude and a photon-wave-function piece including rescattering 
terms. Here the assumption of the existence of a quark mass shell was made. 
\item 
The equivalence of an incoming or outgoing high-energy photon to 
incoming or outgoing quark-antiquark-dipole states was given explicitly 
in section \ref{sec:genphotons} for arbitrary reactions. 
As always, vertex and rescattering terms occurred. 
\item 
The lowest order perturbative wave function of the photon 
was studied in detail. It was shown that the polarisation vectors 
of the virtual photon have to be chosen 
such that they do not contain terms growing indefinitely as 
$|\mathbf {q}|\to \infty$. Gauge invariance was 
discussed in detail and was found to be crucial in this context, 
see section \ref{wavefunctsec}. 
\item 
In section \ref{sec:dippictdis} 
deep inelastic scattering was treated and the assumptions 
needed to arrive at the usual formulae of the dipole model were spelled 
out in detail. A bound on $\sigma_L/\sigma_T$ which followed from 
these assumptions was derived and its consequences were 
discussed in section \ref{sec:phencons}. 
This bound can be used to test the kinematic region in which the 
usual dipole model can be applied. 
\end{itemize}

The dipole picture has been surprisingly successful in describing
the data obtained at HERA and has substantially contributed to the
understanding of photon-nucleon scattering. 
In the near future the LHC will offer the possibility to study 
photon-proton and photon-photon collisions at even higher 
energies for quasi-real photons. Particularly suitable for this purpose 
are highly peripheral heavy ion collisions where the Coulomb field 
of a fast heavy ion acts as a beam of quasi-real photons. 
We expect that the dipole picture will be a very useful tool for 
analysing the data to come from the LHC in this field. 
A thorough study of the dipole formalism, 
its nonperturbative foundations and its necessary 
modifications is therefore very important. With the present study 
we hope to have made some useful steps towards this ambitious goal. 

\section*{Acknowledgements}

We are grateful to J.\ Bartels, J.-P.\ Blaizot, A.\ Donnachie, 
H.\,G.\ Dosch, J.\ Forshaw, S.\ Forte, K.\ Golec-Biernat, 
A.\ Hebecker, G.\ Korchemsky, P.\,V.\ Landshoff, A.\ Mueller, 
H.\,J.\ Pirner, J.\ Raufeisen, D.\ Triantafyllopoulos and S.\ Wallon 
for helpful discussions.
We thank A.\ Bodek and U.\,K.\ Yang for providing data for figure 
\ref{fig:bound}. 
C.\,E.\ was supported by the Bundesministerium f\"ur
Bildung und Forschung, projects HD 05HT1VHA/0
and HD 05HT4VHA/0, and by a Feodor Lynen fellowship of the
Alexander von Humboldt Foundation.

\begin{appendix}
\numberwithin{equation}{section}

\section{Kinematical relations in the high energy limit}
\label{appuseful}

In this appendix we supplement section \ref{highenergysect} 
with a more detailed discussion of the kinematics relevant to 
the pinch condition occurring there and further provide some 
formulae which are used in the calculation of the photon 
wave function in section \ref{wavefunctsec}. 

\subsection{The pinch condition}
\label{appIIApinch}

We start with the discussion of the pinch condition 
$\Delta E\rightarrow 0$, see (\ref{3.5})-(\ref{3.13}). 
The condition (\ref{3.10}) on $\tilde{m}^2(\alpha,\mathbf{k}_T)$ restricts, 
for given $\bar{Q}^2$, the range in $\alpha$ and $\mathbf{k}_T$ 
where $\Delta E$ fulfills (\ref{3.13}). For massless quarks, $m_q=0$, 
the condition (\ref{3.10}) can be fulfilled for all 
$\alpha$ with $0\leq\alpha \leq 1$ with the $\mathbf{k}_T^2$-range 
limited as shown in figure \ref{fig8}. For massive 
quarks, $m_q\neq 0$, the condition (\ref{3.10}) can only be fulfilled for a 
limited range in $\alpha$. Indeed, for $|\mathbf{q}|\gg m_q$ this 
restriction is (as can be seen from (\ref{m2atlargeq}))
\begin{equation}\label{D.6}
\alpha_1\leq\alpha\leq\alpha_2 \,,
\end{equation}
where 
\begin{equation}\label{D.7}
\alpha_{1,2}=
\frac{1}{2}\left(1\mp\sqrt{1-\frac{4m^2_q}{\bar{Q}^2}}\,\right)\,.
\end{equation}
This range in $\alpha$ results from requiring that $\mathbf{k}_T^2=0$ 
becomes possible, and the numbers given in the following will be calculated 
for that case. (For a given $\mathbf{k}_T^2 > 0$ to be possible the range 
in $\alpha$ 
becomes smaller, and its limits are obtained if in (\ref{D.7}) $m_q^2$ is 
replaced by $(\mathbf{k}_T^2 + m_q^2)$.) 
Clearly, for $\bar{Q}^2=4m^2_q$ the allowed range in $\alpha$ shrinks to 
the point $\alpha=1/2$, whereas for $\bar{Q}^2\rightarrow\infty$ the 
whole range $0 \le \alpha \le 1$ is allowed. 

In tables \ref{table1} and \ref{table2} we list some values of 
$\alpha_1$ and $\alpha_2$ for the $c$- and the $b$-quark, respectively. 
We choose as in section \ref{densitysect} 
\begin{eqnarray}\label{D.8}
m_c&=&{} 1.3 \,\mbox{GeV} \,,\nonumber\\
m_b&=&{} 4.6 \,\mbox{GeV} \,,
\end{eqnarray}
which implies $\bar{Q}^2\geq 4 m^2_c=6.76\,\mbox{ GeV}^2$ for the $c$-quark 
and $\bar{Q}^2\geq 4m^2_b=84.6\,\mbox{GeV}^2$ for the $b$-quark. 
We see from the tables that an allowed $\alpha$-range (\ref{D.6}) 
from $0.05$ to $0.95$ requires $\bar{Q}^2>35.6\,\mbox{GeV}^2$ 
for the $c$-quark and $\bar{Q}^2>445 \,\mbox{GeV}^2$ for the $b$-quark. 
\begin{table}[ht]
  \centering
\begin{tabular}{r|l|l}
\multicolumn{1}{c|}{$\bar{Q}^2 [\mbox{GeV}^2]$} &
\multicolumn{1}{c|}{$\alpha_1$} &
\multicolumn{1}{c}{$\alpha_2$}
\\ \hline
20\hspace*{.5cm}&0.093&0.907\\
35.6\hspace*{.2cm}&0.050&0.950\\
100\hspace*{.5cm}&0.017&0.983\\
1000\hspace*{.5cm}&0.002&0.998
\end{tabular}
\caption{Values for $\bar{Q}^2$, $\alpha_1$ and $\alpha_2$ for the $c$-quark
\label{table1}}
\end{table}
\begin{table}[ht]
  \centering
\begin{tabular}{r|l|l}
\multicolumn{1}{c|}{$\bar{Q}^2 [\mbox{GeV}^2]$} &
\multicolumn{1}{c|}{$\alpha_1$} &
\multicolumn{1}{c}{$\alpha_2$}
\\ \hline
100\hspace*{.3cm}&0.30&0.70\\
445\hspace*{.3cm}&0.05&0.95\\
1000\hspace*{.3cm}&0.02&0.98\\
10000\hspace*{.3cm}&0.002&0.998
\end{tabular}
\caption{Values for $\bar{Q}^2$, $\alpha_1$ and $\alpha_2$ for the $b$-quark
\label{table2}}
\end{table}

Finally we want to discuss the actual values of $\Delta E$ in (\ref{3.5}) 
or (\ref{3.8}) that one gets in various kinematic domains. 
For high energies we have in the proton rest system 
\begin{equation}\label{D.9}
q^0+k^0+k'^{0} \cong 
2q^0\equiv 
2\nu=\frac{Q^2}{m_p}\frac{2m_p\nu}{Q^2}
=\frac{Q^2}{m_p x_{Bj}} \,,
\end{equation}
where $x_{\rm Bj}$ is Bjorken's scaling variable. 
Inserting this in (\ref{3.13}) gives 
\begin{equation}\label{D.10}
\frac{\Delta E}{m_p}\leq\frac{Q^2+\bar{Q}^2}{Q^2}\,x_{\rm Bj}\,.
\end{equation}
Let us take now as an example massless quarks and 
$Q^2=1\,\mbox{GeV}^2$, $\bar{Q}^2=4~\mbox{GeV}^2$. 
Then $\Delta E/m_p\leq0.1$ for $x_{\rm Bj}<0.02$. 
For $c$-quarks with $Q^2=4\,\mbox{GeV}^2$ and 
$\bar{Q}^2=36 \,\mbox{GeV}^2$ 
we get $\Delta E/m_p\leq 0.1$ for $x_{\rm Bj}<0.01$. For $b$-quarks 
finally we find with $Q^2=20\,\mbox{GeV}^2$ and $\bar{Q}^2=450\,\mbox{GeV}^2$ 
that $\Delta E/m_p<0.1$ is only reached for $x_{\rm Bj}<0.005$. 

With  these numbers for $x_{\rm Bj}$ we do by no means want to establish 
limits for the validity of the dipole picture. We only want to show how small 
$\Delta E$ is in concrete cases.

\subsection{Further kinematical relations}
\label{appIIAfurther}

In the high energy limit $|\mathbf{q}| \to \infty$ and with 
the conditions (\ref{kinlimitalphaq}) we find 
from (\ref{3.2})-(\ref{3.4}) the expansions
\be
\label{expk0he}
k^0= \alpha\, |\mathbf{q}| + 
\frac{1}{2} \,\frac{\mathbf{k}_T^2 + m_q^2}{\alpha \, |\mathbf{q}|}
+ {\cal{O}} \left(\frac{(\mathbf{k}_T^2 + m_q^2)^2}{|\mathbf{q}|^3} \right)
\ee
and 
\be
\label{expkstr0he}
k'^0= (1-\alpha)\, |\mathbf{q}| + 
\frac{1}{2} \,\frac{\mathbf{k}_T^2 + m_q^2}{(1-\alpha )\, |\mathbf{q}|}
+ {\cal{O}} \left(\frac{(\mathbf{k}_T^2 + m_q^2)^2}{|\mathbf{q}|^3} \right)
\,.
\ee
We further have the products 
\be
\label{k0kstr0he}
k^0 k^{\prime 0}= 
\alpha (1-\alpha) |\mathbf{q}|^2 
+ \frac{1}{2} \left( \frac{\alpha}{1-\alpha} + \frac{1-\alpha}{\alpha} \right)
(\mathbf{k}_T^2 + m_q^2) 
+ {\cal{O}} \left(\frac{(\mathbf{k}_T^2 + m_q^2)^2}{|\mathbf{q}|^2} \right)
\ee
and 
\bea
\label{k0plusmkstr0plusm}
(k^0 +m_q) (k^{\prime 0} +m_q) &=& {} 
\alpha (1-\alpha) |\mathbf{q}|^2 + m_q |\mathbf{q}| 
+ \frac{1}{2} \left( \frac{\alpha}{1-\alpha} + \frac{1-\alpha}{\alpha} \right)
(\mathbf{k}_T^2 + m_q^2) + m_q^2 
\nn \\
&&{} + {\cal{O}} \left(\frac{m_q (\mathbf{k}_T^2 + m_q^2)}{|\mathbf{q}|} \right)
\,.
\eea
The leading contribution to $(\Delta E)^{-1}$ (see (\ref{3.1}) and (\ref{3.5})) 
at high energy is obtained as 
\bea
\label{DeltaElargeq}
\left(\Delta E\right)^{-1} &=& {}
\frac{q^0 + k^0 + k'^0}{\left(k^0 + k'^0\right)^2 - (q^0)^2}
\nn \\
&=& {}
\frac{\alpha (1-\alpha) \left(q^0 + |\mathbf{q}|\right)}
{\alpha (1-\alpha) Q^2 + \mathbf{k}_T^2 + m_q^2} 
\left[1 
+ {\cal{O}} \left(\frac{\mathbf{k}_T^2 + m_q^2}{|\mathbf{q}|^2} \right)
\right] \,,
\eea
and we have 
\be
\label{qfrac}
\frac{q^0+|\mathbf{q}|}{2 \,|\mathbf{q}|} = 
1 + {\cal{O}} \left(\frac{Q^2}{|\mathbf{q}|^2}\right) \,.
\ee
Furthermore, we find 
\be
\label{m2atlargeq}
\tilde{m}^2(\alpha,\mathbf{k}_T) = 
\frac{\mathbf{k}_T^2 + m_q^2}{\alpha (1-\alpha)}
\left[1 
+ {\cal{O}} \left(\frac{\mathbf{k}_T^2 + m_q^2}{|\mathbf{q}|^2}\right) 
\right] \,,
\ee
for which $0<\alpha < 1$ is required, as was assumed throughout 
the discussion of the high energy limit, see (\ref{3.7}) and (\ref{kinlimitalphaq}). 

For the Fourier transformation of the photon wave function 
from transverse momentum space to coordinate space we need 
the well-known integral 
\be
\label{BesselK0}
\int \frac{d^2 k_T}{(2 \pi)^2} \,e^{i \mathbf{k}_T \mathbf{R}_T} \, 
\frac{1}{\mathbf{k}_T^2 + \epsilon_q^2}
= \frac{1}{2 \pi} K_0(\epsilon_q R_T) \,.
\ee
The computation of the wave function of transversely polarised 
photons also requires the integral 
\be
\label{BesselK1}
\int \frac{d^2 k_T}{(2 \pi)^2} \,e^{i \mathbf{k}_T \mathbf{R}_T} \, 
\frac{\mathbf{k}_T}{\mathbf{k}_T^2 + \epsilon_q^2}
= -\frac{1}{2 \pi i} \, \epsilon_q \frac{\mathbf{R}_T}{R_T} 
K_1(\epsilon_q R_T) \,,
\ee
which is obtained from (\ref{BesselK0}) by differentiating with 
respect to the components of $\mathbf{R}_T$. 

\section{Separate gauge invariance of different contributions to 
the Compton amplitude}
\label{appdiffcontgauge}

In this appendix we discuss the separate gauge invariance 
of different contributions to the skeleton decomposition of the 
Compton amplitude, see figure I.\ref{fig2}. Here we deal in particular 
with those contributions that are subleading at high energies. 

We first recall from I that the different contributions in the decomposition 
(I.\ref{2.7a1}) of the Compton amplitude can be expressed as 
(see (I.\ref{A.7a})-(I.\ref{AJg})) 
\begin{eqnarray}
\label{defMdurchJ}
\mathcal{M}^{(j)\mu\nu}_{s^{\prime}s}(p^{\prime},p,q)&=&
- \frac{i}{2\pi m_pZ_p}\int d^4y^{\prime}~d^4y \,
e^{ip^{\prime}y^{\prime}}
\bar{u}_{s^{\prime}}(p^{\prime})(-i\rightslash_{y^{\prime}}
+m_p) 
\nn \\
&& \int d^4x \, e^{-iqx} \left. {\cal J}^{(j)} \right|_{x'=0} 
( i\leftslash_y + m_p ) u_s(p) e^{-ipy} \,,
\end{eqnarray}
where $j=a,\dots,g$, and the explicit expressions for the 
$ {\cal J}^{(j)}$ are given in appendix I.\ref{appA}. 
For $p$ and $p'$ off shell we can integrate in (\ref{defMdurchJ}) 
by parts with respect to $y$ and $y'$ and get 
\bea
\label{Mallgnachpartiell}
\!\!\!\!\!
\mathcal{M}^{(j)\mu\nu}_{s^{\prime}s}(p^{\prime},p,q)&=&
- \frac{i}{2\pi m_pZ_p} \,
\bar{u}_{s^{\prime}}(p^{\prime}) 
\bigg[
(- \! \slash{p}' + m_p ) 
\int d^4y^{\prime}~d^4y \, e^{ip^{\prime}y^{\prime}} 
\nn \\
&&{} 
\left. 
\int d^4x \, e^{-iqx} \left. {\cal J}^{(j)} \right|_{x'=0} 
e^{-ipy} (- \! \slash{p} + m_p ) 
\bigg] \right|_{
\begin{array}{c}
\slash{p}' \!\to  m_p \\[-0.1cm]
\slash{p} \to m_p 
\end{array}
}
\!\!  u_s(p) 
\,.
\eea
Here, as usual, one has to take the limit $\slash{p}' \to m_p$ and 
$\slash{p} \to m_p$ for the expression in square brackets first 
and then to multiply with the Dirac spinors. 
The contraction of $\mathcal{M}^{(j)\mu\nu}$ with $q_\nu$ 
can then be obtained in analogy to (\ref{qtimesAa}) as 
\bea
\label{contrqMallg}
q_\nu \mathcal{M}^{(j)\mu\nu}_{s^{\prime}s}(p^{\prime},p,q)&=&
- \frac{i}{2\pi m_pZ_p} \, 
\bar{u}_{s^{\prime}}(p^{\prime}) 
\Big[
(- \! \slash{p}' + m_p ) \, \mathcal{N}^{(j)} (p',p,q) 
\nn \\
&&{}
(- \! \slash{p} + m_p ) 
\left. 
\Big]
\right|_{
\begin{array}{c}
\slash{p}' \!\to  m_p \\[-0.1cm]
\slash{p} \to m_p 
\end{array}
}
u_s(p) 
\,,
\eea
where we have defined 
\be
\label{defNallg}
\mathcal{N}^{(j)} (p',p,q) = 
- \int d^4y^{\prime}~d^4y \, e^{ip^{\prime}y^{\prime}} e^{-ipy} 
\int d^4x \, e^{-iqx} \,i \frac{\del}{\del x^\nu} \,
\left. \mathcal{J}^{(j)} \right|_{x'=0} 
\,.
\ee
From (\ref{contrqMallg}) we see that a nonvanishing contribution 
to $q_\nu \mathcal{M}^{(j)\mu\nu}$ is obtained only if 
$\mathcal{N}^{(j)}$ has poles both at $\slash{p}' = m_p$ 
and at $\slash{p}= m_p$. 

In section \ref{subsec:gaugefullampl} we have already discussed the 
separate gauge invariance of $\mathcal{M}^{(a)}$ and $\mathcal{M}^{(b)}$. 
Let us now consider the gauge invariance of the amplitude 
$\mathcal{M}^{(f)}$. Following steps similar to those in 
(\ref{qnuAcontr}) we find that 
\be
\label{der2prop}
i \frac{\del}{\del x^\nu} 
\left[ \frac{1}{i} S_F^{(q)}(y',x;G) \gamma^\nu 
\frac{1}{i}S_F^{(q)}(x,y;G) \right]
= 
\frac{1}{i}S_F^{(q)}(y',y;G) \, i \left[ \delta^{(4)}(x-y) 
- \delta^{(4)}(y'-x) \right] 
\,.
\ee
As a consequence we obtain from (I.\ref{AJf}) with (I.\ref{A.10}) 
\bea
\label{delxJf}
i \frac{\del}{\del x^\nu} \, \mathcal{J}^{(f)} &=&{} 
( 2 Q_u + Q_d) \, i \left[ \delta^{(4)}(x-y) - \delta^{(4)}(y'-x) \right] 
\nn \\
&&{}
\Bigg\langle\left[ 
\sum_{q'} Q_q' (-1) \mathrm{Tr} 
\left( 
\gamma^\mu \frac{1}{i}S_F^{(q^{\prime})}(x',x';G) 
\right)
\right]
\wick{2}{<1\psi_p(y^{\prime})>1{\overline{\psi}}_p(y)}
\Bigg\rangle_G
\,,
\eea
and inserting this into (\ref{defNallg}) we arrive at 
\bea
\label{Nfresult}
\mathcal{N}^{(f)} (p',p,q) &=&{}
- i \int d^4y^{\prime}~d^4y \, 
\left\{
e^{i p' y'} e^{-i (p+q) y} - e^{i (p'-q) y'} e^{-i p y} 
\right\}
( 2 Q_u + Q_d) 
\nn \\
&&{}
\Bigg\langle\left[ 
\sum_{q'} Q_q' (-1) \mathrm{Tr} 
\left( 
\gamma^\mu \frac{1}{i}S_F^{(q^{\prime})}(0,0;G) 
\right)
\right]
\wick{2}{<1\psi_p(y^{\prime})>1{\overline{\psi}}_p(y)}
\Bigg\rangle_G
\,.
\eea
We can now interpret the integrals over $y'$ and $y$ as 
Fourier transformations. The first term in the curly brackets 
in (\ref{Nfresult}) leads to a contribution corresponding to 
an incoming proton of momentum $p+q$ and an outgoing 
proton of momentum $p'$. Consequently, this contribution 
has poles at $\slash{p}' = m_p$ and at $\slash{p} \, + \!\slash{q}=m_p$, 
but not at $\slash{p}= m_p$. Similarly, the second term in the curly 
brackets gives rise to a contribution which has poles 
at $\slash{p} = m_p$ and $\slash{p}'- \!\slash{q}=m_p $, but 
not at $\slash{p}'=m_p$. Therefore, neither of the two contributions 
can give a nonvanishing contribution when inserted in 
(\ref{contrqMallg}), and hence 
\be
\label{gaugeMfresultapp}
q_\nu \mathcal{M}^{(f)\mu\nu}_{s^{\prime}s}(p^{\prime},p,q) = 0 
\,. 
\ee
This completes the proof that $\mathcal{M}^{(f)}$ is separately 
gauge invariant. 

In order to discuss the remaining parts of the amplitude it will be 
useful to consider first the $u$- and the $d$-quark contributions 
to the electromagnetic current, 
\bea
\label{currentJu}
J_u^\lambda (x) &=& Q_u \, \bar{u}(x) \gamma^\lambda u(x) \,,
\\
\label{currentJd}
J_d^\lambda (x) &=& Q_d \, \bar{d}(x) \gamma^\lambda d(x) \,,
\eea
and their matrix elements between incoming and outgoing proton 
states. Using the LSZ reduction formula and the methods explained 
in I we get in analogy to (I.\ref{2.6}) for these matrix elements 
\bea
\label{lszforonecurrent}
\langle p(p^{\prime},s^{\prime})|J_{u,d}^\mu(x)|p(p,s)\rangle
&=&
-\frac{1}{Z_p}
\int d^4y^{\prime}~d^4y \, e^{ip^{\prime}y^{\prime}}
\bar{u}_{s^{\prime}}(p^{\prime})
(-i\rightslash_{y'}+m_p)
\\
&&{}
\left\langle\psi_p(y^{\prime}) 
J_{u,d}^{\mu}(x)\overline{\psi}_p(y)\right\rangle_{G,q,\bar{q}} \,
(i\leftslash_y+m_p)u_s(p)e^{-ipy}
\,.
\nn
\eea
In analogy to the treatment of (I.\ref{A.7}) in appendix I.\ref{appA} 
we obtain using (I.\ref{A.6}) and (I.\ref{A.10}) for the functional 
integrals in (\ref{lszforonecurrent}) involving these currents 
\bea
\label{psiJupsibar}
\lefteqn{
\Big\langle\psi_p(y^{\prime})J_u^{\mu}(x) \overline{\psi}_p(y) 
\Big\rangle_{G,q,\bar{q}}
=
- Q_u 
\bigg\langle\wick{2}{<1\psi_p(y^{\prime})>1{\overline{\psi}}_p(y)}
\mathrm{Tr}\left(\frac{1}{i}S^{(u)}_F(x,x;G)\gamma^\mu\right)
\bigg\rangle_G
}
\nonumber \\
&&{}\hspace*{1.65cm}
+ Q_u \,
\Gamma_{\alpha^{\prime}\beta^{\prime}\gamma^{\prime}}
\bar{\Gamma}_{\alpha\beta\gamma} 
\bigg\langle
\frac{1}{i}S^{(d)}_{F\gamma'\gamma}(y',y;G)
\nn \\
&&{}\hspace*{2.1cm}
\bigg[
\left(\frac{1}{i} S^{(u)}_F(y',x;G) \gamma^\mu 
\frac{1}{i} S^{(u)}_F(x,y;G) \right)_{\alpha' \alpha}
\frac{1}{i} S^{(u)}_{F\,\beta' \beta}(y',y;G) 
\nn \\
&&{}\hspace*{2.1cm}
- (\alpha' \leftrightarrow \beta') - (\alpha \leftrightarrow  \beta) 
+ (\alpha' \leftrightarrow \beta', \alpha \leftrightarrow  \beta) 
\bigg] \bigg\rangle_G
\eea
and 
\bea
\label{psiJdpsibar}
\lefteqn{
\Big\langle\psi_p(y^{\prime})J_d^{\mu}(x) \overline{\psi}_p(y) 
\Big\rangle_{G,q,\bar{q}}
=
- Q_d
\bigg\langle\wick{2}{<1\psi_p(y^{\prime})>1{\overline{\psi}}_p(y)}
\mathrm{Tr}\left(\frac{1}{i}S^{(d)}_F(x,x;G)\gamma^\mu\right)
\bigg\rangle_G
}
\nonumber \\
&&{}\hspace*{1.65cm}
+ Q_d \,
\Gamma_{\alpha^{\prime}\beta^{\prime}\gamma^{\prime}}
\bar{\Gamma}_{\alpha\beta\gamma} 
\bigg\langle
\left(\frac{1}{i} S^{(d)}_F(y',x;G) \gamma^\mu 
\frac{1}{i} S^{(d)}_F(x,y;G) \right)_{\gamma' \gamma}
\nn \\
&&{}\hspace*{2.1cm}
\bigg[
\frac{1}{i} S^{(u)}_{F\,\alpha' \alpha}(y',y;G) 
\frac{1}{i} S^{(u)}_{F\,\beta' \beta}(y',y;G) 
- (\alpha \leftrightarrow  \beta) 
\bigg] \bigg\rangle_G \,.
\eea

We now turn to the amplitudes $\mathcal{M}^{(c)}$ and  
$\mathcal{M}^{(d)}$ which are related to $\mathcal{J}^{(c)}$ 
and $\mathcal{J}^{(d)}$, respectively, via (\ref{defMdurchJ}). 
From (I.\ref{A.12}) we find using (\ref{der2prop}) 
\be
\label{derJc}
i \frac{\del}{\del x^\nu} \,\mathcal{J}^{(c)} = 
i \left[ \delta^{(4)}(x-y) - \delta^{(4)}(x'-x) \right]  \mathcal{E}
\,,
\ee
where $\mathcal{E}$ is given by 
\bea
\label{defcalE}
 \mathcal{E} &=&{}
\Gamma_{\alpha^{\prime}\beta^{\prime}\gamma^{\prime}}
\bar{\Gamma}_{\alpha\beta\gamma} 
\bigg\langle 
Q_d^2 
\left(\frac{1}{i} S^{(d)}_F(y',x';G) \gamma^\mu 
\frac{1}{i} S^{(d)}_F(x',y;G) \right)_{\gamma' \gamma}
\nn \\
&&{} \hspace*{2.4cm}
\bigg[ 
\frac{1}{i} S^{(u)}_{F\,\alpha' \alpha}(y',y;G) 
\frac{1}{i} S^{(u)}_{F\,\beta' \beta}(y',y;G) 
- (\alpha \leftrightarrow  \beta) 
\bigg] 
\nn \\
&&{}
+ Q_u^2 \,\frac{1}{i} S^{(d)}_{F\,\gamma' \gamma}(y',y;G) 
\nn \\
&&{}\hspace*{0.6cm}
\bigg[
\left(\frac{1}{i} S^{(u)}_F(y',x';G) \gamma^\mu 
\frac{1}{i} S^{(u)}_F(x',y;G) \right)_{\alpha' \alpha}
\frac{1}{i} S^{(u)}_{F\,\beta' \beta}(y',y;G) 
\nn \\
&&{}\hspace*{0.8cm}
- (\alpha' \leftrightarrow \beta') - (\alpha \leftrightarrow  \beta) 
+ (\alpha' \leftrightarrow \beta', \alpha \leftrightarrow  \beta) 
\bigg] \bigg\rangle_G
\,.
\eea
Again using (\ref{der2prop}) we find from (I.\ref{A.Jd}) that also the 
derivative of $\mathcal{J}^{(d)}$ with respect to $x^\nu$ can 
be expressed in terms of $\mathcal{E}$ in a similar way, 
\be
\label{derJd}
i \frac{\del}{\del x^\nu} \, \mathcal{J}^{(d)} = 
i \left[ \delta^{(4)}(x-x') - \delta^{(4)}(y'-x) \right]  \mathcal{E}
\,.
\ee
We see that in the sum of (\ref{derJc}) and (\ref{derJd}) the terms 
with the delta function involving $x'$ cancel. Hence we have 
\be
\label{derJcplusd}
i \frac{\del}{\del x^\nu} 
\left( \mathcal{J}^{(c)} + \mathcal{J}^{(d)} \right) 
= i \left[ \delta^{(4)}(x-y) - \delta^{(4)}(y'-x) \right]  \mathcal{E}
\,.
\ee
From (\ref{defNallg}) we then find using (\ref{psiJupsibar}) 
and (\ref{psiJdpsibar}) 
\bea
\label{Ncdresult}
\lefteqn{
\mathcal{N}^{(c)} (p',p,q) + \mathcal{N}^{(d)} (p',p,q) 
=
- i \int d^4y^{\prime}~d^4y \, 
\left\{
e^{i p' y'} e^{-i (p+q) y} - e^{i (p'-q) y'} e^{-i p y} 
\right\}
}
\nn \\
&&{}
\bigg[ Q_d 
\bigg(
\Big\langle\psi_p(y^{\prime})J_d^{\mu}(0) \overline{\psi}_p(y) 
\Big\rangle_{G,q,\bar{q}}
+ Q_d
\bigg\langle\wick{2}{<1\psi_p(y^{\prime})>1{\overline{\psi}}_p(y)}
\mathrm{Tr}\left(\frac{1}{i}S^{(d)}_F(0,0;G)\gamma^\mu\right)
\bigg\rangle_G
\bigg)
\nn \\
&&{}
+ Q_u 
\bigg(
\Big\langle\psi_p(y^{\prime})J_u^{\mu}(0) \overline{\psi}_p(y) 
\Big\rangle_{G,q,\bar{q}}
+ Q_u 
\bigg\langle\wick{2}{<1\psi_p(y^{\prime})>1{\overline{\psi}}_p(y)}
\mathrm{Tr}\left(\frac{1}{i}S^{(u)}_F(0,0;G)\gamma^\mu\right)
\bigg\rangle_G
\bigg)
\bigg] \,.
\nn \\
\eea
With this we have expressed the sum of $\mathcal{N}^{(c)}$ 
and $\mathcal{N}^{(d)}$ in terms of the proton field $\psi_p$. 
Hence the four terms in square brackets can be 
understood as contributions with incoming and outgoing protons, 
so that we can apply the same argument as in the case of 
$\mathcal{N}^{(f)}$ above. We again interpret the integrals 
over $y'$ and $y$ as Fourier transformations and find that 
none of the two contributions arising from the two terms 
in curly brackets has simultaneous poles at $\slash{p}' = m_p$ 
and $\slash{p}= m_p$. Consequently, 
$\mathcal{N}^{(c)} +\mathcal{N}^{(d)}$ gives a 
vanishing contribution when inserted in (\ref{contrqMallg}), 
such that the sum of $\mathcal{M}^{(c)}$ and $\mathcal{M}^{(d)}$ 
is gauge invariant by itself, 
\be
\label{gaugeMcdresultapp}
q_\nu \left( 
\mathcal{M}^{(c)\mu\nu}_{s^{\prime}s}(p^{\prime},p,q) 
+ \mathcal{M}^{(d)\mu\nu}_{s^{\prime}s}(p^{\prime},p,q) 
\right)= 0 
\,.
\ee
Note that neither $\mathcal{M}^{(c)}$ nor $\mathcal{M}^{(d)}$ 
are separately gauge invariant. If we consider for example 
$\mathcal{M}^{(c)}$ we see that after inserting (\ref{derJc}) 
into (\ref{defNallg}) the delta function containing $x'$ will 
give rise to a term that has simultaneous poles at $\slash{p}' = m_p$ 
and $\slash{p}= m_p$ and hence will contribute to 
$q_\nu \mathcal{M}^{(c)\mu\nu}$. Only due to the cancellation 
of this contribution with the opposite one in 
$q_\nu \mathcal{M}^{(d)\mu\nu}$ is the sum of $\mathcal{M}^{(c)}$ 
and $\mathcal{M}^{(d)}$ separately gauge invariant. 

Next we consider the gauge invariance of the amplitude 
$\mathcal{M}^{(e)}$. From (I.\ref{AJe}) we find with (\ref{der2prop}) 
\bea
\label{derJe}
i \frac{\del}{\del x^\nu} \,\mathcal{J}^{(e)} &=&{} 
i \left[ \delta^{(4)}(x-y) - \delta^{(4)}(y'-x) \right]  
\Gamma_{\alpha^{\prime}\beta^{\prime}\gamma^{\prime}}
\bar{\Gamma}_{\alpha\beta\gamma} 
\nn \\
&&{}
\bigg\langle 
Q_u (Q_u + Q_d) \frac{1}{i} S^{(d)}_{F\,\gamma' \gamma}(y',y;G) 
\nn \\
&&{}\hspace*{0.6cm}
\bigg[
\left(\frac{1}{i} S^{(u)}_F(y',x';G) \gamma^\mu 
\frac{1}{i} S^{(u)}_F(x',y;G) \right)_{\alpha' \alpha}
\frac{1}{i} S^{(u)}_{F\,\beta' \beta}(y',y;G) 
\nn \\
&&{}\hspace*{0.8cm}
- (\alpha' \leftrightarrow \beta') - (\alpha \leftrightarrow  \beta) 
+ (\alpha' \leftrightarrow \beta', \alpha \leftrightarrow  \beta) 
\bigg] 
\nn \\
&&{}
+ 2 Q_u Q_d 
\left(\frac{1}{i} S^{(d)}_F(y',x';G) \gamma^\mu 
\frac{1}{i} S^{(d)}_F(x',y;G) \right)_{\gamma' \gamma}
\nn \\
&&{} \hspace*{0.6cm}
\bigg[ 
\frac{1}{i} S^{(u)}_{F\,\alpha' \alpha}(y',y;G) 
\frac{1}{i} S^{(u)}_{F\,\beta' \beta}(y',y;G) 
- (\alpha \leftrightarrow  \beta) 
\bigg] \bigg\rangle_G \,.
\eea
Inserting this in (\ref{defNallg}) and expressing it in terms 
of the proton field $\psi_p$ via (\ref{psiJupsibar}) 
and (\ref{psiJdpsibar}) we obtain 
\bea
\label{Neresult}
\lefteqn{
\mathcal{N}^{(e)} (p',p,q) 
=
- i \int d^4y^{\prime}~d^4y \, 
\left\{
e^{i p' y'} e^{-i (p+q) y} - e^{i (p'-q) y'} e^{-i p y} 
\right\}
}
\nn \\
&&{}
\bigg[ (Q_u + Q_d ) 
\bigg(
\Big\langle\psi_p(y^{\prime})J_u^{\mu}(0) \overline{\psi}_p(y) 
\Big\rangle_{G,q,\bar{q}}
\\
&&{} \hspace*{2.5cm}
+ Q_u 
\bigg\langle\wick{2}{<1\psi_p(y^{\prime})>1{\overline{\psi}}_p(y)}
\mathrm{Tr}\left(\frac{1}{i}S^{(u)}_F(0,0;G)\gamma^\mu\right)
\bigg\rangle_G
\bigg)
\nn \\
&&{}
+ 2 Q_u 
\bigg(
\Big\langle\psi_p(y^{\prime})J_d^{\mu}(0) \overline{\psi}_p(y) 
\Big\rangle_{G,q,\bar{q}}
+ Q_d
\bigg\langle\wick{2}{<1\psi_p(y^{\prime})>1{\overline{\psi}}_p(y)}
\mathrm{Tr}\left(\frac{1}{i}S^{(d)}_F(0,0;G)\gamma^\mu\right)
\bigg\rangle_G
\bigg)
\bigg] .
\nn 
\eea
Again in complete analogy to the discussion of $\mathcal{M}^{(f)}$ 
above we find that this expression does not contain terms which 
have simultaneous poles at $\slash{p}' = m_p$ and $\slash{p}= m_p$, 
and hence does not contribute to $q_\nu \mathcal{M}^{(e)\mu \nu}$ 
when inserted in (\ref{contrqMallg}), 
\be
\label{gaugeMeresultapp}
q_\nu \mathcal{M}^{(e)\mu\nu}_{s^{\prime}s}(p^{\prime},p,q) = 0 
\,, 
\ee
so that $\mathcal{M}^{(e)}$ is separately gauge invariant. 

Finally we turn to the amplitude $\mathcal{M}^{(g)}$. Here inspection 
of (I.\ref{AJg}) shows that the $x$-dependence of $\mathcal{J}^{(g)}$ 
and hence of $\mathcal{M}^{(g)}$ is fully contained in the same factor 
\be
\label{relevantfactorinJg}
\sum_{q} Q_{q} (-1) 
\mathrm{Tr}\left(\gamma^\nu\frac{1}{i}S_F^{(q)}(x,x;G)\right) 
\ee
which occurred already in $\mathcal{J}^{(b)}$. We can therefore 
in analogy to (\ref{Abpartial})-(\ref{fullcovderinMb}) derive that 
\be
q_\nu \mathcal{M}^{(g) \mu\nu}_{s^{\prime}s}(p^{\prime},p,q) =0 \,.
\ee
Thus also $\mathcal{M}^{(g)}$ is separately gauge invariant. 

\end{appendix}

\end{document}